\documentclass[a4paper,prl,reprint,superscriptaddress,notitlepage,floatfix,nofootinbib,longbibliography]{revtex4-2}
\usepackage[utf8]{inputenc}
\usepackage[T1]{fontenc}

\usepackage[top=2cm,bottom=2cm,left=1.75cm,right=1.75cm]{geometry}

\usepackage{amsmath}
\usepackage{amssymb}
\usepackage{stmaryrd}
\usepackage{fourier-orns}
\usepackage{commath}
\usepackage{mathrsfs}
\usepackage{braket}
\usepackage{xcolor}
\usepackage{bm}
\usepackage{csquotes}
\usepackage{graphicx}
\usepackage{booktabs}
\usepackage{enumitem}
\usepackage{siunitx}
\usepackage{microtype}
\usepackage[version=4]{mhchem}

\usepackage{tikz}

\usepackage{hyperref}
\usepackage{bookmark}

\definecolor{tab_blue}{HTML}{1F77B4}
\definecolor{tab_orange}{HTML}{FF7F0E}
\definecolor{tab_green}{HTML}{2CA02C}
\definecolor{tab_red}{HTML}{D62728}
\definecolor{tab_purple}{HTML}{9467BD}
\definecolor{tab_brown}{HTML}{8C564B}
\definecolor{tab_pink}{HTML}{E377C2}
\definecolor{tab_gray}{HTML}{7F7F7F}
\definecolor{tab_olive}{HTML}{BCBD22}
\definecolor{tab_cyan}{HTML}{17BECF}

\hypersetup{breaklinks=true,
  colorlinks=true,
  linkcolor=tab_blue,
  filecolor=tab_blue,
  urlcolor=tab_blue,
  citecolor=tab_blue,
}

\newcolumntype{L}[1]{>{\raggedright\let\newline\\\arraybackslash\hspace{0pt}}m{#1}}
\newcolumntype{C}[1]{>{\centering\let\newline\\\arraybackslash\hspace{0pt}}m{#1}}
\newcolumntype{R}[1]{>{\raggedleft\let\newline\\\arraybackslash\hspace{0pt}}m{#1}}

\def\dd{\text{d}}
\def\ii{\text{i}}
\def\ee{\text{e}}

\def\vec#1{\bm{#1}}

\let\strong\textbf

\definecolor{tab_blue}{HTML}{1F77B4}
\definecolor{tab_orange}{HTML}{FF7F0E}
\definecolor{tab_green}{HTML}{2CA02C}
\definecolor{tab_red}{HTML}{D62728}
\definecolor{tab_purple}{HTML}{9467BD}
\definecolor{tab_brown}{HTML}{8C564B}
\definecolor{tab_pink}{HTML}{E377C2}
\definecolor{tab_gray}{HTML}{7F7F7F}
\definecolor{tab_olive}{HTML}{BCBD22}
\definecolor{tab_cyan}{HTML}{17BECF}

\def\odd{\text{odd}}

\def\figurename{Fig.}
\makeatletter
\let\@orig@make@capt@title\@make@capt@title
\def\@make@capt@title#1#2{\@orig@make@capt@title{{\bf #1}}{#2}}
\makeatother

\makeatletter
\begin{document}

\title{Pattern formation by turbulent cascades}
\author{Xander M. de Wit}
\thanks{Both authors contributed equally to this work.}
\affiliation{Department of Applied Physics and Science Education, Eindhoven University of Technology, 5600 MB Eindhoven, Netherlands}
\author{Michel Fruchart}
\thanks{Both authors contributed equally to this work.}
\affiliation{Gulliver, ESPCI Paris, Université PSL, CNRS, 75005 Paris, France}
\affiliation{James Franck Institute, The University of Chicago, Chicago, IL 60637, USA}
\author{Tali Khain}
\affiliation{James Franck Institute, The University of Chicago, Chicago, IL 60637, USA}
\author{Federico Toschi}
\affiliation{Department of Applied Physics and Science Education, Eindhoven University of Technology, 5600 MB Eindhoven, Netherlands}
\affiliation{CNR-IAC, I-00185 Rome, Italy}
\author{Vincenzo Vitelli}
\affiliation{James Franck Institute, The University of Chicago, Chicago, IL 60637, USA}
\affiliation{Kadanoff Center for Theoretical Physics, The University of Chicago, Chicago, IL 60637, USA}

\begin{abstract}

Fully developed turbulence is a universal and scale-invariant chaotic state characterized by an energy cascade from large to small scales where the cascade is eventually arrested by dissipation~\cite{Cardy2008,Davidson2015,Falkovich2001,Alexakis2018,Eyink2006,Frisch1995}. 
In this article, we show how to harness these seemingly structureless turbulent cascades to generate patterns.
Pattern formation entails a process of wavelength selection, which can usually be traced to the linear instability of a homogeneous state~\cite{Cross1993}. By contrast, the mechanism we propose here is fully non-linear. It is triggered by a non-dissipative arrest of turbulent cascades: energy piles up at an intermediate scale, which is neither the system size nor the smallest scales at which energy is usually dissipated. 
Using a combination of theory and large-scale simulations, we show that the tunable wavelength of these cascade-induced patterns can be set by a non-dissipative transport coefficient called odd viscosity, ubiquitous in chiral fluids ranging from bio-active to quantum systems~\cite{Avron1998,oddreview,Berdyugin2019,Morrison1984,Saarloos2023}. 
Odd viscosity, which acts as a scale-dependent Coriolis-like force, leads to a two-dimensionalization of the flow at small scales, in contrast with rotating fluids where a two-dimensionalization occurs at large scales~\cite{Alexakis2018}.
Beyond odd-viscosity fluids, we discuss how cascade-induced patterns can arise in natural systems including atmospheric flows~\cite{Diamond2005,Sukoriansky2007,Berloff2009,Chekhlov1996,Rhines1979,Legras1999,Grianik2004}, stellar plasma such as the solar wind~\cite{Squire2022,Meyrand2021,Miloshevich2021}, or the pulverization and coagulation of objects or droplets where mass rather than energy cascades~\cite{Krapivsky2010,Testik2007,Friedlander2000}.
\end{abstract}

\maketitle

\begin{figure*}
    \centering
    \includegraphics[width=\textwidth]{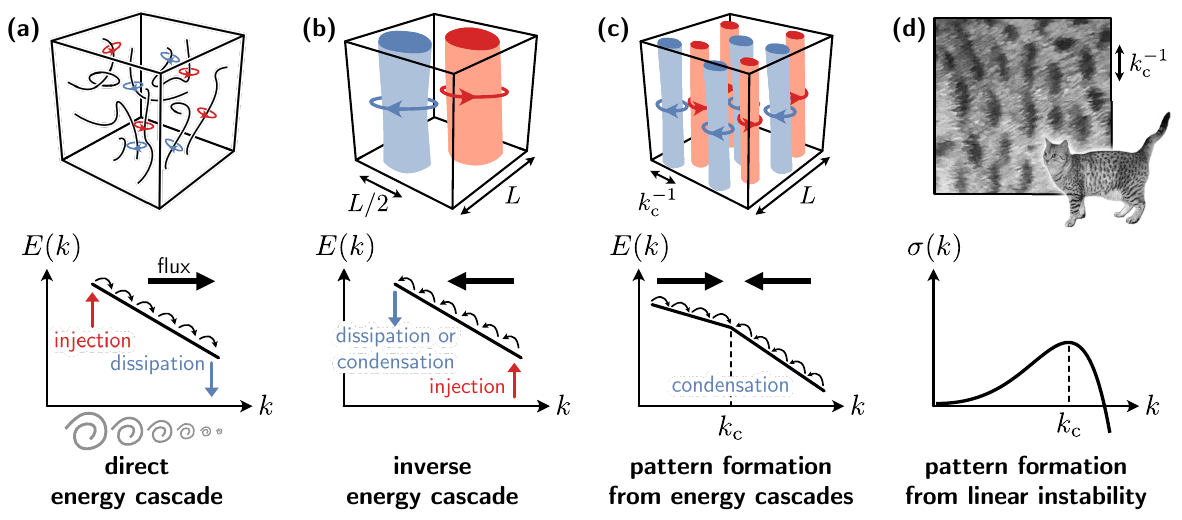}
    \caption{\label{figure_pattern_cascade} \textbf{Cascade-induced pattern formation.}
    (a) Direct energy cascade: in a turbulent 3D fluid, energy injected at large scales (red arrow) is transferred to smaller and smaller length scales (black arrows) to microscopic length scales where dissipation occurs (blue arrow), as captured by the so-called energy spectrum $E(k)$ which describes how much kinetic energy is contained in modes with wavenumber $k$. The energy transfer across scales can be traced to vortices breaking up into smaller and smaller vortices up to dissipative scales.
    This mechanism is intrinsically nonlinear: it relies on triadic couplings between the modes of the system.
    (b) Inverse energy cascade: in a turbulent 2D fluid, or in a rotating 3D fluid, there is instead a transfer of energy from the scale where energy is injected (red arrow) to larger and larger scales, where energy is either dissipated or piles up at the largest scale available (blue arrow), the size of the system. Correspondingly, vortices merge together until only a single positive vortex and a single negative vortex remain, both of which have approximately half the size $L$ of the system.
    Inverse cascades can also arise in 3D from mirror symmetry breaking~\cite{Alexakis2018,Biferale2012,Slomka2017} or by imposing large-scale shear~\cite{Xia2011}.
    (c) In a hypothetical situation where a direct and an inverse cascades can be put together in the right order (black arrows in the figure), energy will be transferred to an intermediate length scale $k_{\text{c}}^{-1}$, leading to the appearance of structures with a characteristic size $k_{\text{c}}^{-1}$ independent of the size $L$ of the system. This non-linear wavelength selection mechanism relying on combined turbulent cascades can be seen as an instance of pattern formation.
    (d) Standard pattern formation from a linear instability: the wavelength $k_{\text{c}}^{-1}$ corresponding to the most unstable linear mode (i.e., the one with the largest growth rate $\sigma(k)$) is selected.  As an example, we have shown the coat pattern of a cat.
    }
\end{figure*}

Fully-developed turbulence is a highly chaotic non-equilibrium state in which energy is transferred across scales through a non-linear mechanism known as a turbulent cascade~\cite{Cardy2008,Davidson2015,Falkovich2001,Alexakis2018,Eyink2006,Frisch1995}. While cascades occur in diverse contexts ranging from optical fibers to solid plates~\cite{Zakharov2012,Nazarenko2011,Galtier2022,Newell2011}, their most iconic manifestation is in fluids. Heuristically, large eddies, typically created by the injection of energy at macroscopic scales, break-up into smaller and smaller eddies. This energy transfer towards small scales, called a direct or forward cascade, is eventually arrested by dissipation (Fig.~\ref{figure_pattern_cascade}a). Away from the scales at which energy is injected and dissipated, turbulence is universal and scale invariant.

We start by asking an almost paradoxical question: can we harness turbulence to generate patterns? Our approach to tackle this task rests on the simple observation that different classes of turbulent cascades exist~\cite{Alexakis2018}. For example, turbulence in two-dimensional and rotating fluids has a tendency to transfer energy towards larger scales in what is known as an inverse cascade (Fig.~\ref{figure_pattern_cascade}b).
Here, we consider what happens when a direct cascade is combined with an inverse cascade as shown in Fig.~\ref{figure_pattern_cascade}c. Energy is transferred to an intermediate length scale $k_{\text{c}}^{-1}$ ($k$ are wavenumbers, so their inverses are lengths) both from smaller and larger scales, depending on where energy is injected. As energy accumulates around that scale, structures emerge with characteristic size $k_{\text{c}}^{-1}$, which is neither the size of the system nor the smallest scales at which dissipation typically occurs. This \enquote{spectral condensation} at intermediate scales requires the mechanism responsible for arresting both cascades to be non-dissipative. As we shall see, Nature has found an elegant solution to this problem: a viscosity that does not dissipate energy~\cite{oddreview,Saarloos2023}, variously known as odd viscosity~\cite{Avron1998}, Hall viscosity~\cite{Berdyugin2019} or gyroviscosity~\cite{Morrison1984}.

Before delving into potential realizations, let us compare and contrast this scenario with the textbook picture of pattern formation represented in Fig.~\ref{figure_pattern_cascade}d. In its simplest incarnation, pattern formation originates from the linear instability of a homogeneous system: the length scale $k_{\text{c}}^{-1}$, corresponding to the maximum of the growth rate $\sigma(k)$, is selected because the corresponding mode grows faster, and sets the characteristic size of the emerging pattern. While non-linearities are important in saturating the growth and selecting the precise shape of the pattern, they only play a role once the linear instability has set in. 
This linear mechanism is at play in many areas of science~\cite{Cross1993,Saarloos2023}.
By contrast, in the mechanism shown in Fig.~\ref{figure_pattern_cascade}c, it is the non-linear interaction between modes that gives rise to the turbulent cascade. 

\begin{figure*}
    \centering
    \includegraphics[width=\textwidth]{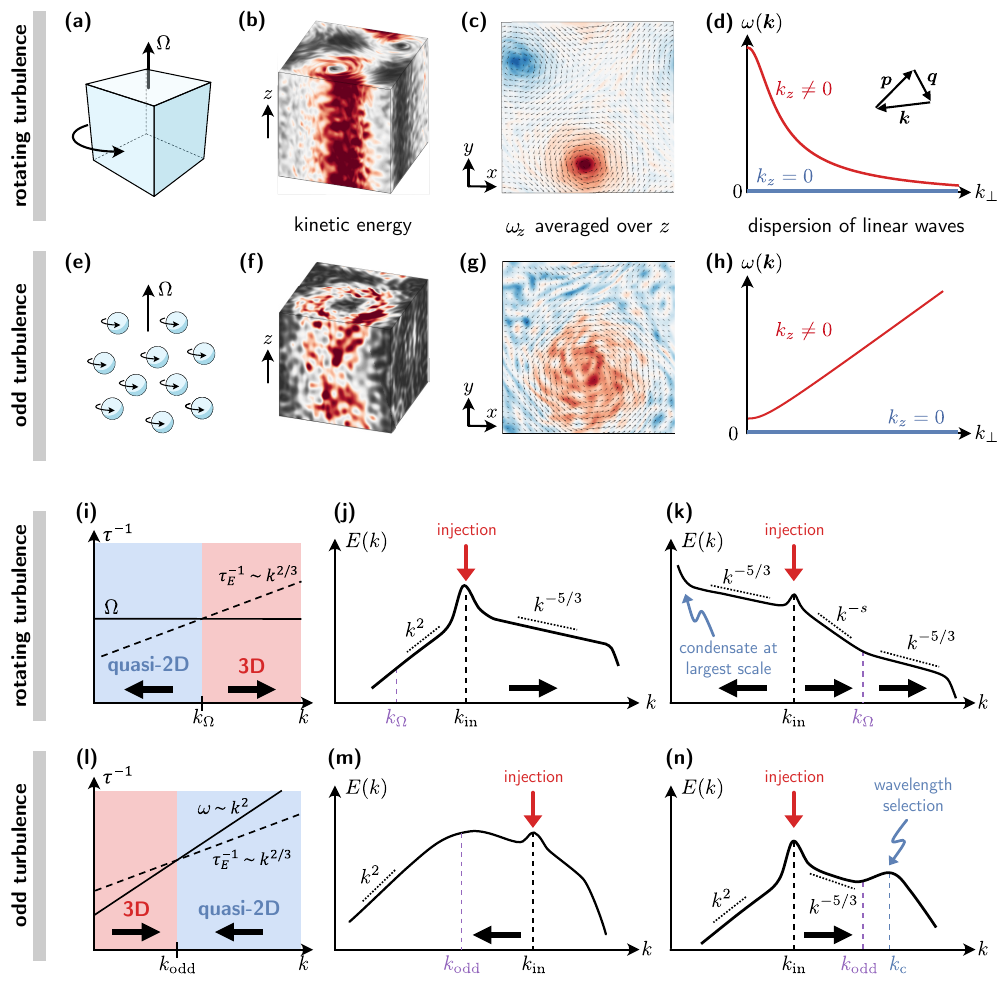}
    \caption{\label{figure_two_dimensionalization} \textbf{Rotating versus odd turbulence.}
    We compare turbulence in a fluid rotating with high frequency $\bm{\Omega}$ (panels a-d and i-k) and a fluid with high odd viscosity (panels e-h and l-n). 
    (a-h) 
    Both fluids are characterized by a rotation direction $\bm{\Omega}$ (along $z$), making them anisotropic and chiral.
    The rotation is global in rotating fluids (panel a). 
    It is induced at microscopic scales in odd fluids, for instance by particles that all spin in the same direction $\bm{\Omega}$ (panel e).
    In both cases, the flow becomes two-dimensional, with column-like structures aligned with $\bm{\Omega}$, as seen in the kinetic energy (panels b and f) and the $z$-averaged vertical vorticity $\langle\omega_z\rangle_z$ (panels c and g) obtained from simulations.
    The two-dimensionalization originates from the decorrelation by waves in the fluid (inertial waves in panel d and odd waves in panel h) of the triads by which energy transfer occurs (inset of panel d).
    Modes with $k_z \neq 0$ have finite frequencies (red lines) and quickly decorrelate, while modes with no vertical variation ($k_z = 0$, blue lines) all have $\omega = 0$.
    (i-n) 
    To predict the direction of the cascades (black arrows), we compare the inverse frequency of waves with the time over which energy transfer takes place (the eddy turnover time $\tau_E^{-1} \sim k^{2/3}$).
    In rotating fluids (panel i), the flow is quasi-2D at small wavenumbers (blue region) and isotropic (3D) at large wavenumbers (once $\tau_E^{-1} > \Omega$, red region). 
    In odd fluids (panel l), we expect the flow to be quasi-2D at large wavenumbers (blue region), and isotropic at low $k$ (once $\tau_E^{-1} > \tau_\odd^{-1}$, red region). 
    The crossover point defines a characteristic scale $k_\odd$, in analogy with the Zeman scale $k_\Omega$ in rotating fluids. 
    We sketch cascades in the energy spectra when the injection scale is smaller (panel j,m) and larger (panel k,n) than the characteristic scale.
    In rotating fluids, there is a direct cascade of energy above the rotation (Zeman) scale (panel j), and an inverse cascade below (panel k). 
    This situation is known as a split cascade~\cite{Alexakis2018}.
    In odd fluids, we expect the situation to be reversed: energy cascades directly for wavenumbers below $k_\odd$ (panel n) and inversely above (panel m), causing a pile up of energy at the odd viscosity length scale and arresting both cascades.
    The pileup is saturated by viscous dissipation, leading to a bump in the energy spectrum at another scale $k_{\text{c}}$.
    }
\end{figure*}

In order to realize the mixed cascade of Fig.~\ref{figure_pattern_cascade}c, one first needs to turn a direct cascade into an inverse cascade. 
This can be achieved by simply rotating the fluid at high velocities~\cite{Davidson2015,Alexakis2018}, as illustrated in Fig.~\ref{figure_two_dimensionalization}. The Coriolis force
$\vec{f}_{\vec{\Omega}} = 2  \rho  \vec{v} \times \vec{\Omega}$ (in which $\vec{v}(t, \vec{x})$ is the velocity field, $\rho$ is the density, $\vec{\Omega}$ is the rotation vector and $\times$ is the vector product) tends to align vortex lines with the rotation axis, without injecting nor dissipating energy.
As the rotation speed increases, the vortex tangle becomes more and more polarized, which induces a two-dimensionalization of the flow. This prevents vortex stretching and leads to an inverse energy cascade similar to the case of 2D fluids. Eventually, the energy condenses into two vortices of opposite vorticity. 
As the inverse cascade proceeds all the way to the largest scales, this condensation occurs only at the size of the system (Fig.~\ref{figure_pattern_cascade}b and Fig.~\ref{figure_two_dimensionalization}b-c). 

\medskip
\noindent\textbf{Turbulence with odd viscosity.} 
For our purposes, we need an inverse cascade at small scales (large wavevectors) only, see Fig.~\ref{figure_pattern_cascade}c. 
This could be produced by a scale-dependent version of the Coriolis force that would involve gradients of the velocity, in a way similar to a viscosity term, so that it is negligible at large scales.
To do so, we consider a situation where rotation is induced at microscopic scales, for instance by spinning particles large enough to be inertial, see Fig.~\ref{figure_two_dimensionalization}e.
It turns out that such a system has an antisymmetric part in its viscosity tensor $\eta_{i j k \ell} \neq \eta_{k \ell i j}$, known as odd viscosity.
Like the Coriolis force, the antisymmetric, or odd, part of the viscosity tensor does not contribute to energy dissipation or injection as it drops out from the energy balance equation \cite{Khain2022}.
Odd viscosities arise in various experimental systems breaking time-reversal and inversion symmetry at the microscopic scale~\cite{Avron1998,oddreview} from magnetized polyatomic gases~\cite{Beenakker1970} to magnetized graphene~\cite{Berdyugin2019} and active colloids~\cite{Soni2019}.

To mathematically account for the effect of odd viscosity, we consider a simple extension of the Navier-Stokes equations
\begin{align}
\label{odd_ns_simplified}
D_t \vec{v} = - \vec{\nabla} {P} + 
\nu \Delta \vec{v} +
\nu_\odd \vec{e}_z \times \Delta \vec{v} + \vec{f}(t, \vec{x})
\end{align}
with the incompressibility condition $\vec{\nabla}\cdot \vec{v} = 0$.
Here, $D_t = \partial_t + \vec{v}\cdot\vec{\nabla}$ is the convective derivative and $\vec{f}$ is an external forcing representing energy injection, $P$ is the pressure, $\nu=\eta/\rho$ is the familiar shear viscosity, $\nu_\odd = \eta_\odd/\rho$ is a particular combination of odd viscosities 
(see SI for the general case),
and $\vec{e}_z$ is the unit vector along $z$ (the direction set by the magnetic field or rotation axis).
Equation~\eqref{odd_ns_simplified} can be seen as a non-linear diffusion equation for momentum with an anti-symmetric cross-diffusion coefficient $\nu_\odd$.
The resulting odd viscosity term $\nu_\odd \vec{e}_z \times \Delta \vec{v}$ (or $-\nu_\odd k^2 \vec{e}_z \times \vec{v}(\vec{k})$ in wavenumber space) is very similar to the Coriolis force. Both are non-dissipative and anisotropic, see Methods.
The additional Laplacian ensures that the action of $\nu_\odd$ vanishes for small wavenumbers, as needed to arrest the turbulent cascade at intermediate scales. 

\medskip
\noindent\textbf{Two-dimensionalization: energy transfer and odd waves.} 
Direct numerical simulations of the Navier-Stokes equations (Methods) in Fig.~\ref{figure_two_dimensionalization} confirm that strong odd viscosity fluids can exhibit features similar to quickly rotating fluids such as Taylor columns and quasi-two-dimensionalization \cite{Davidson2015}, compare Fig.~\ref{figure_two_dimensionalization}a-c and Fig.~\ref{figure_two_dimensionalization}e-g.
The two-dimensionalization of the flow can be heuristically justified using a generalization of the Taylor-Proudman argument to odd fluids, which yields $\partial_z \Delta \vec{v} = \vec{0}$ (see SI). 

To account for the role of the convective term, we 
now turn to the analysis of the non-linear energy transfer, which governs the redistribution of energy among scales~\cite{Davidson2015,Alexakis2018}.
The distribution of energy among scales is described by the energy spectrum $E(k, t) = \frac{1}{2} \left\langle \lVert\vec{v}(t, \vec{k})\rVert^2 \right\rangle_{k\leq\lVert\vec{k}\rVert<k+1}$ averaged over a spherical shell.
Its evolution is captured by the energy balance equation $\partial_t E = - T -\nu k^2 E + F$, in which $F$ represents the forcing and $T$ the non-linear energy transfer between scales.

As odd viscosity is non-dissipative, it does not act as an energy source or sink.
However, it has an indirect effect on the energy transfer, because it induces waves in the fluid, that oscillate at a frequency $\omega(\bm{k}) = \pm \nu_\odd k_z \abs{k}$ (see Fig.~\ref{figure_two_dimensionalization}h and SI). 
The transfer described by $T$ arises through interactions between three modes with wavenumbers $\vec{k}$, $\vec{p}$ and $\vec{q}$ that satisfy $\vec{k}+\vec{p}+\vec{q}=\vec{0}$ (called a triad; see inset of Fig.~\ref{figure_two_dimensionalization}d).
Because of the odd waves described above, the different modes in a triad quickly go out of phase with each other. 
This suppresses the non-linear energy transfer, except for modes with $k_z = 0$ which all have $\omega = 0$ (blue line in Fig.~\ref{figure_two_dimensionalization}h) and therefore do not decorrelate. 
These two-dimensional modes form a so-called slow (or resonant) manifold that contributes to most of the non-linear energy transfer, giving rise to an inverse cascade.
This can be seen from the expression of the energy transfer $T \sim \ee^{\ii[\omega(\bm{k})+\omega(\bm{p})+\omega(\bm{q})]t}$, see Methods for details (recall that the time average of $\ee^{\ii \omega t}$ vanishes when $\omega \neq 0$).

\medskip
\noindent\textbf{Scaling theory of the arrested cascade.} In a turbulent flow, the lifespan of a typical eddy is called the turnover time $\tau_E$, and its inverse is called the eddy turnover frequency. The processes transferring energy across scales occur over a few turnover times.
In order to assess whether odd waves suppress the energy transfer, we compare the eddy turnover frequency $\tau_E^{-1}$ with the frequency $\omega(\vec{k})$ of odd waves.
Assuming $k_z\sim k$ (motivated by the isotropization at small $k$), we look for the scale $k_\odd$ such that $\omega(k = k_\odd) = \tau_E^{-1}(k = k_\odd)$ (Fig.~\ref{figure_two_dimensionalization}l).
We estimate the eddy turnover frequency $\tau_E^{-1} = k v_k \sim k^{2/3} \epsilon^{1/3}$ from the rate of dissipation of energy at small scales $\epsilon$ using the Kolmogorov scaling valid at $k \ll k_\odd$, and find
\begin{equation}
    \label{kodd}
    k_\odd \equiv \epsilon^{1/4} \nu_\odd^{-3/4}.
\end{equation}

\begin{figure*}
    \centering
    \includegraphics[width=0.9\textwidth]{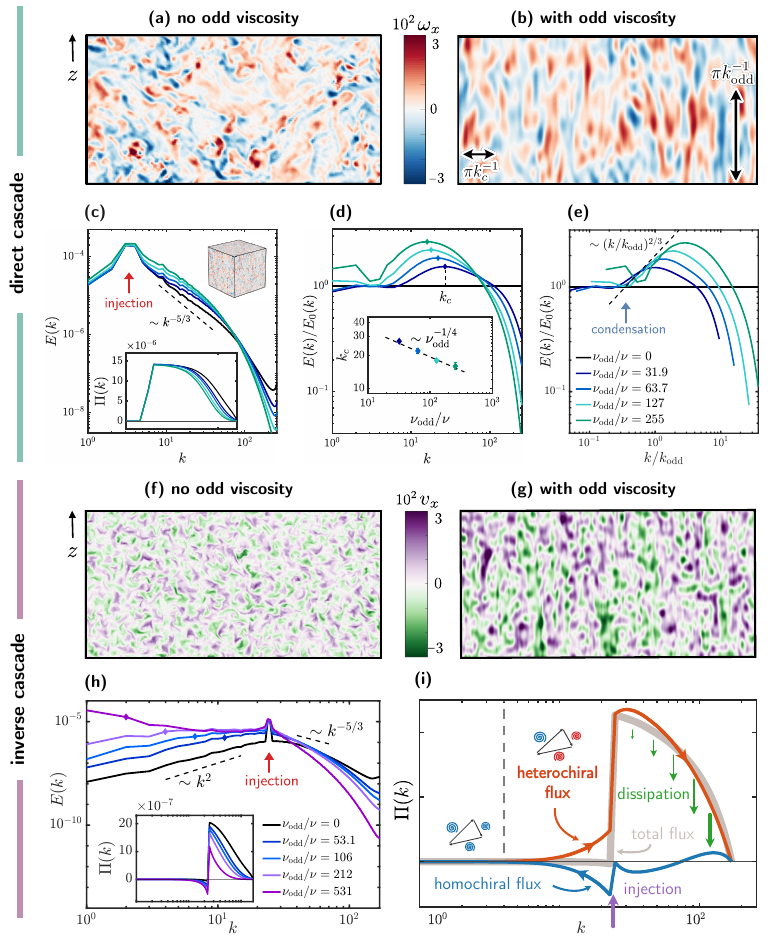}
    \caption{\label{figure_direct_spectra}
    \textbf{Odd waves induce wavelength selection and flux loops.} 
    We perform direct simulations of the Navier-Stokes equation without and with odd viscosity. 
    In panels a-e, energy is injected at wavenumbers $k_{\text{in}}<k_\odd$ and the direct cascade dominates.
    In panels f-i, $k_{\text{in}}>k_\odd$ and the inverse cascade dominates.
    (a-b) Slices of the in-plane component $\omega_x$ of the vorticity with $k_{\text{in}}<k_\odd$.
    Without odd viscosity (panel a), vortices of all sizes are present. 
    With odd viscosity (panel b, where $\nu_\odd/\nu=255$), characteristic horizontal and vertical scales $\sim k_{\text{c}}^{-1}$ and $\sim k_{\text{odd}}^{-1}$ emerge (black arrows). 
    This wavelength selection is due to the arrest of the direct cascade near $k_\odd$.
    (c) Energy spectrum $E(k)$ and flux $\Pi(k)$ (inset) obtained from simulations, for different values of odd viscosity (legend in panel e). 
    Energy flows from the injection scale $k_{\text{in}}$ (red arrow) towards larger $k$, as evidenced by the positive energy flux $\Pi(k)$. 
    The cascade is progressively arrested near $k_\odd$ and energy piles up, triggering viscous dissipation.
    (d) The relative energetic amplification/attenuation due to odd viscosity is measured by the compensated spectrum $E(k)/E_0(k)$ (where $E_0(k)$ is the energy spectrum without odd viscosity), which peaks at a scale $k_{\text{c}}$ (diamonds in panel d). 
    The peak position $k_{\text{c}}$ decreases with odd viscosity (inset of panel d), as predicted by scaling arguments (dashed line, see Eq.~\eqref{scaling_kc}). 
    (e) Plotting the compensated spectra against $k/k_\odd$ confirms that condensation begins near $k_\odd$ (blue arrow) and follows the scaling prediction (dashed line, see Eqs.~\eqref{kodd}-\eqref{Ek_scaling_aboveko}).
    (f-g) Slices of the in-plane velocity component $v_x$ when $k_{\text{in}}>k_\odd$.
    We visualize $v_x$ instead of $\omega_x$ to emphasize the large scales.
    Without odd viscosity (panel f), structures of all scales are present, dominated by the injection scale. 
    With odd viscosity (panel g, where $\nu_\odd/\nu = 212$), secondary features with larger sizes appear due to the arrest of the inverse cascade.
    (h) Energy spectrum $E(k)$ and flux $\Pi(k)$ obtained from the simulations (diamonds indicate $k_\odd$).
    (i) The inverse cascade is arrested by a flux-loop mechanism, as evidenced by a decomposition of the flux in homochiral (blue) and heterochiral (red) channels that correspond respectively to triads with different/same signs of helicity. In this panel, we have used hyperdissipation in the simulations to highlight the flux loop (see Extended Data Fig.~\ref{figure_numerics_extended}a iv).
    }
\end{figure*}

\noindent When $k \gg k_\odd$, the effect of odd viscosity is important: the contribution of 3D triads to the energy transfer averages to zero over the lifespan of a typical eddy, and we expect quasi-2D behavior. In contrast, when $k \ll k_\odd$, the effect of odd viscosity is negligible and we expect normal 3D behavior. This is summarized in Fig.~\ref{figure_two_dimensionalization}l.
As a consequence, both a direct and an inverse cascade are arrested when they approach the odd viscosity wavenumber $k_\odd$, due to the inherent tendency to cascade in the opposite direction beyond that wavenumber (Fig.~\ref{figure_two_dimensionalization}m,n). 
The direct cascade dominates when energy is injected below $k_\odd$ (Fig.~\ref{figure_two_dimensionalization}n), while the inverse cascade dominates when energy is injected above $k_\odd$ (Fig.~\ref{figure_two_dimensionalization}m).

Fig.~\ref{figure_two_dimensionalization} compares the cases of odd and rotating fluids. In the case of rotating turbulence~\cite{Biferale2016,Buzzicotti2018,Deusebio2014,Smith1999,Alexakis2018}, odd waves are replaced by so-called inertial waves with dispersion $\omega_{\vec{\Omega}}(\vec{k}) = \pm 2 \vec{\Omega} \cdot \vec{k}/k$ (Fig.~\ref{figure_two_dimensionalization}d), and the scale $k_\odd$ is replaced by the so-called Zeman scale $k_\Omega=\Omega^{3/2}\epsilon^{-1/2}$ \cite{Zeman1998,Mininni2012}. 
Comparing panels i and l of Fig.~\ref{figure_two_dimensionalization} shows that, crucially, the order of the 3D direct cascade and the quasi-2D inverse cascade are permuted in rotating and odd fluids. 
As a consequence, the fluxes are convergent in the case of odd turbulence, while they are divergent in the case of rotating turbulence, and the pattern formation effect is thus only observed in the former scenario.

\medskip
\noindent\textbf{Wavelength selection in the energy spectrum.}
We now refine the intuitive picture in Fig.~\ref{figure_pattern_cascade}c and show that two length scales, rather than a single one, are implicated in cascade-induced pattern formation.
To do so, we develop a scaling theory based on dimensional analysis~\cite{Kraichnan1965,Zhou1995,Zeman1998,Chakraborty2007,Biferale2016,Zhou2004}, focusing on the case where energy is injected at large scale $k_{\text{in}} < k_\odd$ and the direct cascade dominates.
As the cascade is generated by nonlinear triadic interactions,
one expects that it is related to the correlation time $\tau_{3}(k)$ of the triadic interactions.
Assuming energy conservation and locality in scale of the cascade, dimensional analysis leads to
$E(k) = C \, [\epsilon/\tau_{3}(k)]^{1/2} \, k^{-2}$
in which $C$ is a constant~\cite{Kraichnan1965,Zhou1995,Zhou2004}. 

In the absence of odd viscosity, or when it is negligible ($k \ll k_\odd$), the only time scale available is the eddy turnover time $\tau_{E}(k) = [k v_k]^{-1} = k^{-3/2} E^{-1/2}(k)$, leading to the Kolmogorov spectrum 
\begin{equation}
    \label{Ek_scaling_belowko}
    E(k) \sim \epsilon^{2/3} k^{-5/3}
    \qquad \text{($k \ll k_\odd$).}
\end{equation}
When odd viscosity is dominant ($k \gg k_\odd$), the relevant time scale is given by the frequency of odd waves $\omega (k) = \nu_\odd k^{2}$ (again, we assume $k_z \sim k$), leading to
\begin{equation}
    \label{Ek_scaling_aboveko}
	E(k) \sim \epsilon^{1/2} \nu_\odd^{1/2} \, k^{-1} \qquad \text{($k \gg k_\odd$)}.
\end{equation}
As a point of comparison, the relevant time scale is $\Omega^{-1}$ in rotating turbulence, so this argument leads to a different scaling $E \sim k^{-2}$~\cite{Zhou1995,Biferale2016}.

The preceding argument shows that the cascade starts to get arrested when it reaches $k_\odd$, leading to an amplification of the modes with wavenumbers $k>k_\odd$. The relative amplification due to odd viscosity can be described by the ratio between the modified spectrum $E(k)$ given by Eq.~\eqref{Ek_scaling_aboveko} and the Kolmogorov spectrum $E_0(k)$ given by Eq.~\eqref{Ek_scaling_belowko} that would occur in the absence of odd viscosity. Ignoring first the effect of dissipation, this yields
$E/E_0 = 1$ for $k \ll k_\odd$
and
$E/E_0 \sim (k/k_\odd)^{2/3}$ for $k \gg k_\odd$.
As energy piles up at wavevectors larger than $k_\odd$, it is eventually saturated by viscous dissipation, leading to a maximum in $E/E_0$ after which the spectrum decays dissipatively. 

By balancing energy injection and viscous dissipation, we can find the position $k_{\text{c}}$ of the maximum as (see Methods)
\begin{equation}
    \label{scaling_kc}
    k_{\text{c}} 
    \sim \epsilon^{1/4} 
    \nu^{-1/2}
    \nu_\odd^{-1/4}.
\end{equation}
The magnitude of the spectral condensation can be estimated as the height of the peak $E(k_\text{c})/E_0(k_\text{c}) \sim (\nu_\odd/\nu)^{1/3}$.
The ratio $\nu_\odd/\nu$ thus controls the height of the peak.
According to kinetic theory calculations corroborated by experimental measurements, this ratio increases linearly with the time-reversal breaking field (e.g. the spinning speed in Fig.~\ref{figure_two_dimensionalization}e or the applied magnetic field), see Methods.

The overall picture, summarized in Fig.~\ref{figure_two_dimensionalization}n, involves the two length scales $k_\odd$ and $k_{\text{c}}$ defined in Eqs.~\eqref{kodd} and \eqref{scaling_kc}. 
As the direct cascade (black arrow) approaches $k_\odd$ (purple dashed line), it is gradually arrested: the rate of energy transfer from scale to scale decreases as $k$ increases. 
This leads to the condensation of kinetic energy in wavenumbers $k>k_\odd$. In turn, the amplification of these modes leads to an increase in viscous dissipation, and the energy spectrum exhibits a maximum deviation from the Kolmogorov spectrum at a characteristic wavenumber $k_{\text{c}}$ (blue dashed line).

\medskip
\noindent\textbf{Simulations of the odd Navier-Stokes equations.}
To put this scenario to test, we numerically integrate the Navier-Stokes equation \eqref{odd_ns_simplified} using a parallelized pseudo-spectral solver (see Methods). 
In a normal fluid, eddies of all sizes can be found in the statistical steady-state (Fig.~\ref{figure_direct_spectra}a). In the presence of odd viscosity, the turbulent state selects a dominant scale, as shown in the visualizations of the vorticity field in Fig.~\ref{figure_direct_spectra}b. The features manifest as  vertically aligned, intermediate scale structures, as expected from the quasi-2D nature of the system.
A direct cascade occurs when energy is injected at large scales ($k_{\text{in}} < k_\odd$).
As predicted, we find that this turbulent cascade is arrested due to odd viscosity. This can be seen from the net flux of energy $\Pi(k)=\sum_{k'<k}T(k')$, that gradually decays as $k$ passes $k_\odd$
(Fig.~\ref{figure_direct_spectra}c, inset). 

This gradual arrest of the cascade near $k_\odd$ leads to spectral condensation at intermediate scales. Quantitatively, the spectral condensation and wavelength selection can be better appreciated from the relative energetic amplification of each mode $E(k)/E_0(k)$ shown in Fig.~\ref{figure_direct_spectra}d. Rescaling the wavenumbers by $k_\odd$ (Fig.~\ref{figure_direct_spectra}e), we observe an approximate collapse of the curves compatible with the scaling predicted in the previous paragraph.
The condensation peaks around a wavenumber $k_{\text{c}}$, which we can compare quantitatively with our scaling prediction Eq.~\eqref{scaling_kc}, see Fig.~\ref{figure_direct_spectra}d (inset). 
An extension of our scaling theory taking into account the anisotropy of the flow (Methods) reveals the visual meaning of the two length scales involved in cascade-induced patterns: $k_c^{-1}$ manifests predominantly in the horizontal direction, while the typical vertical scale is mainly given by $k_\odd^{-1}$ (black arrows in Fig.~\ref{figure_direct_spectra}b).  

\medskip
\noindent\textbf{Flux loops and helicity conservation.}
When energy is injected at $k_{\text{in}}>k_\odd$ (Fig.~\ref{figure_direct_spectra}f-i), we expect an inverse cascade to be arrested by odd viscosity. 
This is indeed the case, as evidenced by snapshots of the steady-state, that exhibit scales larger than the injection scale (Fig.~\ref{figure_direct_spectra}g).
In contrast with the case of the arrested direct cascade, here energy gets piled up at large scales, where viscous dissipation is not an effective saturation mechanism.
Instead, what prevents energy blow-up is a mechanism known as flux-loop cascade~\cite{Alexakis2018}: energy goes from the small injection scale to large scales and then back to even smaller scales where it is dissipated.
To see that, we decompose the energy flux into \enquote{heterochiral} (red) and \enquote{homochiral} (blue) channels, that correspond respectively to triads with different/same signs of helicity. Helicity is the volume integral of $\vec{v}\cdot\vec{\omega}$ where $\vec{\omega} = \nabla \times \vec{v}$ is the vorticity and it is an invariant of the inviscid Navier-Stokes equation. The conservation of helicity is not affected by odd viscosity, see Methods.
As shown in Fig.~\ref{figure_direct_spectra}i, the {heterochiral} flux (red) tends to cascade directly, while the {homochiral} flux (blue) tends to cascade inversely. Below the injection scale, both fluxes cancel exactly, leading to a vanishing net flux (gray line). 
In the case of the inverse cascade, the resulting pattern is less visible than in the direct cascade, because the energy is deposited over a more broadband range $k_\odd < k < k_{\text{in}}$.

\begin{figure*}
    \centering
    \includegraphics[width=0.95\textwidth]{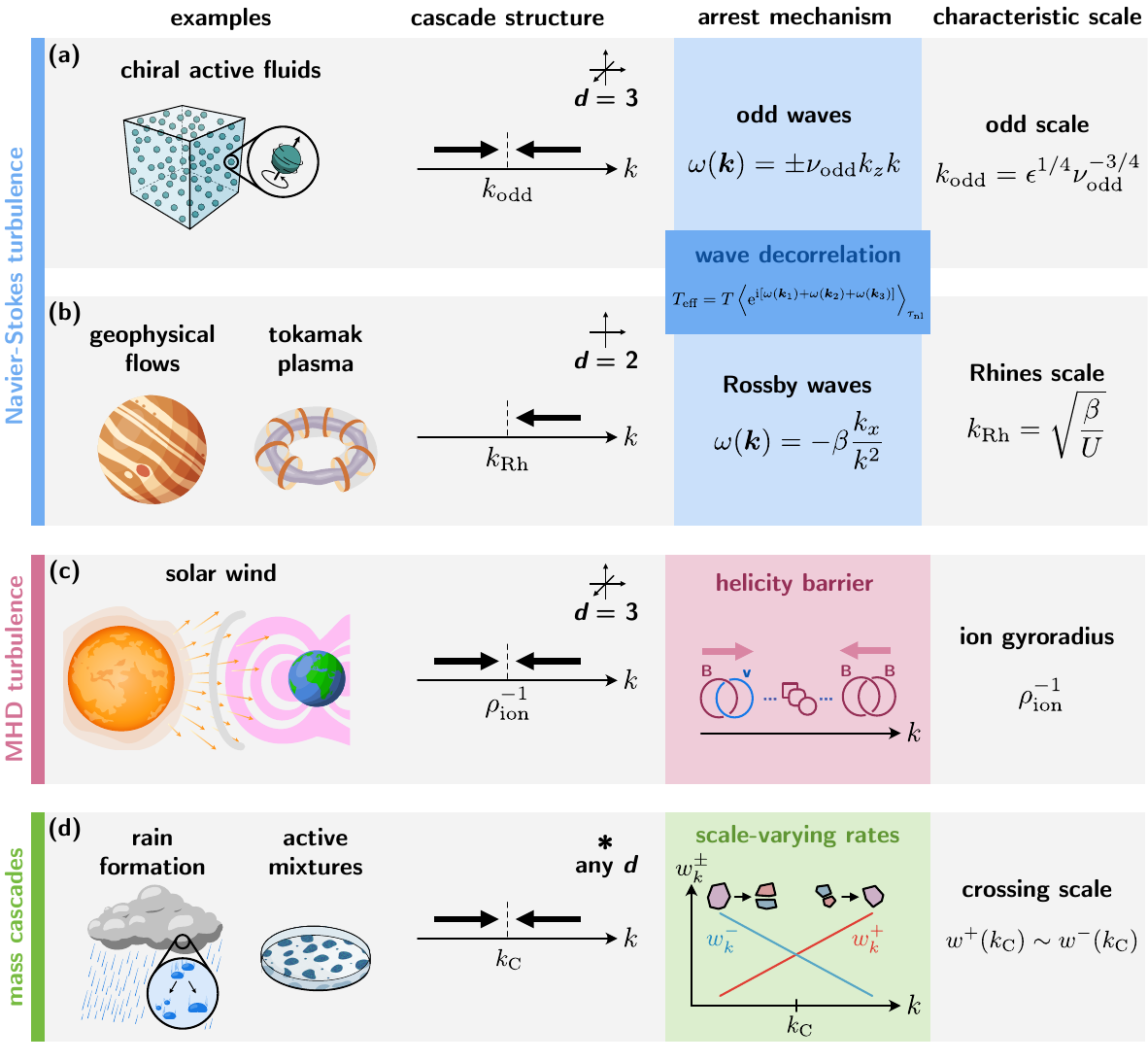}
    \caption{\label{figure_examples_wmc}
    \textbf{Cascade-induced pattern formation across domains.} 
    Cascade-induced scale selection can occur in systems ranging from Navier-Stokes turbulence (panels a-b) and magnetohydrodynamics (MHD) turbulence (panel c) to mass cascades (panel d).
    (a) Chiral active fluids are an example of fluids with odd viscosity. As demonstrated in Fig.~\ref{figure_direct_spectra}, these fluids are expected to exhibit a double arrested cascade at high enough $\nu_\odd/\nu$ and Reynolds number (see Methods for a discussion of orders of magnitude).
    We have interpreted this phenomenon as the result of a decorrelation of wavenumber triads by odd waves.
    (b) In two-dimensional geophysical flows and plasma, an arrested inverse cascade associated with wavelength selection occurs~\cite{Diamond2005,Sukoriansky2007,Berloff2009,Chekhlov1996,Rhines1979,Legras1999,Grianik2004}.
    It can be seen as the consequence of the decorrelation of triads by Rossby (or drift) waves, which set the characteristic scale $k_{\text{Rh}}$ known as the Rhines scale.
    (c) A double arrested cascade has been predicted in the solar wind, based on the properties of inviscid invariants of finite Larmor radius MHD~\cite{Squire2022,Meyrand2021,Miloshevich2021}. 
    This mechanism, known as a helicity barrier, relies on the change of nature of an inviscid invariant, which interpolates between cross-helicity and magnetic helicity (these quantities cascade in opposite directions).
    (d) Scale selection can also occur in mass cascades, ranging from the stationary distribution of rain drop sizes that would occur in steady-state conditions~\cite{Testik2007} to smoke aerosols~\cite{Friedlander2000}.
    This arises from the balance between coalescence and breakup of the droplets, which effectively have scale-varying rates ($w_k^\pm$, red and blue curves in the schematic).  
    Similar phenomenology arises in active mixtures~\cite{Politi2004,Halatek2018,Cates2015,Perlekar2014,Theurkauff2012,vanderLinden2019}, although not necessarily with a flux across scales.
    In the Methods, we provide a minimal model of mass cascade exhibiting scale-selection.
    }
\end{figure*}

\medskip
\noindent\textbf{Pattern-induced cascades beyond odd fluids.} 
Our analysis demonstrates that the non-dissipative arrest of turbulent cascades provides a mechanism of wavelength selection for the emergent pattern. 
The decorrelation of triads by waves and the subsequent emergence of a resonant manifold are not unique to odd fluids (Fig.~\ref{figure_examples_wmc}). It is the first step needed to establish a weak turbulence theory for odd waves, which could also be applied to 
optical or elastic turbulence~\cite{Zakharov2012,Nazarenko2011,Galtier2022}.
For instance, optical and mechanical metamaterials~\cite{oddreview,Miri2019} allow one to design arbitrary dispersion relations, including that of odd waves, e.g. by using a combination of so-called odd and even elastic moduli, which replace viscosities in elastodynamics~\cite{oddreview}.

In two-dimensional atmospheric flows and confined plasmas, for instance, Rossby waves (also called drift waves) are at the origin of the arrest of an inverse cascade (Fig.~\ref{figure_examples_wmc}b), at a scale $k_{\text{Rh}}$ known as the Rhines scale~\cite{Diamond2005,Sukoriansky2007,Berloff2009,Chekhlov1996,Rhines1979,Legras1999,Grianik2004}.
This leads to the appearance of a pattern with characteristic scale $k_{\text{Rh}}$ accompanied by a one-dimensionalization of the flow (Extended Data Figure \ref{figure_rossby}), eventually leading to mean flows known as zonal flows.
Other waves, such as gravity waves in stratified flows, can play a similar role~\cite{Balmforth2002,Boffetta2011}.
In contrast with the case of odd waves in 3D,  there is no arrested direct cascade in these \mbox{(quasi-)2D} systems. 
In space plasma such as the solar corona (Fig.~\ref{figure_examples_wmc}c), the existence of a \enquote{helicity barrier} leading to the arrest of cascades has been proposed, and traced to the change of nature of inviscid invariants. This mechanism is enabled by the existence of additional degrees of freedom in magnetohydrodynamics (MHD) compared to standard hydrodynamics. In the case of odd turbulence, the only inviscid invariants are energy and helicity (see Methods), exactly as in standard turbulence.

\medskip
\noindent\textbf{Scale selection by mass cascades.} Cascade-induced patterns can also occur in systems where it is mass rather than energy that cascades (Fig.~\ref{figure_examples_wmc}d).
Mass cascades can for instance take place in the pulverization of objects into debris or the coalescence and breakup of droplets~\cite{Krapivsky2010,Friedlander2000}.
In this context, a cascade-induced scale selection would manifest in the selection of objects with a preferred scale that is neither the largest nor the smallest possible size.
The existence of a steady-state with such a characteristic scale can be observed in situations ranging from rain formation~\cite{Testik2007} and smoke aerosols~\cite{Friedlander2000} to active mixtures~\cite{Politi2004,Halatek2018,Cates2015,Perlekar2014,Theurkauff2012,vanderLinden2019}.
In the Methods, we present a minimal model of scale selection in the steady-state of a mass cascade, in the spirit of shell models of turbulence~\cite{Biferale2003}. 
The key idea is that large droplets (or clusters) tend to breakup, while small ones tend to coalesce very much like vortices in odd fluids: the rate of aggregation $w^+_k$ increases with $k$ (red curve in Fig.~\ref{figure_examples_wmc}d), while the rate of fragmentation $w^-_k$ decreases (blue curve). 
This can be captured within a population balance model which we analyze in the Methods using numerical simulations and analytical solutions.
As shown in Extended Data Figure~\ref{figure_mass_cascade}, a preferential scale, that is neither the largest nor the smallest droplet size, emerges from the balance between these two physical processes, which play a similar role as the homochiral and heterochiral channels in odd fluid turbulence.
This kind of scale selection can also occur in closed systems where mass is neither injected nor removed (i.e. with no net flux), such as in the arrested or interrupted coarsening of mixtures~\cite{Politi2004,Halatek2018,Cates2015,Perlekar2014,Theurkauff2012,vanderLinden2019}.

\noindent\textbf{Conclusion.} 
We have developed a theory of turbulent cascades modified by odd waves 
which captures how non-linear scale selection emerges due to the arrest of the three-dimensional direct and inverse cascades.
Our work highlights the impact of chiral waves on eddy turbulence.
Beyond fluid turbulence, similar mechanisms of scale selection may occur in domains ranging from wave turbulence in parity-violating optical media or solids with odd elasticity to mass cascades as well as cascades that occur in the time domain~\cite{Ghashghaie1996,Bouchaud2003}.

\medskip
\noindent\textbf{Acknowledgements.}
This paper is dedicated to the memory of Krzysztof Gawędzki.
We thank Luca Biferale and Alex Schekochihin for discussions.
We are grateful for the support of the Netherlands Organisation for Scientific Research (NWO) for the use of supercomputer facilities (Snellius) under Grant No. 2021.035. This publication is part of the project “Shaping turbulence with smart particles” with project number OCENW.GROOT.2019.031 of the research programme Open Competitie ENW XL which is (partly) financed by the Dutch Research Council (NWO).
M.F. acknowledges partial support from the National Science Foundation under grant DMR-2118415, a Kadanoff–Rice fellowship funded by the National Science Foundation under award no. DMR-2011854 and the Simons Foundation.
T.K. acknowledges partial support from the National Science
Foundation Graduate Research Fellowship under Grant No. 1746045.
V.V. acknowledges partial support from the Army Research Office under grant  W911NF-22-2-0109 and W911NF-23-1-0212.
M.F. and V.V acknowledge partial support from the France Chicago center through a FACCTS grant. 
This research was partly supported from the National Science Foundation through the Center for Living Systems (grant no. 2317138).

\strong{Data Availability.} The data generated during the course of this study is available on Zenodo at \href{https://doi.org/10.5281/zenodo.10371195}{10.5281/zenodo.10371195}.

\strong{Code availability.} The code used for processing the data, generating the figures, and for the mass cascade and Rossby wave simulations as well as an executable for the DNS are available on Zenodo at \href{https://doi.org/10.5281/zenodo.10371195}{10.5281/zenodo.10371195} under the 2-clause BSD licence. 

\strong{Authors contributions.} XdW, MF, TK, FT and VV designed the research, performed the research, and wrote the paper. 

\strong{Correspondence and requests for materials} should be addressed to \href{mailto:f.toschi@tue.nl}{f.toschi@tue.nl} and \href{mailto:vitelli@uchicago.edu}{vitelli@uchicago.edu}.

\clearpage

\section*{Materials and Methods}

\renewcommand{\figurename}{Extended Data Figure}
\renewcommand{\tablename}{Extended Data Table}
\setcounter{figure}{0}
\renewcommand{\theHfigure}{EDF{\arabic{figure}}}

\subsection*{Direct numerical simulations of the Navier-Stokes equation with odd viscosity}

Direct numerical simulations of the Navier-Stokes equation with odd viscosity \eqref{odd_ns_simplified} are performed in a cubic box of size $L=2\pi$ with periodic boundary conditions. 
Our results can be reproduced with any Navier-Stokes solver by including a modified Coriolis term modulated by $k^2$ (or, equivalently, by a vector Laplacian for real-space based methods) to account for odd viscosity.
We use a pseudo-spectral method with Adams-Bashforth time-stepping and a 2/3-dealiasing rule \cite{Peyret2002}. Both normal and odd viscosities are integrated exactly using integrating factors. The forcing $\vec{f}(t,\vec{k})$ acts on a band of wavenumbers $k\in [k_\text{in},k_\text{in}+1]$ with random phases that are delta-correlated in space and time, ensuring a constant average energy injection rate $\epsilon = \langle \vec{u} \cdot \vec{f} \rangle$. It has a zero mean component $\langle \vec{f}(t,\vec{k}) \rangle = \vec{0}$
and covariance $\langle \vec{f}(t,\vec{k}) \cdot \vec{f}(t',\vec{k}') \rangle = \epsilon \delta(t-t') \delta(\vec{k}-\vec{k}')$.
The time-step is chosen to resolve the fastest odd wave with frequency $\tau_{\odd,\text{max}}^{-1} = \nu_\odd k_{\text{max}}^2$
where $k_{\text{max}}$ is the highest resolved wavenumber in the domain. We find that stable integration requires a time-step $\Delta t \lesssim 0.1\tau_{\odd,\text{max}}$. A complete overview of the input parameters for the simulations in this work are provided in SI.
Approximately \mbox{$3$ million} CPU hours were required to perform the simulations underlying this work.

\subsection*{Effect of odd waves on the nonlinear energy transfer}

In this section, we describe how the waves induced by odd viscosity (odd waves) affect the non-linear energy transfer. Our analysis closely follows that of rotating turbulence~\cite{Smith1999,Alexakis2018,Davidson2015,Mahalov1996}.

\subsubsection*{Nonlinear energy transfer}

Fourier-transforming the Navier-Stokes equation, multiplying with $\vec{v}^*(t, \vec{k})$ (the star denotes complex conjugation), and adding the complex conjugate, we find the energy balance equation~\cite{Davidson2015,Alexakis2018} 
\begin{equation}
    \label{Ek_dynamics_meth}
    \partial_t E = - 2 \nu k^2 E - T + F
\end{equation}
where $\nu = \eta/\rho$ is the kinematic viscosity, and in which 
\begin{equation}
    \label{nl_transfer}
    T(\bm{k},t)=\textrm{Im} \sum_{\bm{p}+\bm{q}=\bm{k}} {v}_i^*(\bm{k},t) P_{ij}(\bm{k})q_\ell {v}_\ell(\bm{p},t){v}_j(\bm{q},t).
\end{equation}
This term describes the non-linear energy transfer between scales, while $F = \vec{v}^* \cdot \vec{f}$ corresponds to energy injection by the forcing term $\vec{f}$. 
The term $-2 \nu k^2 E$ represents standard viscous dissipation.
In Eq.~\eqref{nl_transfer}, the sum runs on momenta $\vec{p}$ and $\vec{q}$ such that $\vec{p}+\vec{q} = \vec{k}$, and $P_{i j}(\vec{k}) = \delta_{ij} - k_i k_j/k^2$ is the projector on incompressible flows.

At first glance, Eq.~\eqref{Ek_dynamics_meth} is left unchanged by odd viscosity, due to its non-dissipative nature. However, odd viscosity has indirect effects on the energy transfer (in the same way as the non-dissipative Coriolis force has an indirect effect on the energy transfer in rotating turbulence). 

\subsubsection*{Odd waves}

To see that, we first consider the linear and inviscid limit of the Navier-Stokes equation \eqref{odd_ns_simplified} (so we set $\nu = 0$ and $(\vec{u} \cdot \vec{\nabla}) \vec{u} = 0$). As detailed in the section \emph{Linear stability of the fluid and odd waves} of the SI (in which we consider a more general odd viscosity tensor), this equation has wave solutions of the form
\begin{equation}
    \vec{v}(t, \vec{x}) = \vec{h}^{\pm}(\vec{k}) \, \ee^{\ii \omega_{\pm}(\vec{k}) t + \ii \vec{k}\cdot\vec{x}} + \text{c.c.}
\end{equation}
in which $\vec{h}^{\pm}(\vec{k}) = \vec{e}(\vec{k}) \times (\vec{k}/k) \pm \ii \vec{e}(\vec{k})$ with $\vec{e}(\vec{k}) = \vec{\hat{e}}_z \times \vec{k}/\lVert\vec{\hat{e}}_z \times \vec{k}\rVert$ \cite{Waleffe1992} with frequency
\begin{equation}
    \label{dispersion_odd_waves}
    \omega_{\pm}(\bm{k}) = \pm  \nu_\odd k_z \abs{k}.
\end{equation}
Taking into account normal viscosity leads to an additional exponential decay of the waves with rate $-\nu k^2$ (see SI). In particular, we note that the linearized Navier-Stokes equation does not exhibit any linear instability.
By construction, $\vec{k} \cdot \vec{h}^{\pm}(\vec{k}) = 0$, so these modes represent incompressible flows. In addition, $(\vec{h}^{+}(\vec{k}))^* \cdot \vec{h}^{-}(\vec{k}) = 0$ and $(\vec{h}^{\pm}(\vec{k}))^* \cdot \vec{h}^{\pm}(\vec{k}) = 2$. Hence, odd waves provide an orthonormal basis for incompressible flows. As $\vec{k} \times \vec{h}^{\pm} = - k \vec{h}^{\pm}$, the basis functions have a well-defined helicity $\mp 1$.

\subsubsection*{Decomposition of the energy transfer on odd waves}

Expanding the velocity field as a superposition of helical waves
\begin{equation}
    \vec{v}(t, \vec{x}) = 
    \sum_{\vec{k}} 
    \sum_{s = \pm}
    v_{s}(t, \vec{k}) \vec{h}^{s}(\vec{k}) \ee^{\ii \omega t + \ii \vec{k}\cdot\vec{x}}
\end{equation}
in which $v_{s}^*(t, \vec{k}) = v_{s}(t, -\vec{k})$ to ensure the reality of $\vec{v}(t, \vec{x})$, the Navier-Stokes equation becomes
\begin{equation}
    \label{ns_helical}
    \partial_t v_{s_k} = \!\!\!\!
    \sum_{\substack{\vec{k}+\vec{p}+\vec{q}=\vec{0} \\ s_p, s_q = \pm}}
    \!\!\!\!
    C_{k|p,q} 
    \ee^{\ii [\omega(\vec{k})+\omega(\vec{p})+\omega(\vec{q})] t}
    v_{s_p}^*
    v_{s_q}^*
    - \nu k^2  v_{s_k}
    + f_{s_k}
\end{equation}
in which we have used the short $v_{s_k}$ for $v_{s_k}(t, \vec{k})$, the term $f_{s_k}(\vec{k})$ corresponds to the forcing term, and
\begin{equation}
    C_{k|p,q} = - \frac{1}{4} (s_p p - s_q q) 
    [ (\vec{h}^{s_p}(\vec{p}) \times \vec{h}^{s_q}(\vec{q})) \cdot \vec{h}^{s_k}(\vec{k}) ]^*
\end{equation}
satisfy $C_{k|p,q} = C_{k|q,p}$.

\subsubsection*{Inviscid invariants: helicity and energy conservation}

In terms of the components $v_\pm(\vec{k})$, energy and helicity respectively read \cite{Waleffe1992,Alexakis2018}
\begin{align}
E &= \sum_{\vec{k}} (|v_+(\vec{k})|^2 + |v_-(\vec{k})|^2) \\
H &= \sum_{\vec{k}} k (|v_+(\vec{k})|^2 - |v_-(\vec{k})|^2).
\end{align}
A direct calculation shows that \cite{Waleffe1992,Smith1999}
\begin{equation}
    C_{k|p,q} + C_{p|q,k} + C_{q|k,p} = 0
\end{equation}
and
\begin{equation}
    s_k k C_{k|p,q} + s_p p C_{p|q,k} + s_q q C_{q|k,p} = 0
\end{equation}
from which we deduce that energy and helicity are conserved when normal viscosity and the forcing can be neglected ($\nu = 0$ and $\vec{f}=\vec{0}$), even if odd viscosities are present. 
In particular,
\begin{equation}
    \partial_t E(k) = v_{s_k}^* \partial_t v_{s_k} + \text{c.c.}
\end{equation}
so using Eq.~\eqref{ns_helical} we find (when $\nu = 0$ and $\vec{f}=\vec{0}$)
\begin{equation}
    \partial_t E(k) = \!\!\!\!
    \sum_{\substack{\vec{k}+\vec{p}+\vec{q}=\vec{0} \\ s_p, s_q = \pm}}
    \!\!\!\!
    C_{k|p,q} 
    \ee^{\ii [\omega(\vec{k})+\omega(\vec{p})+\omega(\vec{q})] t}
    v_{s_k}^*
    v_{s_p}^*
    v_{s_q}^*
    + \text{c.c.}
\end{equation}
This equation shows that the non-linear energy transfer $T(\vec{k},t)$ in Eq.~\eqref{nl_transfer} is suppressed when averaged over long times compared to $\omega(\vec{k})+\omega(\vec{p})+\omega(\vec{q})$, unless this quantity vanishes exactly, as is the case for 2D modes (blue line in Fig.~\ref{figure_two_dimensionalization}h corresponding to modes with $k_z = 0$).

\subsubsection*{Resonant manifold}

The 2D modes with $k_z = 0$ form a so-called slow manifold, or resonant manifold, that contributes to most of the non-linear energy transfer. 
In addition, isolated triads with $k_z \neq 0$ can also satisfy the resonance condition $\omega(\bm{p})+\omega(\bm{q})+\omega(\bm{k}) = 0$. 
In the case of rotating turbulence, resonant triads primarily transfer energy from the 3D modes to quasi-2D slow manifold with $k_z = 0$, leading to an accumulation of energy in the slow manifold, enhancing the two-dimensionalization of the flow~\cite{Biferale2016,Buzzicotti2018,Deusebio2014,Smith1999,Alexakis2018}. We expect a similar phenomenon to occur in the case of fluids with odd viscosity owing to its similarity to rotating fluids, as is also suggested from the two-dimensionalization observed in our numerical simulations.
As a consequence, the effective spatial dimension of the system depends on the scale at which it is observed (like in rotating turbulence or thick layers~\cite{Alexakis2018,Celani2010,Deusebio2014}).
More insights may be obtained by developing a weak turbulence theory for odd waves, in the same spirit as for rotating flows (we refer to Refs.~\cite{Zakharov2012,Nazarenko2011,Galtier2022} for more details on wave turbulence).

\begin{figure*}
    \centering
    \includegraphics[width=0.95\textwidth]{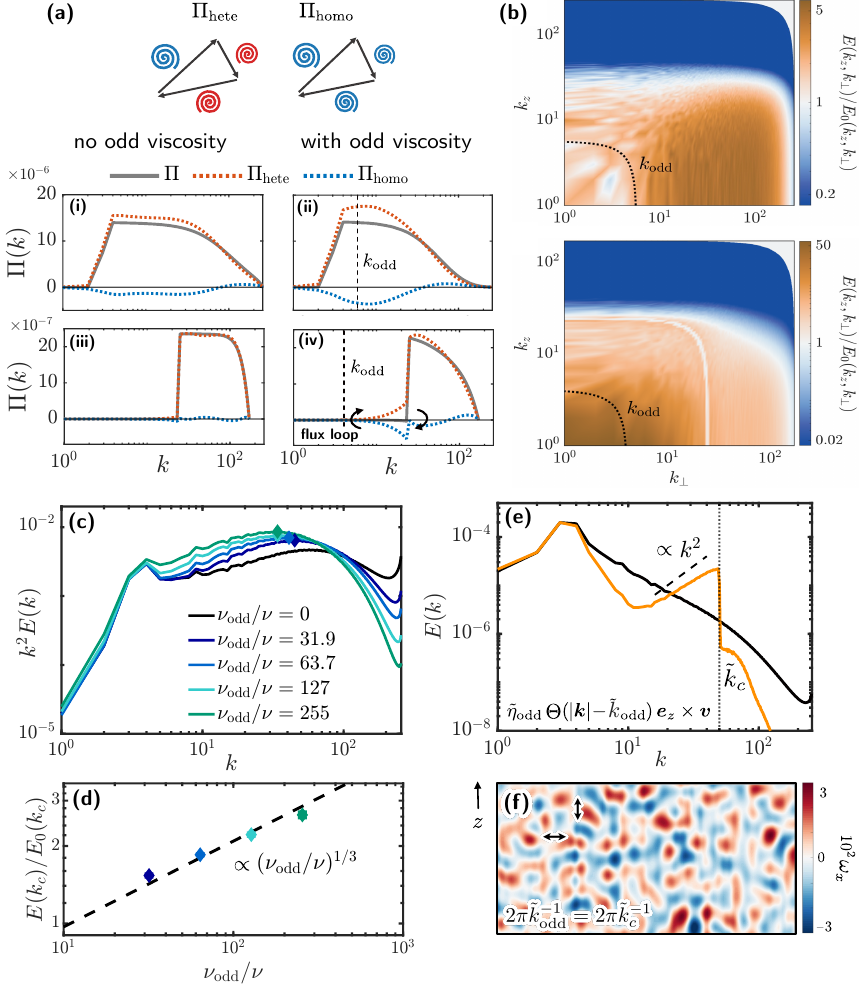}
    \caption{\label{figure_numerics_extended}
    \textbf{Arrest of turbulent cascades in numerical simulations.} (a) The total energy flux $\Pi(k)$ is decomposed into heterochiral $\Pi_\textrm{hete}(k)$ and homochiral $\Pi_\textrm{homo}(k)$ components for direct cascading cases (i,ii) and inverse cascading cases (iii,iv). The cases without odd viscosity (i, iii) are compared to the cases with odd viscosity $\nu_\odd/\nu=255$ (ii) and $\nu_\odd/\nu=212$ (iv). Odd viscosity enhances the homochiral manifold that predominantly cascades inversely, which is in turn balanced by an increased heterochiral flux. For the inverse cascading cases this leads to a flux loop condensate state with vanishing net flux. In (iii, iv) hyperdissipation is used to mimic increased scale separation. See also the section \textit{Helical decomposition} in the SI.
    (b) The anisotropic kinetic energy spectrum $E(k_z,k_\perp)$ with odd viscosity normalized by the case without odd viscosity $E_0(k_z,k_\perp)$ for the forward cascading case with $\nu_\odd/\nu=255$ (top panel) and the inverse cascading case with $\nu_\odd/\nu=212$ (bottom panel). Both panels indicate the regions in $k$-space where the energy condensation due to odd viscosity occurs. 
    (c) In order to determine the characteristic wavelengths in the vorticity field for the direct cascading case, we compute the vorticity spectrum $\lVert\vec{\omega}(\vec{k})\rVert^2$ as $k^2 E(k)$. 
    Without odd viscosity, the maximum of the spectrum is close to the dissipative scale. When odd viscosity is present, a stronger peak emerges in the spectrum as a consequence of the spectral condensation at intermediate scales, evidencing the wavelength selection.
    (d) The lozenges give the value of $h \equiv E(k_{\text{c}})/E_0(k_{\text{c}})$ obtained from the numerical simulations, and are compared with the predicted scaling $h \sim (\nu_\odd/\nu)^{1/3}$ (black dashed line).
    (e) We demonstrate the kinetic energy spectrum $E(k)$ for the case of a sharp transition from a direct cascade at small $k$ to an inverse cascade at large $k$, modeled as a step function of odd viscosity, stepping at $\tilde{k}_\odd$ (orange), compared to the case without odd viscosity (black). Here, $\Theta$ is the Heaviside step function. The sharp transition leads to a sharp condensation at $\tilde{k}_c \equiv \tilde{k}_\odd$ and a diffusive equipartitioned scaling $\sim k^2$ to the left of it. The resulting pattern in $\omega_x$ is shown in (f), with typical wavelength $\sim \tilde{k}_c^{-1} = \tilde{k}^{-1}_\odd$ in both the horizontal and vertical directions. See also the section \textit{Odd hyperviscosity} in the SI.}
\end{figure*}

\subsection*{Scaling relations and wavelength selection}

\subsubsection*{Scaling relation for the energy spectrum}

We first analyze the power spectrum, building on the phenomenological theory of \cite{Kraichnan1965} (see \cite{Zhou2004} for a review). This theory relies on the following hypotheses:
(i) energy is conserved away from injection and dissipative scales,
(ii) the cascade is local, which means that different length scales are only coupled locally (e.g. very large scales are not directly coupled to very small scales), and
(iii) the rate of energy transfer $\varepsilon(k)$ from scales higher than $k$ to scales smaller than $k$ is directly proportional to the triad correlation time $\tau_3$. 
Because of hypotheses (i) and (ii), rate of energy transfer $\varepsilon(k)$ is constant across the scales (i.e. does not depend on $k$), and can be identified with the energy dissipation rate $\epsilon$. In addition, because of hypothesis (ii), $\varepsilon(k)$ should only depend on local quantities $k$ and $E(k)$, in addition to $\tau_3(k)$. 
Therefore, using hypothesis (iii), we write
\begin{equation}
	\epsilon = \varepsilon(k) = A \tau_3(k) k^\alpha E^\beta
\end{equation}
where $A$ is a constant. The exponents are found using dimensional analysis (with $[E] = L^3 T^{-2}$, $[\epsilon] = L^2 T^{-3}$, $[k] = L^{-1}$, $[\tau_3] = T$), which yields $\alpha = 4$ and $\beta = 2$.

In the main text (see Fig.~\ref{figure_two_dimensionalization}), we argue that the eddy turnover time $\tau_{E}$ is the relevant timescale when $k\ll k_\odd$ (so we set $\tau_3 = \tau_E$ in the expression above), while the frequency of odd waves $\omega$ is the relevant timescale when $k \gg k_\odd$ (so we set $\tau_3 = \omega^{-1})$.
The dispersion relation of odd waves is computed in the SI and given in Eq.~\eqref{dispersion_odd_waves}.
The eddy turnover time is $\tau_E(k) = [1/k]/v_k$. As $E(k)$ is the shell-average of $v_{k}$, we have dimensionally $E(k) \sim v_k^2/k$, so $v_k = [k E(k)]^{1/2}$.
Putting everything together, we end up with Eqs.~\eqref{Ek_scaling_belowko} and \eqref{Ek_scaling_aboveko} of the main text.

\subsubsection*{Scaling relation for \texorpdfstring{$k_{\text{c}}$}{kc}}
For the condensation of the forward energy flux, the collapse of numerical results suggests that it can be described by a master scaling law
\begin{equation}
    \frac{E(k)}{E_0(k)} \sim 
    \begin{cases}
    1 \quad &\text{for} \quad k \ll k_\odd,\\
    \left(k/k_\odd\right)^s \quad &\text{for} \quad k \gg k_\odd,
   \end{cases}
\end{equation}
Using the Kolmogorov spectrum for the case without odd viscosity, \mbox{$E_0(k)\sim\epsilon^{2/3}k^{-5/3}$}, we find for the energy spectrum
\begin{equation}
    E(k) \sim
    \begin{cases}
    \epsilon^{2/3}k^{-5/3} \quad &\text{for} \quad k \ll k_\odd,\\
    \epsilon^{2/3}k^{-5/3}\left(k/k_\odd\right)^s \quad &\text{for} \quad k \gg k_\odd.
   \end{cases}
\end{equation}
Using the scaling argument of previous section (see main text Eqs.~\eqref{Ek_scaling_belowko} and \eqref{Ek_scaling_aboveko}), we find $s = 2/3$, which is compatible with the numerical results.
This scaling continues until dissipation saturates the condensation. 
We can thus estimate the location of the condensation peak $k_{\text{c}}$ from the balance between injection and dissipation. Neglecting contributions to the dissipation from wavenumbers $k<k_\odd$ (where there is no significant change from Kolmogorov scaling), we obtain
\begin{equation}
    \epsilon \sim \int_{k_{\odd}}^{k_{\text{c}}} \nu k^2 E(k) dk.
\end{equation}
Assuming $k_\odd\ll k_{\text{c}}$, this yields
\begin{equation}
    \epsilon \sim \nu \, \epsilon^{2/3} \, k_{\text{c}}^{4/3} \left(k_{\text{c}}/k_\odd\right)^s,
\end{equation}
resulting in the scaling relation for the peak condensation
\begin{equation}
    k_{\text{c}} \sim (\epsilon^{1/3}\nu^{-1}k_\odd^s)^{\frac{1}{4/3+s}} \sim (k_\nu^{4/3} k_\odd^s)^{\frac{1}{4/3+s}},
\end{equation}
where in the last relation, we substituted the normal Kolmogorov wavenumber $k_\nu\sim\epsilon^{1/4}\nu^{-3/4}$.

For $s=2/3$, we find
\begin{equation}
    k_{\text{c}} \sim (k_\nu^{4/3} k_\odd^{2/3})^{1/2} \sim \epsilon^{1/4}\nu^{-1/2}\nu_\odd^{-1/4}
\end{equation}
as quoted in the main text.

\subsubsection*{Estimation of the height for the peak}

The mechanism of non-dissipative arrest analyzed in this work is reminiscent of but distinct from the bottleneck effect~\cite{Falkovich1994,Kuchler2019,Lohse1995,Donzis2010,Verma2007,Sreenivasan1997} generated by the usual viscosity. 

A coarse estimate of the height of the peak in $E(k)/E_0(k)$ can be obtained by evaluating Eq.~\eqref{Ek_scaling_aboveko} (to get $E(k)$) and Eq.~\eqref{Ek_scaling_belowko} (to get $E_0(k)$) at $k=k_{\text{c}}$ given by Eq.~\eqref{scaling_kc}, yielding $h \equiv E(k_{\text{c}})/E_0(k_{\text{c}}) \sim (\nu_\odd/\nu)^{1/3}$ (see ED Fig.~\ref{figure_numerics_extended}d for a comparison with numerical data).  Notably, this suggests that $h$ depends only on the ratio of odd to normal viscosity. We also note that $h$ increases as normal viscosity $\nu$ decreases (i.e., when the Reynolds number increases), in contrast with the bottleneck effect due to dissipative viscosity~\cite{Falkovich1994,Kuchler2019,Lohse1995,Donzis2010,Verma2007,Sreenivasan1997} in which the magnitude of the effect decreases as viscosity decreases.

\subsubsection*{Wavelength selection}

In ED Fig.~\ref{figure_numerics_extended}c, we plot an estimate of the power spectrum of the vorticity, evidencing wavelength selection in the vorticity. This suggests that the characteristic wavelength $2\pi/k_{\text{c}}$ should be directly visible in snapshots of the vorticity field. This can be seen in Fig.~\ref{figure_direct_spectra}b. The width of the peak leads to a wide distribution of structure sizes in the image.

We expect the wavelength selection mechanism due to the arrested cascade to persist at arbitrarily long times and to resist small perturbations, in contrast with metastable patterns arising from kinetic effects~\cite{ClarkDiLeoni2020} where the system resides in metastable states for long but finite periods (see SI for convergence plots).

The wavelength selection mechanism we have described can be compared to that in so-called active turbulence, which occurs in bacterial suspensions and self-propelled colloids~\cite{Dunkel2013,Wensink2012,Alert2022,MartinezPrat2019,Alert2020,Carenza2020}.
In active turbulence, however, it has been reported that there is no energy transfer across scales (and hence no cascade): 
energy is typically dissipated at the same scale as it is injected, and it is believed that the wavelength selection is the result of a scale-by-scale balance [see for instance Figs. 3d and 4g and \S 3.2.2 and 4.2.3 in Ref.~\cite{Alert2022} and references therein]. We note, however, that finite energy fluxes have been reported in certain cases~\cite{Slomka2015,Slomka2017,Rorai2022,Bratanov2015,Mukherjee2023,Linkmann2019,Kiran2023}.

In these systems, wavelength selection can be understood as the result of a Swift-Hohenberg-type term included in the stress tensor (leading to a finite-wavelength linear instability), to which noise is added~\cite{Alert2022}. 
In contrast, cascade-induced pattern formation cannot be directly traced to a linear instability of Navier-Stokes equation \eqref{odd_ns_simplified} (see section \emph{Effect of odd waves on the nonlinear energy transfer} as well as SI section \emph{Linear stability of the fluid and odd waves} for a linear stability analysis).
Indeed, the linear stability analysis does not predict any instability, neither to a stable branch with a particular wavelength, nor to an unstable branch that could itself bifurcate to the state of interest as part of a subcritical bifurcation.

An analogy with similar situations such as Rossby/drift wave turbulence~\cite{Marston2023,Parker2014,Constantinou2018,Gurcan2015,Parker2013,Constantinou2014} and laminar/turbulent patterns in wall-bounded shear flows~\cite{Tuckerman2020,Prigent2002,Duguet2010,Kashyap2022} suggests that the wavelength selection may be described by considering the linear stability of the statistically averaged Navier-Stokes equation, for instance using an appropriate turbulence closure model.

\subsubsection*{Anisotropic energy spectra}
In line with the inherent symmetry of the system, we can also consider cylindrically-averaged energy spectra $E(k_\perp, k_z)$, which distinguish the horizontal (perpendicular) directions from the vertical direction~\cite{Vallis1993,Galtier2003,Galperin2008,Meyrand2016}. In order to reveal in which part of the $k$-space the energetic condensation occurs, we compute the cylindrically averaged spectrum of the cases with odd viscosity normalized by the spectrum of the reference case without odd viscosity, as shown in ED Fig.~\ref{figure_numerics_extended}b.
Starting with the direct cascading case in the top panel of ED Fig.~\ref{figure_numerics_extended}b, we see that indeed the flow remains mostly 3D isotropic for $k<k_\odd$ and then proceeds to condensate anisotropically into the low-$k_z$ manifold due to the quasi-2-dimensionalization effect of the odd viscosity. As detailed in the main text, the condensation is saturated by dissipation, leading to a peak condensation wavelength $k_c$, which is thus primarily visible in the perpendicular directions due to the anisotropic condensation. The dominant vertical scale hence remains closer to $k_\odd$. This leads to a crude estimate for the aspect ratio $\gamma$ of the features in the pattern produced by the odd viscosity as
\begin{equation}\label{eq:aspect_ratio}
    \gamma = \frac{k_c}{k_\odd} \sim \nu^{-1/2}\nu_\odd^{1/2}.
\end{equation}
For the case presented in Fig.~\ref{figure_direct_spectra}b in the main text, this leads to an aspect ratio $\gamma \simeq 3$.

For the inverse cascading case (bottom panel of ED Fig.~\ref{figure_numerics_extended}b), we again observe anisotropic condensation in the region $k>k_\odd$. In the region $k<k_\odd$, however, the kinetic energy for the case with odd viscosity is larger than the case without odd viscosity, as indicated by the dark orange color. This is due to the fact that in this range, we expect the same diffusive equipartitioned scaling $E(k)\sim k^2$ for both cases with and without odd viscosity, and there is no active dissipative mechanism to deplete the excess energy that has accumulated at higher wavenumbers in the case with odd viscosity.

\subsection*{Experimental considerations}

In this section, we discuss the conditions required to observe the wavelength selection described in the main text in a fluid with odd viscosity.
In a nutshell, we expect this effect to occur, for instance, in a fluid of self-spinning particles large enough to be inertial (not overdamped). 

First, the Reynolds number $\text{Re} = U L/\nu$ has to be large enough. This puts constraints on the viscosity $\nu$ of the fluid, details of which depend on the experimental setup considered.
The current experimental systems we are aware of in which explicit measurements of odd viscosities were reported (active spinning colloids \cite{Soni2019}, magnetized graphene \cite{Berdyugin2019}, and magnetized polyatomic gases~\cite{Beenakker1970,Mccourt1990}) are all in a regime where the non-linear advective term in the Navier-Stokes equation can be neglected, either because $\nu$ is large enough or for geometric reasons; effectively, $\text{Re} \ll 1$. 
Note also that experimental instances of (especially two-dimensional) odd fluids may include a substrate, on top of which the active particles move. This can lead to the addition of an effective drag force $-\gamma \vec{v}$ in the Navier-Stokes equation describing the odd fluid made of these particles. If such a term is large, it would prevent the existence of an inertial regime, and likely spoil the phenomenology discussed here.

Second, the ratio $\nu^{\odd}/\nu$ has to be large enough for the effect to be visible.
When $\nu^{\odd} \lesssim \nu$, energy is dissipated as soon as, or before any effect of odd waves can arise.
Henceforth, observing effects of odd waves on turbulence would require $\nu^{\odd} > \nu$. 
Odd viscosities ($\nu^{\odd} \neq 0$) typically arise in systems breaking time-reversal and inversion symmetry at the microscopic scale~\cite{Avron1995,Ganeshan2017,oddreview}. They have been experimentally measured in polyatomic gases under magnetic fields~\cite{Beenakker1970,Mccourt1990}, spinning colloids~\cite{Soni2019}, and magnetized electron fluids~\cite{Berdyugin2019}. They have also been predicted in systems including fluids under rotation~\cite{Nakagawa1956}, magnetized plasma~\cite{ChapmanCowling,Lingam2020,Morrison1984}, quantum fluids~\cite{Hoyos2012,Read2009,Vollhardt2013,Avron1995}, vortex matter~\cite{Wiegmann2014}, sheared granular gases~\cite{Zhao2022}, assemblies of spinning objects~\cite{Banerjee2017,Han2021,Markovich2021,Fruchart2022,Tsai2005,Grzybowski2000,Yan2015,Bililign2021,Tan2021,Dunkel2013,Petroff2015,Ivlev2012,Ivlev2015}, and circle swimming bacteria~\cite{Denk2016,Liebchen2017}. 
In the systems mentioned above, where experimental measurements of odd viscosity have been reported, $\nu_\odd/\nu$ reaches at most 1/3 (in active spinning colloids \cite{Soni2019} and magnetized graphene \cite{Berdyugin2019}). 
From a theoretical point of view, the ratio $\nu_\odd/\nu$ is expected to increase linearly with the time-reversal breaking field. For instance, ideal vortex fluids are predicted to have a finite $\nu_\odd$ but a vanishing $\nu$ \cite{Wiegmann2014}, leading to an infinite value of $\nu_\odd/\nu$.
Kinetic theory calculations for magnetized plasma (Ref.~\cite[\S~19.44]{ChapmanCowling}) predict $\nu = \nu_0 /[1+x^2]$ and $\nu_\odd = \nu_0 x/[1+x^2]$ in which $x = 2 \omega \tau$ with $\tau$ is a collision time and $\omega \propto B$ is a frequency proportional to the magnetic field $B$, while $\nu_0$ is the value of normal shear viscosity when $B=0$. 
Similarly, kinetic theory in rotating gases lead to an identical result where $x \propto \Omega$ is proportional to the rotation speed~\cite{Nakagawa1956}.
Finally, in electron gases in graphene, experiments have been performed that validate these theoretical calculations, see Ref.~\cite{Berdyugin2019} (with $x=B/B_0$).
This results into a ratio $\nu_\odd/\nu = x \propto B$.
Similarly, in active fluids, theoretical works suggest that $\nu_\odd$ is proportional to the rotation speed of the spinning particles \cite{Banerjee2017}.

\subsection*{Rossby/drift wave turbulence}

\begin{figure*}
    \centering
    \includegraphics[width=\textwidth]{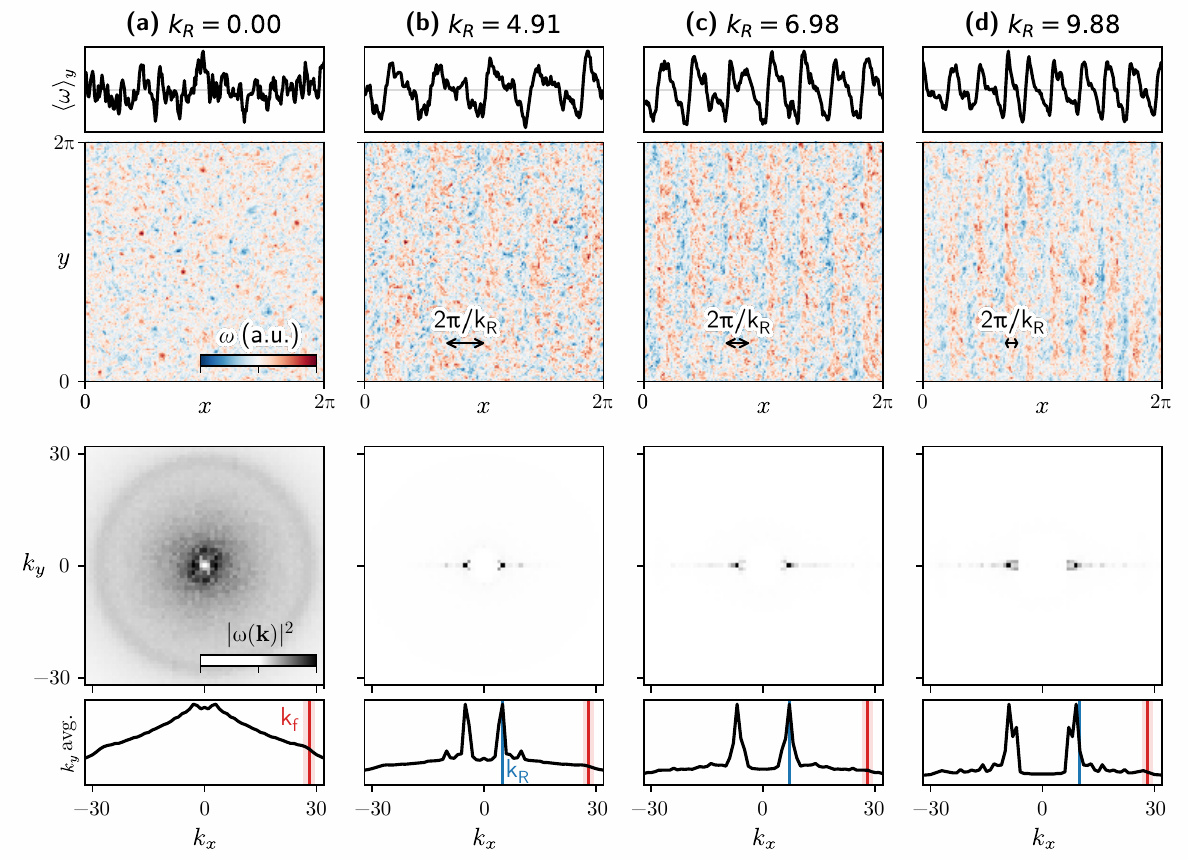}
    \caption{\label{figure_rossby} \textbf{Rossby/drift wave turbulence.}
    Simulations of Eq.~\eqref{chm} describing Rossby/drift wave turbulence demonstrates one-dimensionalization of the 2D flow and the appearance of a pattern with characteristic scale given by the Rhines scale $1/k_{\text{R}}$ \cite{Diamond2005,Sukoriansky2007,Berloff2009,Chekhlov1996,Rhines1975,Rhines1979,Legras1999,Grianik2004}.
    Each column shows, from top to bottom: (i) the vorticity averaged along the $y$ direction at final time, (ii) the vorticity field at final time, (iii) the power spectrum of the vorticity averaged over the last $1/6$ of the simulation and (iv) the $k_y$ average of this quantity.
    The equation is integrated using the pseudospectral solver Dedalus \cite{Burns2020} on a $L\times L$ square domain with size $L=2\pi$ discretized with $N=256$ Fourier harmonics per dimension using a 3rd-order 4-stage Diagonally Implicit/Explicit Runge–Kutta scheme (RK433 in Dedalus) \cite{Ascher1997} with an adaptive timestep for \num{1500.} simulation time units.
    The forcing is taken to be a Gaussian random field concentrated on a ring of radius $k_{\text{f}} = 28$ (red line) and bandwidth $k_{\text{fw}} = 1.5$ (light red rectangle) in Fourier space, scaled by the forcing rate $\epsilon = \num{0.001}$.
    We take a linear drag $\alpha = \num{0.01}$, a viscosity $\nu = \num{0.00001}$.
    The $\beta$ parameter is 
    (a) $\beta=\num{0}$, 
    (b) $\beta=\num{20}$, 
    (c) $\beta=\num{40}$, 
    (d) $\beta=\num{80}$, leading to the measured values of the Rhines wavenumber $k_{\text{R}} =1/2 \sqrt{\beta/U_{\text{rms}}}$ given in the figure (blue line in the bottom plots) in which $U_{\text{rms}}$ is obtained from the measured energy spectrum.
    }
\end{figure*}

Extended Data Fig.~\ref{figure_rossby} shows examples of simulations of the Rossby/drift wave turbulence mentioned in Fig.~\ref{figure_examples_wmc}b.
A brief review is contained in the SI, and we refer the reader to Refs.~\cite{Connaughton2015,Boffetta2002,Tassi2009,Hasegawa1977,Charney1971,Horton1999,Pedlosky1979,Galperin2008,Diamond2005,Sukoriansky2007,Berloff2009,Chekhlov1996,Rhines1975,Rhines1979,Legras1999,Grianik2004} for more details.
In the figure, we simulate the Charney-Hasegawa-Mima (CHM) equation~\cite{Connaughton2015,Boffetta2002,Tassi2009,Hasegawa1977,Charney1971,Horton1999,Pedlosky1979,Galperin2008}
\begin{equation}
\label{chm}
\partial_t \omega + J(\psi, \omega) + \beta \partial_x \psi = - \alpha \omega + \nu \Delta \omega + f_{\omega}
\end{equation}
in which 
$J(a,b) = (\partial_x a)(\partial_y b) - (\partial_y a) (\partial_x b)$,
$\omega = \Delta \psi$, and $\psi$ is the stream function, defined such that the velocity field is $\vec{v} = - \epsilon \cdot \vec{\nabla} \psi$. 
The parameter $\beta$ represents the gradient of the Coriolis force in a $\beta$-plane approximation; $\alpha$ represents large-scale friction and $\nu$ is viscosity, while $f_\omega$ is a vorticity forcing.
Simulations are performed using the open-source pseudospectral solver Dedalus~\cite{Burns2020}.

Note that in Rossby wave turbulence, the only \emph{exact} inviscid invariants are energy and helicity. However, it has been established that a quantity dubbed zonostrophy evolves slowly enough to be considered as an invariant for practical purposes~\cite{Balk1991b,Balk1991a,Nazarenko2009,Diamond2005,Galperin2008}. This raises the question of whether such an \enquote{adiabatic invariant} may exist for odd turbulence, and whether it can predict the direction of the cascades (see Refs.~\cite{Alexakis2018,Sahoo2017} for discussions of the relation between inviscid invariants and the direction of turbulent cascades).

\subsection*{Minimal model of mass cascade with scale selection}

In this section, we consider a simple model of mass cascade that exhibits wavelength selection.

Mass cascades can, for instance, occur in the pulverization of objects into debris or in the coalescence and breakup of droplets~\cite{Smoluchowski1916,Kolmogorov1941,Krapivsky2010,Gorokhovski2008,Brilliantov2015,Gorokhovski2008,Cheng1990,Brilliantov2021}. 
These processes can be modelled by the aggregation and fragmentation of clusters composed of monomers linked together: two clusters that collide may merge into a larger cluster; and a given cluster may split into smaller ones, spontaneously or upon collision. 
The mean-field kinetics of these processes is described by a population balance equation generalizing the so-called Smoluchowski equation~\cite{Krapivsky2010,Leyvraz2003,Wattis2006,Ramkrishna2014} which can exhibit scale-invariant cascades, akin to that present in the Navier-Stokes equation~\cite{Connaughton2004,Connaughton2006,Connaughton2018}.
This kinetic equation may describe two classes of situations: (i) closed systems where mass is conserved and (ii) open systems where particles are injected and removed from the system. Case (i) may somehow be compared to freely decaying turbulence, while (ii) may be compared to driven turbulence in which energy is injected and dissipated.

Intuitively, we expect that the balance between aggregation and fragmentation will lead to a preferred size if large clusters tend to breakup, while small clusters tend to coalesce. Such a preferred size should manifest as a peak in the distribution of aggregate sizes.
This has been reported, for instance, in the case of raindrop sizes~\cite{
Testik2007,Srivastava1971,Testik2013}, in which the distribution originates from complex mechanisms including air turbulence and fluid fragmentation~\cite{Pumir2016,Babler2012,Grabowski2013,Villermaux2007,Falkovich2002,Babler2012,Bec2010}. 

In our toy model, we consider clusters $M_n$ composed of $2^{n-1}$ monomers $M_1$, with $n=1,\dots,N$. 
This is reminiscent of what is done in {shell models} of turbulence~\cite{Biferale2003}, in which the wavenumbers are chosen in geometric progression. 
We assume that (i) there are interactions only between clusters of the same size, and (ii) there is a maximum cluster size $N$.
The first assumption ensures that the mass fluxes are local, and the second allows us to consider a finite number of equations. 
We include a constant source of monomers, and a sink that removes the largest clusters $M_N$.
In the case of raindrops in a cloud, for instance, the source may describe the condensation of droplets from vapor, and the sink may describe the precipitation of large droplets out of the cloud.
The model is summarized by the set of reactions
\begin{subequations}
\label{reactions_mass_cascade}
\begin{align}
&\ce{
{\varnothing} ->[{$J_{\text{in}}$}] M_1
}
\\
&\ce{
2 M_n
<=>[{$k_n^+$}][{$k_{n+1}^-$}] 
M_{n+1}
}
\\
&\ce{
M_N ->[{$J_{\text{out}}$}] {\varnothing}
}
\end{align}
\end{subequations}
in which $M_{n}$ ($n=1,\dots,N$) represents a clusters of size $2^{n-1}$ ($M_{1}$ represents a monomer), while $J_{\text{in}}$, $J_{\text{out}}$, and $k_n^\pm$ are the rates of the corresponding reactions.

The number densities $c_n$ of clusters then follow the dynamical equation
\begin{equation}
\label{frag_eom}
\frac{d c_n}{dt} = 2 k^{-}_{n+1} c_{n+1} - k^{-}_{n} c_{n} + \frac{1}{2} k^{+}_{n-1} c_{n-1}^2 - k^{+}_{n} c_{n}^2 + J_{\text{ext}}
\end{equation}
where 
\begin{equation}
    J_{\text{ext}} = \delta_{n,1} J_{\text{in}} - \delta_{n,N} J_{\text{out}} c_{N}
\end{equation}
in which it is implied that $c_{n} \equiv 0$ for $n < 1$ and $n > N$.

We can also consider the mass density $\rho_{n} = 2^{n-1} m_0 c_{n}$, in which $m_0$ is the mass of a monomer. Multiplying Eq.~\eqref{frag_eom} with $2^{n-1} m_0$, we find that the terms with prefactors $k^{\pm}_{n}$ cancel as in a telescoping series.
This manifests that Eq.~\eqref{frag_eom} with $J_{\text{in}} = J_{\text{out}} = 0$ conserves mass. 
It is therefore convenient to introduce the fluxes
\begin{equation}
	\label{Jplus}
	J^{+}(n) = - \int_{1}^n \dd n' \left[ \frac{1}{2} k^{+}_{n'-1} c_{n'-1}^2 - k^{+}_{n'} c_{n'}^2 \right]
\end{equation}
and
\begin{equation}
	\label{Jminus}
	J^{-}(n) = - \int_{1}^n \dd n' \left[ 2 k^{-}_{n'+1} c_{n'+1} - k^{-}_{n'} c_{n'} \right]
\end{equation}
corresponding to the reactions with rates $k_n^\pm$, and such that
\begin{equation}
	\frac{d c_n}{dt} = - \partial_n [ J^{+}(n) + J^{-}(n) ] + \delta_{n,1} J_{\text{in}} - \delta_{n,N} J_{\text{out}} c_{N}.
\end{equation}

In order to induce wavelength selection, we choose particular forms for $k_n^{\pm}$. The basic idea is the forward flux $k_n^{+}$ should decrease with $n$, while the backward flux $k_n^{-}$ should increase with $n$. Experimentation suggests that various strictly increasing functions of $(N-n)/(N-1)$ and $(n-1)/(N-1)$, respectively, lead to similar results. We choose 
\begin{equation}
    \label{kpmn_simple}
	k_n^{+} = \kappa_0^+ + \kappa_1^+ \frac{N-n}{N-1}
	\quad
	\text{and}
	\quad
	k_n^{-} = \kappa_0^- + \kappa_1^- \frac{n-1}{N-1}.
\end{equation}

\begin{figure*}[htbp]
    \centering
    \includegraphics[width=0.8\textwidth]{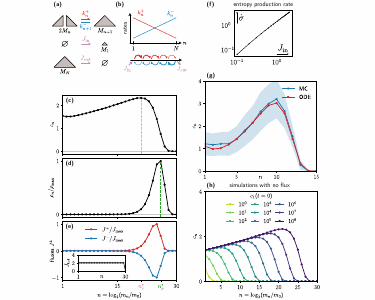}
    \caption{\label{figure_mass_cascade}
    \textbf{Minimal model of mass cascade with scale selection.} 
    (a) In the model, two clusters $M_n$ of size $2^{n-1}$ can merge into a cluster $M_{n+1}$ of size $2^n$. 
    This aggregation process occurs with a rate $k^+_n$.
    Conversely, a cluster $M_{n+1}$ can split into two clusters $M_n$ (except for monomers $M_1$).
    This fragmentation process occurs with a rate $k^-_{n+1}$.
    (b) The rates $k^\pm_n$ of aggregation/fragmentation depend on the size of the cluster, so that (i) large clusters are more likely to fragment than small ones (blue curve) and (ii) small clusters are more likely to aggregate than large ones (red curve).
    (c-e) Equations \eqref{frag_eom} are numerically solved starting from the initial condition $c_n = 0$ for all $n$.
    We have set $\kappa_0^\pm = 2$ and $\kappa_1^\pm = 1$, as well as  $J_{\text{in}} = J_{\text{out}} = 1$.
    The number distribution $c_n$ is plotted in panel c, while the mass distribution $\rho_n = 2^{n-1} m_0 c_n$ (normalized by its maximum value) is plotted in panel d. In panel e, we show the fluxes $J^{+}_n$ (red) and $J^{-}_n$ (blue) defined in Eqs.~(\ref{Jplus}-\ref{Jminus}). The inset of panel e shows the total flux $J_n = J_n^+ + J_n^-$, which is constant away from the boundaries.
    (f) Entropy production in the mass cascade.
    We plot the rate of entropy production (computed within the mean-field model) as a function of the flux $J_{\text{in}}$ through the system, in logarithmic scale. (As the distribution is stationary, there is a constant flux equal to the input flux.) 
    We have set $N=15$, $\kappa_0^\pm = 0.1$, $\kappa_1^\pm = 1$, $J_{\text{out}} = 1$ and $c_n(t=0) = 0$ for all $n$.
    (g) Comparison between mean-field and Monte-Carlo simulations.
    The red curve (labelled ODE) shows the mean-field solution of Eqs.~\eqref{frag_eom}, while the blue curve (labelled MC) shows the solution of the kinetic Monte-Carlo simulations (average plus or minus half a standard deviation over \num{1000} samples).
    We have set $N=15$, $\kappa_0^\pm = 0.1$, $\kappa_1^\pm = 1$, $J_{\text{in}} = J_{\text{out}} = 1$ and $c_n(t=0) = 0$ for all $n$.
    (h) Equations \eqref{frag_eom} with no influx and outflux ($J_{\text{in}} = J_{\text{out}} = 0$) are numerically solved starting from the initial condition with only monomers $c_n(t=0) = c_0(0) \delta_{n,0}$, with different values of the number of monomers $c_0(0)$.
    We observe that (i) a peak in the steady-state distribution only arises when the initial number of monomers $c_0(0)$ is large enough and (ii) the position of the peak moves as $c_0(0)$ increases. 
    We have set $\kappa_0^\pm = 2$ and $\kappa_1^\pm = 1$.
    }
\end{figure*}

Equations \eqref{frag_eom} are then solved starting from the initial condition $c_n = 0$ for all $n$ using \texttt{DifferentialEquations.jl} \cite{Rackauckas2017} with a 4th order A-stable stiffly stable Rosenbrock method (\texttt{Rodas4P}) until a steady-state is reached. The resulting steady-state is shown in ED Fig.~\ref{figure_mass_cascade}. In ED Fig.~\ref{figure_mass_cascade}c, we observe that the density $c_n$ is peaked at an intermediate value $n_c^*$ (pink dashed line), which is neither the maximum cluster size $N$, nor the monomer size $1$, demonstrating wavelength selection. 
Similarly, ED Fig.~\ref{figure_mass_cascade}d shows that the mass density $\rho_n$ is peaked around a (different) scale $n_\rho^*$ (red dashed line).
As we have considered a mean-field description that does not take space into account, there is no proper \enquote{pattern} -- only wavelength selection.

We observe in ED Fig.~\ref{figure_mass_cascade}e that the flux $J_{\text{tot}} \equiv J^+ + J^-$ (black curve in inset) is constant and nonzero for $1 < n < N$. Indeed, in 1D, the existence of a steady-state is equivalent to a constant flux. (Note that certain models of aggregation-fragmentation may exhibit oscillations, i.e. limit cycles instead of fixed points~\cite{Ball2012,Matveev2017}.)
The total flux can be decomposed into the forward flux $J^+$ associated to reactions with rates $k^+_n$, and the backward flux $J^-$ associated to reactions with rates $k^-_n$, respectively defined in Eqs.~\eqref{Jplus} and \eqref{Jminus}, and plotted in ED Fig.~\ref{figure_mass_cascade}e (red and blue curves, respectively).

In ED Fig.~\ref{figure_mass_cascade}h, we analyze the initial value problem obtained by setting $J_{\text{in}}=J_{\text{out}}=0$ in Eq.~\eqref{frag_eom}.
An exact solution of this model is given in the SI.
Wavelength selection may occur, even though there is no net flux. 
This can be compared to the arrest of coarsening that can arise in mixtures and similar mass-conserving systems, even if mass is not injected and removed from the system~\cite{Politi2006,Politi2004,Ginot2018,Halatek2018,Brauns2021,Cates2015,Perlekar2014,Theurkauff2012,vanderLinden2019}.
We also observe that wavelength selection only occurs when the total number of mononers is large enough, which is reminiscent of what happens in so-called beam self-cleaning in optics, where light in an optical waveguide at sufficiently high power may undergo a nonlinear redistribution of the mode powers that favors the fundamental, similar to an inverse cascade~\cite{Ferraro2023}.

Equations~\eqref{frag_eom} describe the mean-field dynamics of the reactions \eqref{reactions_mass_cascade}. To check whether the effect is still present beyond mean-field, we solve the corresponding Doob-Gillespie kinetic Monte Carlo problem using the package \texttt{Catalyst.jl} \cite{Loman2022,Rackauckas2017}. The result of the simulation is shown in ED Fig.~\ref{figure_mass_cascade}g, and compared with mean-field simulations, with excellent agreement.

Finally, we discuss the rate of entropy production in the system.
To do so, it is convenient to introduce the rates $k^{+,n} = k^{+}_{n}/2$ and $k^{-,n} = k^{-}_{n+1}$  to match the notations used in the literature on chemical reaction networks \cite{Gaspard2022,Kondepudi2014,Schnakenberg1976,Rao2016}.
We identify the forward and backward fluxes corresponding to the reaction with rates $k^{\pm,n}$ as $J^{+,n} = k^{+,n} c_n^2$ and $J^{-,n} = k^{-,n} c_{n+1}$. The rate of entropy production corresponding to the reaction is then $\dot{\sigma}_n = (J^{+,n} - J^{-,n}) \log(J^{+,n}/J^{-,n})$ \cite{Gaspard2022,Kondepudi2014,Schnakenberg1976,Rao2016}.
We can then evaluate this quantity and the total rate of entropy production $\dot{\sigma} = \sum_n \dot{\sigma}_n$ from the steady-state distributions $c_n$ obtained numerically (ED Fig.~\ref{figure_mass_cascade}f).
The rate of entropy production vanishes when the system is isolated ($J_{\text{in}}=J_{\text{out}}=0$), and increases as a function of the flux going through the system (which is equal to $J_{\text{in}}$ as long as there is a stationary state).

\clearpage

\begin{table*}[ht]
\caption{\label{tab:input}Parameters that are used for the simulations in this work for the forward cascade runs (I) and inverse cascade runs (II). Listed are the average energy injection rate $\langle \epsilon \rangle$, normal shear viscosity $\nu$, odd viscosity $\nu_\odd$, injection wavenumber $k_\text{in}$, odd viscosity wavenumber $k_\odd=\epsilon^{1/4}\nu_\odd^{-3/4}$, grid resolution $N^3$, Kolmogorov length $\ell_\nu=\epsilon^{-1/4}\nu^{3/4}$, total simulation time $T$, time-step $\Delta t$, Kolmogorov time $\tau_\nu=\epsilon^{-1/2}\nu^{1/2}$ and the \textit{a posteriori} integral scale Reynolds number without odd viscosity $\text{Re}$.}
\begin{tabular}{cccccccccccc}
\toprule
&$\langle\epsilon\rangle$&$\nu$&$\nu_\odd$&$k_\text{in}$&$k_\odd$&$N^3$&$\ell_\nu/(L/N)$&$T$&$\Delta t$&$\tau_\nu/\Delta t$&$\text{Re}$\\
\midrule
I \rule{0pt}{10pt} & $1.4\times10^{-5}$ & $9.4\times10^{-6}$ & $[0.3 - 2.4]\times10^{-3}$ & $3$ & $[27-6]$ & $768^3$ & $0.34$ & $2\times10^3$ & $[1.0-0.2]\times10^{-2}$ & $[82-410]$ & $1.3\times10^3$\\
II & $2.1\times10^{-6}$ & $9.4\times10^{-6}$ & $[0.5 - 5.0]\times10^{-3}$ & $24$ & $[11-2]$ & $512^3$ & $0.37$ & $8\times10^3$ & $[1.0-0.5]\times10^{-2}$ & $[210-420]$ & $6.2\times10^1$ \\
\bottomrule
\end{tabular}
\end{table*}

\section*{Supplementary Information}

\setcounter{figure}{0}

\renewcommand{\theequation}{S.\arabic{equation}}
\renewcommand{\theHfigure}{S{\arabic{figure}}}

\renewcommand{\figurename}{Supplementary Figure}
\renewcommand{\tablename}{Supplementary Table}

\subsection*{Navier-Stokes equation with odd viscosity}

In this section, we describe the most general odd viscosity terms in a 3D incompressible fluid with cylindrical symmetry. In an incompressible fluid, any gradient term in the Navier-Stokes equation can be absorbed in the pressure, without modifying the flow. A generic cylindrically symmetric viscosity tensor can have eight independent non-dissipative (odd) viscosities~\cite{Khain2022}. Only two independent odd viscosity terms remain in the incompressible Navier-Stokes equation:
\begin{align}
    \label{odd_ns_full}
        \rho_0 D_t \bm{v} &= -\nabla P + \eta \Delta \bm{v}\\ \nonumber
    &+ 
    \eta_1^{\text{o}}
    \begin{bmatrix}
        (\partial_x^2 + \partial_y^2)v_y\\
        \\
        -(\partial_x^2 + \partial_y^2)v_x\\
        \\
        0
    \end{bmatrix} 
    +
     \eta_2^{\text{o}} 
     \begin{bmatrix}
         -\partial_z^2 v_y - \partial_y \partial_z v_z\\
        \\
         \partial_z^2 v_x + \partial_x \partial_z v_z\\
        \\
        \partial_z(\partial_y v_x - \partial_x v_y)
    \end{bmatrix}\\ \nonumber
\end{align}
The forces due to the remaining odd viscosity coefficients can be expressed as linear combinations of the $\eta_1^{\text{o}}$ and $\eta_2^{\text{o}}$ terms and gradients of functions (see~\cite{Khain2022} for further details). In the main text, we consider the limit $\eta_1^{\text{o}} = - 2\eta_2^{\text{o}}$, where we define $\eta_\odd \equiv \eta_2^{\text{o}}$. In this case, the force due to the odd viscosities is $\eta_\odd (\vec{e}_z \times \Delta \vec{v} - \nabla \omega_z)$, where the second (gradient) term can be absorbed into the pressure. In wavenumber space, the odd viscosity term in this limit reads $\eta_\odd k^2 \vec{v}(\vec{k}) \times \vec{e}_z$ and can be thought of as a wavenumber-dependent rotation.

\subsection*{Parameters of the direct numerical simulations}

Parameters used in the direct numerical simulations are given in Extended Data Table~\ref{tab:input}.

\subsection*{Modified Taylor-Proudman theorem}

The Taylor-Proudman theorem applied to a fluid under rotation implies $\partial_z \bm{v} = 0$ \cite{Taylor1923,Proudman1916}. In the main text, we have used an extension of this result to the case of a fluid with non-dissipative viscosities. Here, we give the detailed proof of this result for all non-dissipative viscosities compatible with cylindrical symmetry.

We first rewrite the odd force terms in the Navier-Stokes equation Eq.~\eqref{odd_ns_full} in terms of a cross product with $\bm{e}_z$:
\begin{align}
    &\eta_1^{\text{o}}\begin{bmatrix}
        (\partial_x^2 + \partial_y^2)v_y\\
        \\
        -(\partial_x^2 + \partial_y^2)v_x\\
        \\
        0
    \end{bmatrix}
    = -\eta_1^{\text{o}} \bm{e}_z \times \Delta_{\perp} \bm{v},
\end{align}
\begin{align}
    &\eta_2^{\text{o}} \begin{bmatrix}
         -\partial_z^2 v_y - \partial_y \partial_z v_z\\
        \\
         \partial_z^2 v_x + \partial_x \partial_z v_z\\
        \\
        \partial_z(\partial_y v_x - \partial_x v_y)
    \end{bmatrix} + \nabla \omega_z
    =\eta_2^{\text{o}} \bm{e}_z \times (\Delta \bm{v} - 2\Delta_{\perp} \bm{v}),
\end{align}
where $\Delta_{\perp} \equiv \partial_x^2 + \partial_y^2$. Note that we can add gradient terms to these expressions without modifying the flow.
Assuming that the dominant balance is between the odd viscosity term and the pressure gradient term, we write for $\eta_1^{\text{o}}$:
\begin{align}
-\eta_1^{\text{o}} \bm{e}_z \times \Delta_{\perp} \bm{v} &= \nabla P.
\end{align}
Next, we take the curl of this equation to remove the pressure term, and simplify the resulting expression by applying the vector calculus identity
\begin{equation*}
\label{vec_identity}
 \vec{\nabla} \times(\vec{A} \times \vec{B})=\vec{A}(\vec{\nabla} \cdot \vec{B})-(\vec{A} \cdot \vec{\nabla}) \vec{B}+(\vec{B} \cdot \vec{\nabla}) \vec{A}-\vec{B}(\vec{\nabla} \cdot \vec{A}).
\end{equation*}

\noindent Enforcing incompressibility ($\nabla \cdot \vec{v} = 0$), we obtain
\begin{align}
\eta_1^{\text{o}}(\bm{e}_z \cdot \nabla) \Delta_{\perp} \bm{v} &= 0, 
\end{align}
which can be simplified into a modified version of the Taylor-Proudman theorem:
\begin{align}
\label{TP_eta1}
\partial_z \Delta_{\perp} \bm{v} &= 0.
\end{align}
Repeating for $\eta_2^{\text{o}}$, we obtain
\begin{align}
\label{TP_eta2}
\partial_z (\Delta - 2\Delta_{\perp}) \bm{v} &= 0.
\end{align}
In the main text, we consider the simplifying limit $\eta_1^{\text{o}} = -2\eta_2^{\text{o}}$, in which case the theorem takes the form 
\begin{align}
\partial_z \Delta \bm{v} &= 0.
\end{align}

In an incompressible flow, all the odd viscosity coefficients compatible with cylindrical symmetry can be expressed as combinations of $\eta_1^{\text{o}}$ and $\eta_2^{\text{o}}$ (plus gradient terms that can be absorbed into the pressure). As the (modified) Taylor-Proudman theorem relies on the linearized Navier-Stokes equation (i.e., the Stokes equation), we can take linear combinations of Eqs.~\eqref{TP_eta1}-\eqref{TP_eta2} above. Therefore, modified versions of the Taylor-Proudman theorem hold for all odd viscosities compatible with cylindrical symmetry.

\subsection*{Linear stability of the fluid and odd waves}

We start with the Navier-Stokes equation
\begin{equation}
    \label{ns}
    \rho D_t v_i = - \partial_i P + \eta_{i j k \ell} \partial_j \partial_\ell v_k 
\end{equation}
with the incompressibility condition $\partial_i v_i = 0$, in which $\vec{v}(t, \vec{x})$ is the velocity field, $\rho$ is the density, $P$ is the pressure, $\eta_{i j k \ell}$ is the viscosity tensor, $D_t = (\partial_t + v_k \partial_k)$ is the convective derivative, $\partial_k = \partial/\partial x_k$. 

To analyze the linear stability of a fluid with odd viscosity, we linearize the Navier-Stokes equations about $\rho = \rho_0$ and $\vec{v}=\vec{0}$. We end up with the incompressible Stokes equations
\begin{align}
\rho_0 \partial_t v_i &= -\partial_i P + \eta_{ijk\ell} \partial_j \partial_\ell v_k\\
\partial_i v_i &= 0.
\end{align}
Considering solutions of the form $\ee^{s t + \ii \vec{k} \cdot \vec{x}}$, where $s=\sigma + \ii \omega$ is a complex growth rate, we rewrite them in Fourier space:
\begin{align}
s \rho_0 v_i &= -\ii k_i P - \eta_{ijk\ell} k_j k_\ell v_k\\
k_i v_i &= 0.
\end{align}
Using the incompressibility condition, we solve for the pressure by multiplying by $k_i$, and find
\begin{align}
    P &= \ii \eta_{ijk\ell}\frac{k_i k_j k_\ell}{k^2} v_k.
\end{align}
Plugging this back into the Stokes equation, we arrive at an equation just for the velocity:
\begin{align}
    \rho_0 s v_n = \eta_{ijk\ell} \frac{k_n k_i k_j k_\ell}{k^2} v_k - \eta_{njk\ell} k_j k_\ell v_k
    \equiv M_{nk}(\vec{k}) v_k.
\end{align}
The dispersion relation $s(\vec{k}) = \sigma(\vec{k}) + \ii \omega(\vec{k})$ is then given by the eigenvalues of $M_{nk}(\vec{k})/\rho_0$.

In the case of Eq.~\eqref{odd_ns_full} where the two independent odd viscosities $\eta_1^{\text{o}}$ and $\eta_2^{\text{o}}$ are present in addition to the normal shear viscosity $\eta$, we obtain 
\begin{align}
\label{dispersion_methods}
\rho_0\omega(\vec{k}) &= \pm  \frac{\eta_1^{\text{o}} k_z(k_x^2 + k_y^2) + \eta_2^{\text{o}} k_z (k_x^2 + k_y^2 -k_z^2)}{|k|}\\
\rho_0 \sigma(\vec{k}) &= -\eta k^2.
\end{align}
In the main text, we consider the case $\eta_1^{\text{o}} = -2\eta_2^{\text{o}}$, in which case the above dispersion relations reduce to the dispersion given in the main text, with $\nu_\odd \equiv \eta^{\text{o}}_2 / \rho_0$.

\subsection*{Helical decomposition}

In this section, we describe the decomposition of the fluxes into different manifolds according to their helical signature used in the main text. Following \cite{Waleffe1992}, we decompose the total net flux $\Pi(k)=\Pi_\textrm{homo}(k)+\Pi_\textrm{hete}(k)$ into a component due to triads with the same sign of helicity $\Pi_\textrm{homo}$ and triads with opposing signs of helicity $\Pi_\textrm{hete}$. To that extent, we decompose the flow into positively and negatively helical modes (as is also laid out in the section \textit{Effect of odd waves on the non-linear energy transfer})
\begin{align}
    \bm{v}(\bm{k},t) &= \bm{v}_+(\bm{k},t) + \bm{v}_-(\bm{k},t) \\\nonumber&= v_+(\bm{k},t)\bm{h}^+(\bm{k})+v_-(\bm{k},t)\bm{h}^-(\bm{k}),
\end{align}
Then the homochiral energy flux is obtained as
\begin{equation}
    \Pi_\textrm{homo}(k) \equiv \Pi_{(+,+,+)}(k) + \Pi_{(-,-,-)}(k),
\end{equation}
with
\begin{align}
    &\Pi_{(\pm,\pm,\pm)}(k) \\\nonumber&= \textrm{Im} \sum_{|\bm{k}|<k} \sum_{\bm{p}+\bm{q}=\bm{k}} {v}_{i,\pm}^*(\bm{k},t) P_{ij}(\bm{k})q_\ell {v}_{\ell,\pm}(\bm{p},t){v}_{j,\pm}(\bm{q},t),
\end{align}
and the other contributions to total flux are from heterochiral triads
\begin{equation}
    \Pi_\textrm{hete}(k) = \Pi(k) - \Pi_\textrm{homo}(k).
\end{equation}
It is known that in 3D turbulence, the heterochiral triads that dominate the energy flux primarily cascade energy directly, while the typically subleading homochiral triads primarily cascade energy inversely \cite{Waleffe1992,Biferale2012}.

Since we have seen that odd viscosity can manipulate the direction of the turbulent cascade, the question naturally arises how the odd viscosity modifies the relative importance of both helical manifolds. For the case of the direct cascade, we see in ED Fig.~\ref{figure_numerics_extended}a(i,ii) that as the cascade approaches $k_\odd$, the homochiral flux becomes stronger, in agreement with the notion that the flow develops a tendency to cascade energy inversely. Consequentially, as the energy starts to pile up, also the heterochiral flux is enhanced. This once more showcases the competition between the inverse cascading tendency and direct cascade that leads to the observed energetic condensation. For the inverse cascading cases in ED Fig.~\ref{figure_numerics_extended}a(iii,iv), this becomes even more striking. There, we observe that the odd viscosity produces an inverse cascade in the homochiral manifold, which is in turn balanced by a forward cascade in the heterochiral manifold, such that the total net flux vanishes in the inverse range. This self-regulating mechanism is referred to as a flux loop state \cite{Alexakis2018}, where kinetic energy is consistently looping into the condensate through the homochiral flux and then back again through the heterochiral flux. Note that to evidence this, we have employed hyperdissipation to mimic more scale separation between the dissipative scales and injection scales. 

This marks an important difference between the odd viscosity condensates resulting from a direct cascade and those resulting from an inverse cascade: while the former are saturated by dissipation as treated in the main text, the condensates from the inverse cascade are saturated by the flux balance between the helical manifolds. However, the significant magnitude of the flux in both helical manifolds indicates that also the condensates in the inverse cascading case are strongly out-of-equilibrium states.

\subsection*{Statistical steady-state of the cascade}

To confirm that we have obtained a statistical steady-state in our simulations, we assess the transient energetic content of the first modes, which are the slowest modes to equilibrate, shown in Supplementary Fig. \ref{fig:inverse_spectra_trans}. This confirms that the dynamics in the averaging window is indeed governed by fluctuations around a statistically steady state.

\begin{figure}[ht!]
    \centering
    \includegraphics[width=0.5\textwidth]{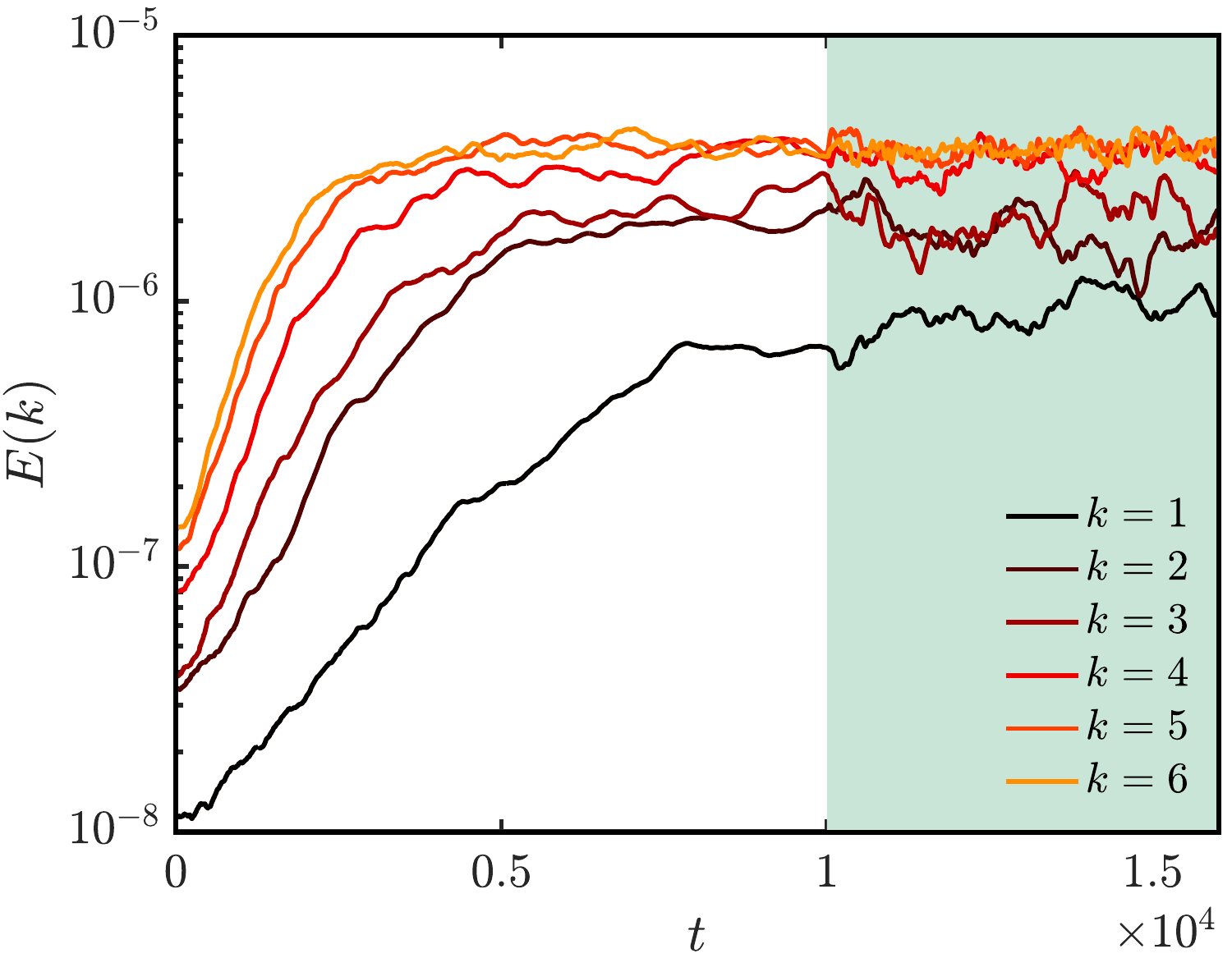}
    \caption{Transient energetic content of the first six modes for the inverse cascade case with $\nu_\odd/\nu=212$. The green highlighted interval denotes the averaging window over which the spectra are obtained.}
    \label{fig:inverse_spectra_trans}
\end{figure}

\subsection*{Odd hyperviscosity}
The crucial ingredient to observe the described turbulent pattern formation is the competing direct cascade (at small $k$) and inverse cascade (at large $k$), and a non-dissipative mechanism that discriminates between the two. In that regard, the scaling theories presented in this work can be straightforwardly extended to odd hyperviscosities of the form $\tilde{\eta}_\odd \vec{e}_z \times \Delta^p \vec{v}$ with power $p>1$, which would make the transition from 3D to quasi-2D arbitrarily sharp/smooth, rendering also the region of spectral condensation arbitrarily peaked. This provides an ideal model system to study more complex natural flow systems in which a variety of different multi-scale mechanisms could exists that produce competing cascades, with varying sharp or smooth transitions between them that are different from the $\sim k^2$ scaling.

For arbitrary $p$, we find for the dispersion of odd waves
\begin{equation}
    \tilde{\omega}(\bm{k})
    =\pm \tilde\nu_\odd k_z \abs{k}^{2p-1},
\end{equation}
yielding for the odd viscosity scale, where condensation begins
\begin{equation}
    \tilde{k}_\odd=\epsilon^{1/(6p-2)}\tilde{\nu}_\odd^{-3/(6p-2)},
\end{equation}
and for the condensation spectrum
\begin{equation}
    \tilde{E}(k) \sim \epsilon^{1/2} \tilde{\nu}_\odd^{1/2} \, k^{p-2} \qquad \text{($k \gg \tilde{k}_\odd$)},
\end{equation}
until the condensation peak is attained at
\begin{equation}
    \tilde{k}_c \sim \left( \epsilon^{1/2} \nu^{-1} \tilde{\nu}_\odd^{-1/2} \right)^{1/(p+1)}.
\end{equation}

While the main text treats the case $p=1$ where the transition from a direct cascade at small $k$ to an inverse cascade at large $k$ is smooth, we can also consider the other extreme, where the transition is fully sharp (semantically, the limit $p\to\infty$). Here the odd viscosity term $\eta_\odd \vec{e}_z \times k^2 \vec{v}$ in the Navier-Stokes equations is replaced by 
$\tilde{\eta}_\odd \theta(|\bm{k}|-|\tilde{\bm{k}}_\odd|) \vec{e}_z \times \vec{v}$
where $\tilde{k}_\odd$ is a set wavenumber in the inertial range and $\theta(|\bm{k}|-|\tilde{\bm{k}}_\odd|)$ is a step function which is non-zero for $|\bm{k}| > |\tilde{\bm{k}}_\odd|$. Since the transition is now sharp, the start of condensation and peak condensation wavenumbers coincide $\tilde{k}_c=\tilde{k}_\odd$, leading to maximum condensation sharply at a single wavelength. Numerical results for this case are provided in ED Fig.~\ref{figure_numerics_extended}e, which indeed shows a sharp condensation spectrum around $\tilde{k}_c$, as well as a diffusive equipartitioned scaling $\sim k^2$ to the left of the condensation peak, akin to what is observed in 3D turbulence on wavenumbers smaller than the injection scale. As a result, the associated pattern in the vorticity field in ED Fig.~\ref{figure_numerics_extended}f is more isotropic than in Fig.~\ref{figure_direct_spectra}b, which is consistent with the prediction of aspect ratio $\gamma=1$ from Eq.~\eqref{eq:aspect_ratio} in the Methods.

\subsection*{Rossby wave turbulence}

In this section, we review the system of 2D turbulence under the influence of Rossby waves and point out the similarities and differences with the case of odd fluids.

We consider an incompressible fluid with Cartesian coordinates ($x$, $y$). 
In the so-called $\beta$-plane approximation, the Coriolis force depends linearly on the coordinate $y$ with a coefficient $\beta$. Namely, it is $\vec{f}_\text{c} = - \beta \, y \, \epsilon \cdot \vec{v}$.
The Navier-Stokes equation reads
\begin{equation}
\partial_t \vec{v} + (\vec{v} \cdot \vec{\nabla}) \vec{v} - \beta \, y \, \epsilon \cdot \vec{v} + \vec{\nabla} p = - \alpha \vec{v} + \nu \Delta \vec{v} + \vec{f}
\end{equation}
in which $\vec{v}$ is the velocity field, $p$ the pressure field, $\epsilon$ the Levi-Civita symbol, $\nu$ the kinematic viscosity (a diffusion coefficient), $\alpha$ a friction term, and $\vec{f}$ a body force representing the velocity forcing.

Taking the curl of the Navier-Stokes equation, we find
\begin{equation}
\partial_t \omega + [\epsilon_{ij} (\partial_i v_k) (\partial_k v_j) + v_k \partial_k \omega] + \beta v_i (\partial_i y) = - \alpha \omega  + \nu \Delta \omega + f_{\omega}
\end{equation}
where $\omega = \nabla \times v = \epsilon_{ij} \partial_i v_j = \partial_x v_y - \partial_y v_x$ is the 2D vorticity, and $f_{\omega} = \epsilon_{i j} \partial_i f_j$ is a vorticity forcing.
The term $\epsilon_{ij} (\partial_i v_k) (\partial_k v_j)$ vanishes because of incompressibility, so we end up with
\begin{equation}
\partial_t \omega + (\vec{v} \cdot \nabla) \omega + \beta v_y = - \alpha \omega + \nu \Delta \omega + f_{\omega}
\end{equation}
in which we have used $\vec{v} \cdot (\vec{\nabla} y) = v_y$.

As we have an incompressible 2D flow, it is convenient to introduce a streamfunction $\psi$ such that
\begin{equation}
\vec{v} = \begin{pmatrix} v_x \\ v_y \end{pmatrix}
= \begin{pmatrix} -\partial_y \psi \\ \partial_x \psi \end{pmatrix}
= - \epsilon \cdot \vec{\nabla} \psi
\end{equation}
In addition, the 2D vorticity is $\omega = \Delta \psi$. 

We introduce the Jacobian (Poisson bracket)
\begin{equation}
J(a,b) = \frac{\partial a}{\partial x} \frac{\partial b}{\partial y} - \frac{\partial a}{\partial y} \frac{\partial b}{\partial x}.
\end{equation}
It allows us to express the convective term in terms of the streamfunction $\psi$ as
\begin{equation}
J(\psi, b) = (\vec{v} \cdot \nabla) b
\end{equation}
Then, the Navier-Stokes equation becomes
\begin{equation}
\label{chmsi}
\partial_t \omega + J(\psi, \omega) + \beta \partial_x \psi = - \alpha \omega + \nu \Delta \omega + f_{\omega}.
\end{equation}
As $\omega = \Delta \psi$, this equation is entirely written in terms of the stream function. Equation \eqref{chmsi} is known as the Charney-Hasegawa-Mima (CHM) equation~\cite{Connaughton2015,Boffetta2002,Tassi2009,Hasegawa1977,Charney1971,Horton1999,Pedlosky1979,Galperin2008}.

Removing the driving and dissipative terms, we get
\begin{equation}
\partial_t \omega + J(\psi, \omega) = - \beta \partial_x \psi
\end{equation}
with $\omega = \Delta \psi$.
Let us first consider the linearized equation
\begin{equation}
\partial_t \omega = - \beta \partial_x \psi.
\end{equation}
It has wave solutions called Rossby waves or drift waves, with dispersion relation
\begin{equation}
\def\vec#1{\mathbf{#1}}
\Omega(\vec{k}) = - \beta \frac{k_x}{k^2}.
\end{equation}

Now, we consider the decomposition of the streamfunction into the corresponding modes
\begin{equation}
\psi(t, x) = \sum_{\vec{k}} \psi(t,\vec{k}) \, e^{i [\Omega(\vec{k}) t - \vec{k} \cdot \vec{r}]}
\end{equation}
with a reality condition so that this is real (this is the equivalent of the decomposition in helical waves used in the Methods section \emph{Effect of odd waves on the non-linear energy
transfer}).

We get
\begin{equation}
\def\vec#1{\mathbf{#1}}
\partial_t \psi_{\vec{k}} = \sum_{\vec{k}+\vec{p}+\vec{q}=\vec{0}} C_{k|p,q} e^{-i[\Omega(p)+\Omega(q)+\Omega(k)]t} \psi_{\vec{p}}^*  \psi_{\vec{q}}^* 
\end{equation}
in which
\begin{equation}
C_{k|p,q}  = \frac{1}{2}\, \frac{(p_{x} q_{y} - p_{y} q_{x}) (q^2 - p^2)}{k^2}
\end{equation}

A characteristic advective timescale is $\tau^{-1}(k) = k v_k \sim k U$ where $U$ is a characteristic velocity.
Comparing with the Rossby wave frequency $\Omega(\vec{k})$ with $k_x \sim k$, we find a critical wavenumber $k_{\rm R}$ such that $\Omega \sim \tau^{-1}$ leading to $k_{\rm R} \equiv \sqrt{\beta/U}$ (called Rhines wavenumber). 
For wavenumbers $k \lesssim k_{\rm R}$, the non-linear energy transfer is reduced because of the decorrelation of the triads by Rossby waves, except for modes with $\Omega = 0$. This leads to a one-dimensionalization of the flow, and to the appearance of a scale $\sim 1/k_{\rm R}$ \cite{Diamond2005,Sukoriansky2007,Berloff2009,Chekhlov1996,Rhines1975,Rhines1979,Legras1999,Grianik2004}.
This is illustrated in ED Fig.~\ref{figure_rossby}, in which we simulate Eq.~\eqref{chmsi} using the open-source pseudospectral solver Dedalus~\cite{Burns2020}.

\subsection*{Exact solution of the equilibrium mass transfer model}

In this section, we determine an exact solution of mass transfer model with no flux ($J_{\text{in}}=J_{\text{out}}=0$). 
Setting $J_n \equiv J^{+,n} - J^{-,n} = 0$ yields the recurrence relation
\begin{equation}
    \label{mass_rec}
     c_{n+1} = k^{+,n}/k^{-,n} c_n^2.
\end{equation}
Defining $\alpha_n \equiv k^{+,n}/k^{-,n}$ and iterating, we find that
\begin{equation}
    \label{mass_sol_prod}
     c_{n+1} = \left[ \prod_{k=0}^{n-1} \alpha_{n-k}^{2^k} \right] c_1^{2^n}.
\end{equation}
With our choice Eq.~\eqref{kpmn_simple}, we find
\begin{equation}
    \alpha_{n} \equiv \frac{k^{+,n}}{k^{-,n}} = \frac{k^{+}_{n}/2}{k^{-}_{n+1}} = 
    \frac{1}{2} \, \frac{\kappa_0^+ + \kappa_1^+ \frac{N-n}{N-1}}{\kappa_0^- + \kappa_1^- \frac{n}{N-1}}
\end{equation}
We then use the identity
\begin{equation}
    \prod_{k=0}^{n-1} (a+k)^{2^k}
    =
    a \ee^{2^n \Phi^{(0,1,0)}(2,0,a+n)-2 \Phi^{(0,1,0)}(2,0,a+1)}
\end{equation}
in which $\Phi(z,s,a)$ is Lerch's Transcendent \cite[\href{https://dlmf.nist.gov/25.14}{\S 25.14}]{NIST_DLMF} and $\Phi^{(0,1,0)}$ its first derivative with respect to the second argument to find \begin{equation}
\label{mass_cascade_sol}
\begin{split}
&c_{n+1} = - \frac{a^{+}_{n}}{a^{-}_{n}}
\,
c_1^{2^n} 
\, 
\left(\frac{\kappa_1^+}{2 \kappa_1^-}\right)^{2^n - 1}
\, \times
\\
&\exp \Bigg[
2^n [\Phi^{(0,1,0)}(2,0,n+a_{+})-\Phi^{(0,1,0)}(2,0,n+a_{-})]
\\
&+2 [\Phi^{(0,1,0)}(2,0,1+a_{-})-\Phi^{(0,1,0)}(2,0,1+a_{+})]
\Bigg]
\end{split}
\end{equation}
in which 
$a^{+}_{n} = (N-1) (\kappa_0^+/\kappa_1^+) - n + N$
and
$a^{-}_{n} = - (N-1) (\kappa_0^-/\kappa_1^-) - n$.
(Numerically, iterating the recurrence relation \eqref{mass_rec} is more convenient than evaluating directly Eq.~\eqref{mass_sol_prod} or Eq.~\eqref{mass_cascade_sol}, as doing so requires some care to avoid floating point issues.)


\clearpage

\begin{thebibliography}{168}\makeatletter
\providecommand \@ifxundefined [1]{\@ifx{#1\undefined}
}\providecommand \@ifnum [1]{\ifnum #1\expandafter \@firstoftwo
 \else \expandafter \@secondoftwo
 \fi
}\providecommand \@ifx [1]{\ifx #1\expandafter \@firstoftwo
 \else \expandafter \@secondoftwo
 \fi
}\providecommand \natexlab [1]{#1}\providecommand \enquote  [1]{``#1''}\providecommand \bibnamefont  [1]{#1}\providecommand \bibfnamefont [1]{#1}\providecommand \citenamefont [1]{#1}\providecommand \href@noop [0]{\@secondoftwo}\providecommand \href [0]{\begingroup \@sanitize@url \@href}\providecommand \@href[1]{\@@startlink{#1}\@@href}\providecommand \@@href[1]{\endgroup#1\@@endlink}\providecommand \@sanitize@url [0]{\catcode `\\12\catcode `\$12\catcode
  `\&12\catcode `\#12\catcode `\^12\catcode `\_12\catcode `\%12\relax}\providecommand \@@startlink[1]{}\providecommand \@@endlink[0]{}\providecommand \url  [0]{\begingroup\@sanitize@url \@url }\providecommand \@url [1]{\endgroup\@href {#1}{\urlprefix }}\providecommand \urlprefix  [0]{URL }\providecommand \Eprint [0]{\href }\providecommand \doibase [0]{https://doi.org/}\providecommand \selectlanguage [0]{\@gobble}\providecommand \bibinfo  [0]{\@secondoftwo}\providecommand \bibfield  [0]{\@secondoftwo}\providecommand \translation [1]{[#1]}\providecommand \BibitemOpen [0]{}\providecommand \bibitemStop [0]{}\providecommand \bibitemNoStop [0]{.\EOS\space}\providecommand \EOS [0]{\spacefactor3000\relax}\providecommand \BibitemShut  [1]{\csname bibitem#1\endcsname}\let\auto@bib@innerbib\@empty
\bibitem [{\citenamefont {Cardy}\ \emph {et~al.}(2008)\citenamefont {Cardy},
  \citenamefont {Falkovich}, \citenamefont {Gaw{\k{e}}dzki}, \citenamefont
  {Nazarenko},\ and\ \citenamefont {Zaboronski}}]{Cardy2008}\BibitemOpen
  \bibfield  {author} {\bibinfo {author} {\bibfnamefont {J.}~\bibnamefont
  {Cardy}}, \bibinfo {author} {\bibfnamefont {G.}~\bibnamefont {Falkovich}},
  \bibinfo {author} {\bibfnamefont {K.}~\bibnamefont {Gaw{\k{e}}dzki}},
  \bibinfo {author} {\bibfnamefont {S.}~\bibnamefont {Nazarenko}},\ and\
  \bibinfo {author} {\bibfnamefont {O.}~\bibnamefont {Zaboronski}},\
  }\href@noop {} {\emph {\bibinfo {title} {Non-equilibrium Statistical
  Mechanics and Turbulence}}},\ London Mathematical Society Lecture Note
  Series\ (\bibinfo  {publisher} {Cambridge University Press},\ \bibinfo {year}
  {2008})\BibitemShut {NoStop}\bibitem [{\citenamefont {Davidson}(2015)}]{Davidson2015}\BibitemOpen
  \bibfield  {author} {\bibinfo {author} {\bibfnamefont {P.}~\bibnamefont
  {Davidson}},\ }\href@noop {} {\emph {\bibinfo {title} {Turbulence: An
  Introduction for Scientists and Engineers}}}\ (\bibinfo  {publisher} {Oxford
  University Press},\ \bibinfo {year} {2015})\BibitemShut {NoStop}\bibitem [{\citenamefont {Falkovich}\ \emph {et~al.}(2001)\citenamefont
  {Falkovich}, \citenamefont {Gaw{\k{e}}dzki},\ and\ \citenamefont
  {Vergassola}}]{Falkovich2001}\BibitemOpen
  \bibfield  {author} {\bibinfo {author} {\bibfnamefont {G.}~\bibnamefont
  {Falkovich}}, \bibinfo {author} {\bibfnamefont {K.}~\bibnamefont
  {Gaw{\k{e}}dzki}},\ and\ \bibinfo {author} {\bibfnamefont {M.}~\bibnamefont
  {Vergassola}},\ }\href {https://doi.org/10.1103/revmodphys.73.913} {\bibfield
   {journal} {\bibinfo  {journal} {Reviews of Modern Physics}\ }\textbf
  {\bibinfo {volume} {73}},\ \bibinfo {pages} {913} (\bibinfo {year}
  {2001})}\BibitemShut {NoStop}\bibitem [{\citenamefont {Alexakis}\ and\ \citenamefont
  {Biferale}(2018)}]{Alexakis2018}\BibitemOpen
  \bibfield  {author} {\bibinfo {author} {\bibfnamefont {A.}~\bibnamefont
  {Alexakis}}\ and\ \bibinfo {author} {\bibfnamefont {L.}~\bibnamefont
  {Biferale}},\ }\href {https://doi.org/10.1016/j.physrep.2018.08.001}
  {\bibfield  {journal} {\bibinfo  {journal} {Physics Reports}\ }\textbf
  {\bibinfo {volume} {767}},\ \bibinfo {pages} {1} (\bibinfo {year}
  {2018})}\BibitemShut {NoStop}\bibitem [{\citenamefont {Eyink}\ and\ \citenamefont
  {Sreenivasan}(2006)}]{Eyink2006}\BibitemOpen
  \bibfield  {author} {\bibinfo {author} {\bibfnamefont {G.~L.}\ \bibnamefont
  {Eyink}}\ and\ \bibinfo {author} {\bibfnamefont {K.~R.}\ \bibnamefont
  {Sreenivasan}},\ }\href {https://doi.org/10.1103/revmodphys.78.87} {\bibfield
   {journal} {\bibinfo  {journal} {Reviews of Modern Physics}\ }\textbf
  {\bibinfo {volume} {78}},\ \bibinfo {pages} {87} (\bibinfo {year}
  {2006})}\BibitemShut {NoStop}\bibitem [{\citenamefont {Frisch}\ and\ \citenamefont
  {Kolmogorov}(1995)}]{Frisch1995}\BibitemOpen
  \bibfield  {author} {\bibinfo {author} {\bibfnamefont {U.}~\bibnamefont
  {Frisch}}\ and\ \bibinfo {author} {\bibfnamefont {A.~N.}\ \bibnamefont
  {Kolmogorov}},\ }\href@noop {} {\emph {\bibinfo {title} {Turbulence: the
  legacy of AN Kolmogorov}}}\ (\bibinfo  {publisher} {Cambridge university
  press},\ \bibinfo {year} {1995})\BibitemShut {NoStop}\bibitem [{\citenamefont {Cross}\ and\ \citenamefont
  {Hohenberg}(1993)}]{Cross1993}\BibitemOpen
  \bibfield  {author} {\bibinfo {author} {\bibfnamefont {M.~C.}\ \bibnamefont
  {Cross}}\ and\ \bibinfo {author} {\bibfnamefont {P.~C.}\ \bibnamefont
  {Hohenberg}},\ }\href {https://doi.org/10.1103/revmodphys.65.851} {\bibfield
  {journal} {\bibinfo  {journal} {Reviews of Modern Physics}\ }\textbf
  {\bibinfo {volume} {65}},\ \bibinfo {pages} {851–1112} (\bibinfo {year}
  {1993})}\BibitemShut {NoStop}\bibitem [{\citenamefont {Avron}(1998)}]{Avron1998}\BibitemOpen
  \bibfield  {author} {\bibinfo {author} {\bibfnamefont {J.~E.}\ \bibnamefont
  {Avron}},\ }\href {https://doi.org/10.1023/a:1023084404080} {\bibfield
  {journal} {\bibinfo  {journal} {Journal of Statistical Physics}\ }\textbf
  {\bibinfo {volume} {92}},\ \bibinfo {pages} {543–557} (\bibinfo {year}
  {1998})}\BibitemShut {NoStop}\bibitem [{\citenamefont {Fruchart}\ \emph {et~al.}(2023)\citenamefont
  {Fruchart}, \citenamefont {Scheibner},\ and\ \citenamefont
  {Vitelli}}]{oddreview}\BibitemOpen
  \bibfield  {author} {\bibinfo {author} {\bibfnamefont {M.}~\bibnamefont
  {Fruchart}}, \bibinfo {author} {\bibfnamefont {C.}~\bibnamefont
  {Scheibner}},\ and\ \bibinfo {author} {\bibfnamefont {V.}~\bibnamefont
  {Vitelli}},\ }\href
  {https://doi.org/10.1146/annurev-conmatphys-040821-125506} {\bibfield
  {journal} {\bibinfo  {journal} {Annual Review of Condensed Matter Physics}\
  }\textbf {\bibinfo {volume} {14}},\ \bibinfo {pages} {471} (\bibinfo {year}
  {2023})}\BibitemShut {NoStop}\bibitem [{\citenamefont {Berdyugin}\ \emph {et~al.}(2019)\citenamefont
  {Berdyugin}, \citenamefont {Xu}, \citenamefont {Pellegrino}, \citenamefont
  {Kumar}, \citenamefont {Principi}, \citenamefont {Torre}, \citenamefont
  {Shalom}, \citenamefont {Taniguchi}, \citenamefont {Watanabe}, \citenamefont
  {Grigorieva}, \citenamefont {Polini}, \citenamefont {Geim},\ and\
  \citenamefont {Bandurin}}]{Berdyugin2019}\BibitemOpen
  \bibfield  {author} {\bibinfo {author} {\bibfnamefont {A.~I.}\ \bibnamefont
  {Berdyugin}}, \bibinfo {author} {\bibfnamefont {S.~G.}\ \bibnamefont {Xu}},
  \bibinfo {author} {\bibfnamefont {F.~M.~D.}\ \bibnamefont {Pellegrino}},
  \bibinfo {author} {\bibfnamefont {R.~K.}\ \bibnamefont {Kumar}}, \bibinfo
  {author} {\bibfnamefont {A.}~\bibnamefont {Principi}}, \bibinfo {author}
  {\bibfnamefont {I.}~\bibnamefont {Torre}}, \bibinfo {author} {\bibfnamefont
  {M.~B.}\ \bibnamefont {Shalom}}, \bibinfo {author} {\bibfnamefont
  {T.}~\bibnamefont {Taniguchi}}, \bibinfo {author} {\bibfnamefont
  {K.}~\bibnamefont {Watanabe}}, \bibinfo {author} {\bibfnamefont {I.~V.}\
  \bibnamefont {Grigorieva}}, \bibinfo {author} {\bibfnamefont
  {M.}~\bibnamefont {Polini}}, \bibinfo {author} {\bibfnamefont {A.~K.}\
  \bibnamefont {Geim}},\ and\ \bibinfo {author} {\bibfnamefont {D.~A.}\
  \bibnamefont {Bandurin}},\ }\href {https://doi.org/10.1126/science.aau0685}
  {\bibfield  {journal} {\bibinfo  {journal} {Science}\ }\textbf {\bibinfo
  {volume} {364}},\ \bibinfo {pages} {162–165} (\bibinfo {year}
  {2019})}\BibitemShut {NoStop}\bibitem [{\citenamefont {Morrison}\ \emph {et~al.}(1984)\citenamefont
  {Morrison}, \citenamefont {Caldas},\ and\ \citenamefont
  {Tasso}}]{Morrison1984}\BibitemOpen
  \bibfield  {author} {\bibinfo {author} {\bibfnamefont {P.~J.}\ \bibnamefont
  {Morrison}}, \bibinfo {author} {\bibfnamefont {I.~L.}\ \bibnamefont
  {Caldas}},\ and\ \bibinfo {author} {\bibfnamefont {H.}~\bibnamefont
  {Tasso}},\ }\href {https://doi.org/10.1515/zna-1984-1102} {\bibfield
  {journal} {\bibinfo  {journal} {Zeitschrift für Naturforschung A}\ }\textbf
  {\bibinfo {volume} {39}},\ \bibinfo {pages} {1023–1027} (\bibinfo {year}
  {1984})}\BibitemShut {NoStop}\bibitem [{\citenamefont {Van~Saarloos}\ \emph {et~al.}(2023)\citenamefont
  {Van~Saarloos}, \citenamefont {Vitelli},\ and\ \citenamefont
  {Zeravcic}}]{Saarloos2023}\BibitemOpen
  \bibfield  {author} {\bibinfo {author} {\bibfnamefont {W.}~\bibnamefont
  {Van~Saarloos}}, \bibinfo {author} {\bibfnamefont {V.}~\bibnamefont
  {Vitelli}},\ and\ \bibinfo {author} {\bibfnamefont {Z.}~\bibnamefont
  {Zeravcic}},\ }\href@noop {} {\emph {\bibinfo {title} {Soft Matter: Concepts,
  Phenomena and Applications.}}}\ (\bibinfo  {publisher} {Princeton University
  Press},\ \bibinfo {year} {2023})\BibitemShut {NoStop}\bibitem [{\citenamefont {Diamond}\ \emph {et~al.}(2005)\citenamefont
  {Diamond}, \citenamefont {Itoh}, \citenamefont {Itoh},\ and\ \citenamefont
  {Hahm}}]{Diamond2005}\BibitemOpen
  \bibfield  {author} {\bibinfo {author} {\bibfnamefont {P.~H.}\ \bibnamefont
  {Diamond}}, \bibinfo {author} {\bibfnamefont {S.-I.}\ \bibnamefont {Itoh}},
  \bibinfo {author} {\bibfnamefont {K.}~\bibnamefont {Itoh}},\ and\ \bibinfo
  {author} {\bibfnamefont {T.~S.}\ \bibnamefont {Hahm}},\ }\href
  {https://doi.org/10.1088/0741-3335/47/5/r01} {\bibfield  {journal} {\bibinfo
  {journal} {Plasma Physics and Controlled Fusion}\ }\textbf {\bibinfo {volume}
  {47}},\ \bibinfo {pages} {R35} (\bibinfo {year} {2005})}\BibitemShut
  {NoStop}\bibitem [{\citenamefont {Sukoriansky}\ \emph {et~al.}(2007)\citenamefont
  {Sukoriansky}, \citenamefont {Dikovskaya},\ and\ \citenamefont
  {Galperin}}]{Sukoriansky2007}\BibitemOpen
  \bibfield  {author} {\bibinfo {author} {\bibfnamefont {S.}~\bibnamefont
  {Sukoriansky}}, \bibinfo {author} {\bibfnamefont {N.}~\bibnamefont
  {Dikovskaya}},\ and\ \bibinfo {author} {\bibfnamefont {B.}~\bibnamefont
  {Galperin}},\ }\href {https://doi.org/10.1175/jas4013.1} {\bibfield
  {journal} {\bibinfo  {journal} {Journal of the Atmospheric Sciences}\
  }\textbf {\bibinfo {volume} {64}},\ \bibinfo {pages} {3312} (\bibinfo {year}
  {2007})}\BibitemShut {NoStop}\bibitem [{\citenamefont {Berloff}\ \emph {et~al.}(2009)\citenamefont
  {Berloff}, \citenamefont {Kamenkovich},\ and\ \citenamefont
  {Pedlosky}}]{Berloff2009}\BibitemOpen
  \bibfield  {author} {\bibinfo {author} {\bibfnamefont {P.}~\bibnamefont
  {Berloff}}, \bibinfo {author} {\bibfnamefont {I.}~\bibnamefont
  {Kamenkovich}},\ and\ \bibinfo {author} {\bibfnamefont {J.}~\bibnamefont
  {Pedlosky}},\ }\href {https://doi.org/10.1017/s0022112009006375} {\bibfield
  {journal} {\bibinfo  {journal} {Journal of Fluid Mechanics}\ }\textbf
  {\bibinfo {volume} {628}},\ \bibinfo {pages} {395} (\bibinfo {year}
  {2009})}\BibitemShut {NoStop}\bibitem [{\citenamefont {Chekhlov}\ \emph {et~al.}(1996)\citenamefont
  {Chekhlov}, \citenamefont {Orszag}, \citenamefont {Sukoriansky},
  \citenamefont {Galperin},\ and\ \citenamefont {Staroselsky}}]{Chekhlov1996}\BibitemOpen
  \bibfield  {author} {\bibinfo {author} {\bibfnamefont {A.}~\bibnamefont
  {Chekhlov}}, \bibinfo {author} {\bibfnamefont {S.~A.}\ \bibnamefont
  {Orszag}}, \bibinfo {author} {\bibfnamefont {S.}~\bibnamefont {Sukoriansky}},
  \bibinfo {author} {\bibfnamefont {B.}~\bibnamefont {Galperin}},\ and\
  \bibinfo {author} {\bibfnamefont {I.}~\bibnamefont {Staroselsky}},\ }\href
  {https://doi.org/10.1016/0167-2789(96)00102-9} {\bibfield  {journal}
  {\bibinfo  {journal} {Physica D: Nonlinear Phenomena}\ }\textbf {\bibinfo
  {volume} {98}},\ \bibinfo {pages} {321} (\bibinfo {year} {1996})}\BibitemShut
  {NoStop}\bibitem [{\citenamefont {Rhines}(1979)}]{Rhines1979}\BibitemOpen
  \bibfield  {author} {\bibinfo {author} {\bibfnamefont {P.~B.}\ \bibnamefont
  {Rhines}},\ }\href {https://doi.org/10.1146/annurev.fl.11.010179.002153}
  {\bibfield  {journal} {\bibinfo  {journal} {Annual Review of Fluid
  Mechanics}\ }\textbf {\bibinfo {volume} {11}},\ \bibinfo {pages} {401}
  (\bibinfo {year} {1979})}\BibitemShut {NoStop}\bibitem [{\citenamefont {Legras}\ \emph {et~al.}(1999)\citenamefont {Legras},
  \citenamefont {Villone},\ and\ \citenamefont {Frisch}}]{Legras1999}\BibitemOpen
  \bibfield  {author} {\bibinfo {author} {\bibfnamefont {B.}~\bibnamefont
  {Legras}}, \bibinfo {author} {\bibfnamefont {B.}~\bibnamefont {Villone}},\
  and\ \bibinfo {author} {\bibfnamefont {U.}~\bibnamefont {Frisch}},\ }\href
  {https://doi.org/10.1103/physrevlett.82.4440} {\bibfield  {journal} {\bibinfo
   {journal} {Physical Review Letters}\ }\textbf {\bibinfo {volume} {82}},\
  \bibinfo {pages} {4440} (\bibinfo {year} {1999})}\BibitemShut {NoStop}\bibitem [{\citenamefont {Grianik}\ \emph {et~al.}(2004)\citenamefont
  {Grianik}, \citenamefont {Held}, \citenamefont {Smith},\ and\ \citenamefont
  {Vallis}}]{Grianik2004}\BibitemOpen
  \bibfield  {author} {\bibinfo {author} {\bibfnamefont {N.}~\bibnamefont
  {Grianik}}, \bibinfo {author} {\bibfnamefont {I.~M.}\ \bibnamefont {Held}},
  \bibinfo {author} {\bibfnamefont {K.~S.}\ \bibnamefont {Smith}},\ and\
  \bibinfo {author} {\bibfnamefont {G.~K.}\ \bibnamefont {Vallis}},\ }\href
  {https://doi.org/10.1063/1.1630054} {\bibfield  {journal} {\bibinfo
  {journal} {Physics of Fluids}\ }\textbf {\bibinfo {volume} {16}},\ \bibinfo
  {pages} {73} (\bibinfo {year} {2004})}\BibitemShut {NoStop}\bibitem [{\citenamefont {Squire}\ \emph {et~al.}(2022)\citenamefont {Squire},
  \citenamefont {Meyrand}, \citenamefont {Kunz}, \citenamefont {Arzamasskiy},
  \citenamefont {Schekochihin},\ and\ \citenamefont {Quataert}}]{Squire2022}\BibitemOpen
  \bibfield  {author} {\bibinfo {author} {\bibfnamefont {J.}~\bibnamefont
  {Squire}}, \bibinfo {author} {\bibfnamefont {R.}~\bibnamefont {Meyrand}},
  \bibinfo {author} {\bibfnamefont {M.~W.}\ \bibnamefont {Kunz}}, \bibinfo
  {author} {\bibfnamefont {L.}~\bibnamefont {Arzamasskiy}}, \bibinfo {author}
  {\bibfnamefont {A.~A.}\ \bibnamefont {Schekochihin}},\ and\ \bibinfo {author}
  {\bibfnamefont {E.}~\bibnamefont {Quataert}},\ }\href
  {https://doi.org/10.1038/s41550-022-01624-z} {\bibfield  {journal} {\bibinfo
  {journal} {Nature Astronomy}\ }\textbf {\bibinfo {volume} {6}},\ \bibinfo
  {pages} {715} (\bibinfo {year} {2022})}\BibitemShut {NoStop}\bibitem [{\citenamefont {Meyrand}\ \emph {et~al.}(2021)\citenamefont
  {Meyrand}, \citenamefont {Squire}, \citenamefont {Schekochihin},\ and\
  \citenamefont {Dorland}}]{Meyrand2021}\BibitemOpen
  \bibfield  {author} {\bibinfo {author} {\bibfnamefont {R.}~\bibnamefont
  {Meyrand}}, \bibinfo {author} {\bibfnamefont {J.}~\bibnamefont {Squire}},
  \bibinfo {author} {\bibfnamefont {A.}~\bibnamefont {Schekochihin}},\ and\
  \bibinfo {author} {\bibfnamefont {W.}~\bibnamefont {Dorland}},\ }\href
  {https://doi.org/10.1017/s0022377821000489} {\bibfield  {journal} {\bibinfo
  {journal} {Journal of Plasma Physics}\ }\textbf {\bibinfo {volume} {87}},\
  \bibinfo {pages} {535870301} (\bibinfo {year} {2021})}\BibitemShut {NoStop}\bibitem [{\citenamefont {Miloshevich}\ \emph {et~al.}(2021)\citenamefont
  {Miloshevich}, \citenamefont {Laveder}, \citenamefont {Passot},\ and\
  \citenamefont {Sulem}}]{Miloshevich2021}\BibitemOpen
  \bibfield  {author} {\bibinfo {author} {\bibfnamefont {G.}~\bibnamefont
  {Miloshevich}}, \bibinfo {author} {\bibfnamefont {D.}~\bibnamefont
  {Laveder}}, \bibinfo {author} {\bibfnamefont {T.}~\bibnamefont {Passot}},\
  and\ \bibinfo {author} {\bibfnamefont {P.~L.}\ \bibnamefont {Sulem}},\ }\href
  {https://doi.org/10.1017/s0022377820001531} {\bibfield  {journal} {\bibinfo
  {journal} {Journal of Plasma Physics}\ }\textbf {\bibinfo {volume} {87}},\
  \bibinfo {pages} {905870201} (\bibinfo {year} {2021})}\BibitemShut {NoStop}\bibitem [{\citenamefont {Krapivsky}\ \emph {et~al.}(2010)\citenamefont
  {Krapivsky}, \citenamefont {Redner},\ and\ \citenamefont
  {Ben-Naim}}]{Krapivsky2010}\BibitemOpen
  \bibfield  {author} {\bibinfo {author} {\bibfnamefont {P.}~\bibnamefont
  {Krapivsky}}, \bibinfo {author} {\bibfnamefont {S.}~\bibnamefont {Redner}},\
  and\ \bibinfo {author} {\bibfnamefont {E.}~\bibnamefont {Ben-Naim}},\
  }\href@noop {} {\emph {\bibinfo {title} {A Kinetic View of Statistical
  Physics}}}\ (\bibinfo  {publisher} {Cambridge University Press},\ \bibinfo
  {year} {2010})\BibitemShut {NoStop}\bibitem [{\citenamefont {Testik}\ and\ \citenamefont
  {Barros}(2007)}]{Testik2007}\BibitemOpen
  \bibfield  {author} {\bibinfo {author} {\bibfnamefont {F.~Y.}\ \bibnamefont
  {Testik}}\ and\ \bibinfo {author} {\bibfnamefont {A.~P.}\ \bibnamefont
  {Barros}},\ }\bibfield  {journal} {\bibinfo  {journal} {Reviews of
  Geophysics}\ }\textbf {\bibinfo {volume} {45}},\ \href
  {https://doi.org/10.1029/2005rg000182} {10.1029/2005rg000182} (\bibinfo
  {year} {2007})\BibitemShut {NoStop}\bibitem [{\citenamefont {Friedlander}\ \emph {et~al.}(2000)\citenamefont
  {Friedlander} \emph {et~al.}}]{Friedlander2000}\BibitemOpen
  \bibfield  {author} {\bibinfo {author} {\bibfnamefont {S.~K.}\ \bibnamefont
  {Friedlander}} \emph {et~al.},\ }\href@noop {} {\emph {\bibinfo {title}
  {Smoke, dust, and haze}}},\ Vol.\ \bibinfo {volume} {198}\ (\bibinfo
  {publisher} {Oxford university press New York},\ \bibinfo {year}
  {2000})\BibitemShut {NoStop}\bibitem [{\citenamefont {Zakharov}\ \emph {et~al.}(2012)\citenamefont
  {Zakharov}, \citenamefont {L'vov},\ and\ \citenamefont
  {Falkovich}}]{Zakharov2012}\BibitemOpen
  \bibfield  {author} {\bibinfo {author} {\bibfnamefont {V.~E.}\ \bibnamefont
  {Zakharov}}, \bibinfo {author} {\bibfnamefont {V.~S.}\ \bibnamefont
  {L'vov}},\ and\ \bibinfo {author} {\bibfnamefont {G.}~\bibnamefont
  {Falkovich}},\ }\href@noop {} {\emph {\bibinfo {title} {Kolmogorov Spectra Of
  Turbulence I - Wave Turbulence}}}\ (\bibinfo  {publisher} {Springer},\
  \bibinfo {year} {2012})\BibitemShut {NoStop}\bibitem [{\citenamefont {Nazarenko}(2011)}]{Nazarenko2011}\BibitemOpen
  \bibfield  {author} {\bibinfo {author} {\bibfnamefont {S.}~\bibnamefont
  {Nazarenko}},\ }\href@noop {} {\emph {\bibinfo {title} {Wave Turbulence}}}\
  (\bibinfo  {publisher} {Springer Science and Business Media LLC},\ \bibinfo
  {year} {2011})\BibitemShut {NoStop}\bibitem [{\citenamefont {Galtier}(2022)}]{Galtier2022}\BibitemOpen
  \bibfield  {author} {\bibinfo {author} {\bibfnamefont {S.}~\bibnamefont
  {Galtier}},\ }\href@noop {} {\emph {\bibinfo {title} {Physics Of Wave
  Turbulence}}}\ (\bibinfo  {publisher} {Cambridge University Press},\ \bibinfo
  {year} {2022})\BibitemShut {NoStop}\bibitem [{\citenamefont {Newell}\ and\ \citenamefont
  {Rumpf}(2011)}]{Newell2011}\BibitemOpen
  \bibfield  {author} {\bibinfo {author} {\bibfnamefont {A.~C.}\ \bibnamefont
  {Newell}}\ and\ \bibinfo {author} {\bibfnamefont {B.}~\bibnamefont {Rumpf}},\
  }\href {https://doi.org/10.1146/annurev-fluid-122109-160807} {\bibfield
  {journal} {\bibinfo  {journal} {Annual Review of Fluid Mechanics}\ }\textbf
  {\bibinfo {volume} {43}},\ \bibinfo {pages} {59–78} (\bibinfo {year}
  {2011})}\BibitemShut {NoStop}\bibitem [{\citenamefont {Biferale}\ \emph {et~al.}(2012)\citenamefont
  {Biferale}, \citenamefont {Musacchio},\ and\ \citenamefont
  {Toschi}}]{Biferale2012}\BibitemOpen
  \bibfield  {author} {\bibinfo {author} {\bibfnamefont {L.}~\bibnamefont
  {Biferale}}, \bibinfo {author} {\bibfnamefont {S.}~\bibnamefont
  {Musacchio}},\ and\ \bibinfo {author} {\bibfnamefont {F.}~\bibnamefont
  {Toschi}},\ }\href {https://doi.org/10.1103/physrevlett.108.164501}
  {\bibfield  {journal} {\bibinfo  {journal} {Physical Review Letters}\
  }\textbf {\bibinfo {volume} {108}},\ \bibinfo {pages} {164501} (\bibinfo
  {year} {2012})}\BibitemShut {NoStop}\bibitem [{\citenamefont {S{\l}omka}\ and\ \citenamefont
  {Dunkel}(2017)}]{Slomka2017}\BibitemOpen
  \bibfield  {author} {\bibinfo {author} {\bibfnamefont {J.}~\bibnamefont
  {S{\l}omka}}\ and\ \bibinfo {author} {\bibfnamefont {J.}~\bibnamefont
  {Dunkel}},\ }\href {https://doi.org/10.1073/pnas.1614721114} {\bibfield
  {journal} {\bibinfo  {journal} {Proceedings of the National Academy of
  Sciences}\ }\textbf {\bibinfo {volume} {114}},\ \bibinfo {pages} {2119}
  (\bibinfo {year} {2017})}\BibitemShut {NoStop}\bibitem [{\citenamefont {Xia}\ \emph {et~al.}(2011)\citenamefont {Xia},
  \citenamefont {Byrne}, \citenamefont {Falkovich},\ and\ \citenamefont
  {Shats}}]{Xia2011}\BibitemOpen
  \bibfield  {author} {\bibinfo {author} {\bibfnamefont {H.}~\bibnamefont
  {Xia}}, \bibinfo {author} {\bibfnamefont {D.}~\bibnamefont {Byrne}}, \bibinfo
  {author} {\bibfnamefont {G.}~\bibnamefont {Falkovich}},\ and\ \bibinfo
  {author} {\bibfnamefont {M.}~\bibnamefont {Shats}},\ }\href
  {https://doi.org/10.1038/nphys1910} {\bibfield  {journal} {\bibinfo
  {journal} {Nature Physics}\ }\textbf {\bibinfo {volume} {7}},\ \bibinfo
  {pages} {321–324} (\bibinfo {year} {2011})}\BibitemShut {NoStop}\bibitem [{\citenamefont {Khain}\ \emph {et~al.}(2022)\citenamefont {Khain},
  \citenamefont {Scheibner}, \citenamefont {Fruchart},\ and\ \citenamefont
  {Vitelli}}]{Khain2022}\BibitemOpen
  \bibfield  {author} {\bibinfo {author} {\bibfnamefont {T.}~\bibnamefont
  {Khain}}, \bibinfo {author} {\bibfnamefont {C.}~\bibnamefont {Scheibner}},
  \bibinfo {author} {\bibfnamefont {M.}~\bibnamefont {Fruchart}},\ and\
  \bibinfo {author} {\bibfnamefont {V.}~\bibnamefont {Vitelli}},\ }\href
  {https://doi.org/10.1017/jfm.2021.1079} {\bibfield  {journal} {\bibinfo
  {journal} {Journal of Fluid Mechanics}\ }\textbf {\bibinfo {volume} {934}},\
  \bibinfo {pages} {A23} (\bibinfo {year} {2022})}\BibitemShut {NoStop}\bibitem [{\citenamefont {Beenakker}\ and\ \citenamefont
  {McCourt}(1970)}]{Beenakker1970}\BibitemOpen
  \bibfield  {author} {\bibinfo {author} {\bibfnamefont {J.~J.~M.}\
  \bibnamefont {Beenakker}}\ and\ \bibinfo {author} {\bibfnamefont {F.~R.}\
  \bibnamefont {McCourt}},\ }\href
  {https://doi.org/10.1146/annurev.pc.21.100170.000403} {\bibfield  {journal}
  {\bibinfo  {journal} {Annual Review of Physical Chemistry}\ }\textbf
  {\bibinfo {volume} {21}},\ \bibinfo {pages} {47} (\bibinfo {year}
  {1970})}\BibitemShut {NoStop}\bibitem [{\citenamefont {Soni}\ \emph {et~al.}(2019)\citenamefont {Soni},
  \citenamefont {Bililign}, \citenamefont {Magkiriadou}, \citenamefont
  {Sacanna}, \citenamefont {Bartolo}, \citenamefont {Shelley},\ and\
  \citenamefont {Irvine}}]{Soni2019}\BibitemOpen
  \bibfield  {author} {\bibinfo {author} {\bibfnamefont {V.}~\bibnamefont
  {Soni}}, \bibinfo {author} {\bibfnamefont {E.~S.}\ \bibnamefont {Bililign}},
  \bibinfo {author} {\bibfnamefont {S.}~\bibnamefont {Magkiriadou}}, \bibinfo
  {author} {\bibfnamefont {S.}~\bibnamefont {Sacanna}}, \bibinfo {author}
  {\bibfnamefont {D.}~\bibnamefont {Bartolo}}, \bibinfo {author} {\bibfnamefont
  {M.~J.}\ \bibnamefont {Shelley}},\ and\ \bibinfo {author} {\bibfnamefont
  {W.~T.~M.}\ \bibnamefont {Irvine}},\ }\href
  {https://doi.org/10.1038/s41567-019-0603-8} {\bibfield  {journal} {\bibinfo
  {journal} {Nature Physics}\ }\textbf {\bibinfo {volume} {15}},\ \bibinfo
  {pages} {1188–1194} (\bibinfo {year} {2019})}\BibitemShut {NoStop}\bibitem [{\citenamefont {Biferale}\ \emph {et~al.}(2016)\citenamefont
  {Biferale}, \citenamefont {Bonaccorso}, \citenamefont {Mazzitelli},
  \citenamefont {van Hinsberg}, \citenamefont {Lanotte}, \citenamefont
  {Musacchio}, \citenamefont {Perlekar},\ and\ \citenamefont
  {Toschi}}]{Biferale2016}\BibitemOpen
  \bibfield  {author} {\bibinfo {author} {\bibfnamefont {L.}~\bibnamefont
  {Biferale}}, \bibinfo {author} {\bibfnamefont {F.}~\bibnamefont
  {Bonaccorso}}, \bibinfo {author} {\bibfnamefont {I.~M.}\ \bibnamefont
  {Mazzitelli}}, \bibinfo {author} {\bibfnamefont {M.~A.~T.}\ \bibnamefont {van
  Hinsberg}}, \bibinfo {author} {\bibfnamefont {A.~S.}\ \bibnamefont
  {Lanotte}}, \bibinfo {author} {\bibfnamefont {S.}~\bibnamefont {Musacchio}},
  \bibinfo {author} {\bibfnamefont {P.}~\bibnamefont {Perlekar}},\ and\
  \bibinfo {author} {\bibfnamefont {F.}~\bibnamefont {Toschi}},\ }\href
  {https://doi.org/10.1103/physrevx.6.041036} {\bibfield  {journal} {\bibinfo
  {journal} {Physical Review X}\ }\textbf {\bibinfo {volume} {6}},\ \bibinfo
  {pages} {041036} (\bibinfo {year} {2016})}\BibitemShut {NoStop}\bibitem [{\citenamefont {Buzzicotti}\ \emph {et~al.}(2018)\citenamefont
  {Buzzicotti}, \citenamefont {Aluie}, \citenamefont {Biferale},\ and\
  \citenamefont {Linkmann}}]{Buzzicotti2018}\BibitemOpen
  \bibfield  {author} {\bibinfo {author} {\bibfnamefont {M.}~\bibnamefont
  {Buzzicotti}}, \bibinfo {author} {\bibfnamefont {H.}~\bibnamefont {Aluie}},
  \bibinfo {author} {\bibfnamefont {L.}~\bibnamefont {Biferale}},\ and\
  \bibinfo {author} {\bibfnamefont {M.}~\bibnamefont {Linkmann}},\ }\href
  {https://doi.org/10.1103/physrevfluids.3.034802} {\bibfield  {journal}
  {\bibinfo  {journal} {Physical Review Fluids}\ }\textbf {\bibinfo {volume}
  {3}},\ \bibinfo {pages} {034802} (\bibinfo {year} {2018})}\BibitemShut
  {NoStop}\bibitem [{\citenamefont {Deusebio}\ \emph {et~al.}(2014)\citenamefont
  {Deusebio}, \citenamefont {Boffetta}, \citenamefont {Lindborg},\ and\
  \citenamefont {Musacchio}}]{Deusebio2014}\BibitemOpen
  \bibfield  {author} {\bibinfo {author} {\bibfnamefont {E.}~\bibnamefont
  {Deusebio}}, \bibinfo {author} {\bibfnamefont {G.}~\bibnamefont {Boffetta}},
  \bibinfo {author} {\bibfnamefont {E.}~\bibnamefont {Lindborg}},\ and\
  \bibinfo {author} {\bibfnamefont {S.}~\bibnamefont {Musacchio}},\ }\href
  {https://doi.org/10.1103/physreve.90.023005} {\bibfield  {journal} {\bibinfo
  {journal} {Physical Review E}\ }\textbf {\bibinfo {volume} {90}},\ \bibinfo
  {pages} {023005} (\bibinfo {year} {2014})}\BibitemShut {NoStop}\bibitem [{\citenamefont {Smith}\ and\ \citenamefont
  {Waleffe}(1999)}]{Smith1999}\BibitemOpen
  \bibfield  {author} {\bibinfo {author} {\bibfnamefont {L.~M.}\ \bibnamefont
  {Smith}}\ and\ \bibinfo {author} {\bibfnamefont {F.}~\bibnamefont
  {Waleffe}},\ }\href {https://doi.org/10.1063/1.870022} {\bibfield  {journal}
  {\bibinfo  {journal} {Physics of Fluids}\ }\textbf {\bibinfo {volume} {11}},\
  \bibinfo {pages} {1608} (\bibinfo {year} {1999})}\BibitemShut {NoStop}\bibitem [{\citenamefont {Zeman}(1994)}]{Zeman1998}\BibitemOpen
  \bibfield  {author} {\bibinfo {author} {\bibfnamefont {O.}~\bibnamefont
  {Zeman}},\ }\href {https://doi.org/10.1063/1.868053} {\bibfield  {journal}
  {\bibinfo  {journal} {Physics of Fluids}\ }\textbf {\bibinfo {volume} {6}},\
  \bibinfo {pages} {3221} (\bibinfo {year} {1994})}\BibitemShut {NoStop}\bibitem [{\citenamefont {Mininni}\ \emph {et~al.}(2012)\citenamefont
  {Mininni}, \citenamefont {Rosenberg},\ and\ \citenamefont
  {Pouquet}}]{Mininni2012}\BibitemOpen
  \bibfield  {author} {\bibinfo {author} {\bibfnamefont {P.~D.}\ \bibnamefont
  {Mininni}}, \bibinfo {author} {\bibfnamefont {D.}~\bibnamefont {Rosenberg}},\
  and\ \bibinfo {author} {\bibfnamefont {A.}~\bibnamefont {Pouquet}},\ }\href
  {https://doi.org/10.1017/JFM.2012.99} {\bibfield  {journal} {\bibinfo
  {journal} {Journal of Fluid Mechanics}\ }\textbf {\bibinfo {volume} {699}},\
  \bibinfo {pages} {263} (\bibinfo {year} {2012})}\BibitemShut {NoStop}\bibitem [{\citenamefont {Kraichnan}(1965)}]{Kraichnan1965}\BibitemOpen
  \bibfield  {author} {\bibinfo {author} {\bibfnamefont {R.~H.}\ \bibnamefont
  {Kraichnan}},\ }\href {https://doi.org/10.1063/1.1761412} {\bibfield
  {journal} {\bibinfo  {journal} {Physics of Fluids}\ }\textbf {\bibinfo
  {volume} {8}},\ \bibinfo {pages} {1385} (\bibinfo {year} {1965})}\BibitemShut
  {NoStop}\bibitem [{\citenamefont {Zhou}(1995)}]{Zhou1995}\BibitemOpen
  \bibfield  {author} {\bibinfo {author} {\bibfnamefont {Y.}~\bibnamefont
  {Zhou}},\ }\href {https://doi.org/10.1063/1.868457} {\bibfield  {journal}
  {\bibinfo  {journal} {Physics of Fluids}\ }\textbf {\bibinfo {volume} {7}},\
  \bibinfo {pages} {2092} (\bibinfo {year} {1995})}\BibitemShut {NoStop}\bibitem [{\citenamefont {Chakraborty}\ and\ \citenamefont
  {Bhattacharjee}(2007)}]{Chakraborty2007}\BibitemOpen
  \bibfield  {author} {\bibinfo {author} {\bibfnamefont {S.}~\bibnamefont
  {Chakraborty}}\ and\ \bibinfo {author} {\bibfnamefont {J.~K.}\ \bibnamefont
  {Bhattacharjee}},\ }\href {https://doi.org/10.1103/physreve.76.036304}
  {\bibfield  {journal} {\bibinfo  {journal} {Physical Review E}\ }\textbf
  {\bibinfo {volume} {76}},\ \bibinfo {pages} {036304} (\bibinfo {year}
  {2007})}\BibitemShut {NoStop}\bibitem [{\citenamefont {Zhou}\ \emph {et~al.}(2004)\citenamefont {Zhou},
  \citenamefont {Matthaeus},\ and\ \citenamefont {Dmitruk}}]{Zhou2004}\BibitemOpen
  \bibfield  {author} {\bibinfo {author} {\bibfnamefont {Y.}~\bibnamefont
  {Zhou}}, \bibinfo {author} {\bibfnamefont {W.}~\bibnamefont {Matthaeus}},\
  and\ \bibinfo {author} {\bibfnamefont {P.}~\bibnamefont {Dmitruk}},\ }\href
  {https://doi.org/10.1103/revmodphys.76.1015} {\bibfield  {journal} {\bibinfo
  {journal} {Reviews of Modern Physics}\ }\textbf {\bibinfo {volume} {76}},\
  \bibinfo {pages} {1015–1035} (\bibinfo {year} {2004})}\BibitemShut
  {NoStop}\bibitem [{\citenamefont {Politi}\ and\ \citenamefont
  {Misbah}(2004)}]{Politi2004}\BibitemOpen
  \bibfield  {author} {\bibinfo {author} {\bibfnamefont {P.}~\bibnamefont
  {Politi}}\ and\ \bibinfo {author} {\bibfnamefont {C.}~\bibnamefont
  {Misbah}},\ }\href {https://doi.org/10.1103/physrevlett.92.090601} {\bibfield
   {journal} {\bibinfo  {journal} {Physical Review Letters}\ }\textbf {\bibinfo
  {volume} {92}},\ \bibinfo {pages} {090601} (\bibinfo {year}
  {2004})}\BibitemShut {NoStop}\bibitem [{\citenamefont {Halatek}\ and\ \citenamefont
  {Frey}(2018)}]{Halatek2018}\BibitemOpen
  \bibfield  {author} {\bibinfo {author} {\bibfnamefont {J.}~\bibnamefont
  {Halatek}}\ and\ \bibinfo {author} {\bibfnamefont {E.}~\bibnamefont {Frey}},\
  }\href {https://doi.org/10.1038/s41567-017-0040-5} {\bibfield  {journal}
  {\bibinfo  {journal} {Nature Physics}\ }\textbf {\bibinfo {volume} {14}},\
  \bibinfo {pages} {507} (\bibinfo {year} {2018})}\BibitemShut {NoStop}\bibitem [{\citenamefont {Cates}\ and\ \citenamefont
  {Tailleur}(2015)}]{Cates2015}\BibitemOpen
  \bibfield  {author} {\bibinfo {author} {\bibfnamefont {M.~E.}\ \bibnamefont
  {Cates}}\ and\ \bibinfo {author} {\bibfnamefont {J.}~\bibnamefont
  {Tailleur}},\ }\href
  {https://doi.org/10.1146/annurev-conmatphys-031214-014710} {\bibfield
  {journal} {\bibinfo  {journal} {Annual Review of Condensed Matter Physics}\
  }\textbf {\bibinfo {volume} {6}},\ \bibinfo {pages} {219} (\bibinfo {year}
  {2015})}\BibitemShut {NoStop}\bibitem [{\citenamefont {Perlekar}\ \emph {et~al.}(2014)\citenamefont
  {Perlekar}, \citenamefont {Benzi}, \citenamefont {Clercx}, \citenamefont
  {Nelson},\ and\ \citenamefont {Toschi}}]{Perlekar2014}\BibitemOpen
  \bibfield  {author} {\bibinfo {author} {\bibfnamefont {P.}~\bibnamefont
  {Perlekar}}, \bibinfo {author} {\bibfnamefont {R.}~\bibnamefont {Benzi}},
  \bibinfo {author} {\bibfnamefont {H.~J.~H.}\ \bibnamefont {Clercx}}, \bibinfo
  {author} {\bibfnamefont {D.~R.}\ \bibnamefont {Nelson}},\ and\ \bibinfo
  {author} {\bibfnamefont {F.}~\bibnamefont {Toschi}},\ }\href
  {https://doi.org/10.1103/physrevlett.112.014502} {\bibfield  {journal}
  {\bibinfo  {journal} {Physical Review Letters}\ }\textbf {\bibinfo {volume}
  {112}},\ \bibinfo {pages} {014502} (\bibinfo {year} {2014})}\BibitemShut
  {NoStop}\bibitem [{\citenamefont {Theurkauff}\ \emph {et~al.}(2012)\citenamefont
  {Theurkauff}, \citenamefont {Cottin-Bizonne}, \citenamefont {Palacci},
  \citenamefont {Ybert},\ and\ \citenamefont {Bocquet}}]{Theurkauff2012}\BibitemOpen
  \bibfield  {author} {\bibinfo {author} {\bibfnamefont {I.}~\bibnamefont
  {Theurkauff}}, \bibinfo {author} {\bibfnamefont {C.}~\bibnamefont
  {Cottin-Bizonne}}, \bibinfo {author} {\bibfnamefont {J.}~\bibnamefont
  {Palacci}}, \bibinfo {author} {\bibfnamefont {C.}~\bibnamefont {Ybert}},\
  and\ \bibinfo {author} {\bibfnamefont {L.}~\bibnamefont {Bocquet}},\ }\href
  {https://doi.org/10.1103/physrevlett.108.268303} {\bibfield  {journal}
  {\bibinfo  {journal} {Physical Review Letters}\ }\textbf {\bibinfo {volume}
  {108}},\ \bibinfo {pages} {268303} (\bibinfo {year} {2012})}\BibitemShut
  {NoStop}\bibitem [{\citenamefont {van~der Linden}\ \emph {et~al.}(2019)\citenamefont
  {van~der Linden}, \citenamefont {Alexander}, \citenamefont {Aarts},\ and\
  \citenamefont {Dauchot}}]{vanderLinden2019}\BibitemOpen
  \bibfield  {author} {\bibinfo {author} {\bibfnamefont {M.~N.}\ \bibnamefont
  {van~der Linden}}, \bibinfo {author} {\bibfnamefont {L.~C.}\ \bibnamefont
  {Alexander}}, \bibinfo {author} {\bibfnamefont {D.~G. A.~L.}\ \bibnamefont
  {Aarts}},\ and\ \bibinfo {author} {\bibfnamefont {O.}~\bibnamefont
  {Dauchot}},\ }\href {https://doi.org/10.1103/physrevlett.123.098001}
  {\bibfield  {journal} {\bibinfo  {journal} {Physical Review Letters}\
  }\textbf {\bibinfo {volume} {123}},\ \bibinfo {pages} {098001} (\bibinfo
  {year} {2019})}\BibitemShut {NoStop}\bibitem [{\citenamefont {Miri}\ and\ \citenamefont {Alù}(2019)}]{Miri2019}\BibitemOpen
  \bibfield  {author} {\bibinfo {author} {\bibfnamefont {M.-A.}\ \bibnamefont
  {Miri}}\ and\ \bibinfo {author} {\bibfnamefont {A.}~\bibnamefont {Alù}},\
  }\bibfield  {journal} {\bibinfo  {journal} {Science}\ }\textbf {\bibinfo
  {volume} {363}},\ \href {https://doi.org/10.1126/science.aar7709}
  {10.1126/science.aar7709} (\bibinfo {year} {2019})\BibitemShut {NoStop}\bibitem [{\citenamefont {Balmforth}\ and\ \citenamefont
  {Young}(2002)}]{Balmforth2002}\BibitemOpen
  \bibfield  {author} {\bibinfo {author} {\bibfnamefont {N.~J.}\ \bibnamefont
  {Balmforth}}\ and\ \bibinfo {author} {\bibfnamefont {Y.-n.}\ \bibnamefont
  {Young}},\ }\href {https://doi.org/10.1017/s0022111002006371} {\bibfield
  {journal} {\bibinfo  {journal} {Journal of Fluid Mechanics}\ }\textbf
  {\bibinfo {volume} {450}},\ \bibinfo {pages} {131–167} (\bibinfo {year}
  {2002})}\BibitemShut {NoStop}\bibitem [{\citenamefont {Boffetta}\ \emph {et~al.}(2011)\citenamefont
  {Boffetta}, \citenamefont {De~Lillo}, \citenamefont {Mazzino},\ and\
  \citenamefont {Musacchio}}]{Boffetta2011}\BibitemOpen
  \bibfield  {author} {\bibinfo {author} {\bibfnamefont {G.}~\bibnamefont
  {Boffetta}}, \bibinfo {author} {\bibfnamefont {F.}~\bibnamefont {De~Lillo}},
  \bibinfo {author} {\bibfnamefont {A.}~\bibnamefont {Mazzino}},\ and\ \bibinfo
  {author} {\bibfnamefont {S.}~\bibnamefont {Musacchio}},\ }\href
  {https://doi.org/10.1209/0295-5075/95/34001} {\bibfield  {journal} {\bibinfo
  {journal} {EPL (Europhysics Letters)}\ }\textbf {\bibinfo {volume} {95}},\
  \bibinfo {pages} {34001} (\bibinfo {year} {2011})}\BibitemShut {NoStop}\bibitem [{\citenamefont {Biferale}(2003)}]{Biferale2003}\BibitemOpen
  \bibfield  {author} {\bibinfo {author} {\bibfnamefont {L.}~\bibnamefont
  {Biferale}},\ }\href {https://doi.org/10.1146/annurev.fluid.35.101101.161122}
  {\bibfield  {journal} {\bibinfo  {journal} {Annual Review of Fluid
  Mechanics}\ }\textbf {\bibinfo {volume} {35}},\ \bibinfo {pages} {441–468}
  (\bibinfo {year} {2003})}\BibitemShut {NoStop}\bibitem [{\citenamefont {Ghashghaie}\ \emph {et~al.}(1996)\citenamefont
  {Ghashghaie}, \citenamefont {Breymann}, \citenamefont {Peinke}, \citenamefont
  {Talkner},\ and\ \citenamefont {Dodge}}]{Ghashghaie1996}\BibitemOpen
  \bibfield  {author} {\bibinfo {author} {\bibfnamefont {S.}~\bibnamefont
  {Ghashghaie}}, \bibinfo {author} {\bibfnamefont {W.}~\bibnamefont
  {Breymann}}, \bibinfo {author} {\bibfnamefont {J.}~\bibnamefont {Peinke}},
  \bibinfo {author} {\bibfnamefont {P.}~\bibnamefont {Talkner}},\ and\ \bibinfo
  {author} {\bibfnamefont {Y.}~\bibnamefont {Dodge}},\ }\href
  {https://doi.org/10.1038/381767a0} {\bibfield  {journal} {\bibinfo  {journal}
  {Nature}\ }\textbf {\bibinfo {volume} {381}},\ \bibinfo {pages} {767–770}
  (\bibinfo {year} {1996})}\BibitemShut {NoStop}\bibitem [{\citenamefont {Bouchaud}\ and\ \citenamefont
  {Muzy}(2003)}]{Bouchaud2003}\BibitemOpen
  \bibfield  {author} {\bibinfo {author} {\bibfnamefont {J.-P.}\ \bibnamefont
  {Bouchaud}}\ and\ \bibinfo {author} {\bibfnamefont {J.-F.}\ \bibnamefont
  {Muzy}},\ }\href {https://doi.org/10.1007/978-3-540-39668-0_11} {\bibfield
  {journal} {\bibinfo  {journal} {The Kolmogorov Legacy in Physics}\ ,\
  \bibinfo {pages} {229–246}} (\bibinfo {year} {2003})}\BibitemShut {NoStop}\bibitem [{\citenamefont {Peyret}(2002)}]{Peyret2002}\BibitemOpen
  \bibfield  {author} {\bibinfo {author} {\bibfnamefont {R.}~\bibnamefont
  {Peyret}},\ }\href@noop {} {\emph {\bibinfo {title} {Spectral Methods for
  Incompressible Viscous Flow}}}\ (\bibinfo  {publisher} {Springer},\ \bibinfo
  {year} {2002})\BibitemShut {NoStop}\bibitem [{\citenamefont {Mahalov}\ and\ \citenamefont
  {Zhou}(1996)}]{Mahalov1996}\BibitemOpen
  \bibfield  {author} {\bibinfo {author} {\bibfnamefont {A.}~\bibnamefont
  {Mahalov}}\ and\ \bibinfo {author} {\bibfnamefont {Y.}~\bibnamefont {Zhou}},\
  }\href {https://doi.org/10.1063/1.868988} {\bibfield  {journal} {\bibinfo
  {journal} {Physics of Fluids}\ }\textbf {\bibinfo {volume} {8}},\ \bibinfo
  {pages} {2138} (\bibinfo {year} {1996})}\BibitemShut {NoStop}\bibitem [{\citenamefont {Waleffe}(1992)}]{Waleffe1992}\BibitemOpen
  \bibfield  {author} {\bibinfo {author} {\bibfnamefont {F.}~\bibnamefont
  {Waleffe}},\ }\href {https://doi.org/10.1063/1.858309} {\bibfield  {journal}
  {\bibinfo  {journal} {Physics of Fluids A: Fluid Dynamics}\ }\textbf
  {\bibinfo {volume} {4}},\ \bibinfo {pages} {350–363} (\bibinfo {year}
  {1992})}\BibitemShut {NoStop}\bibitem [{\citenamefont {Celani}\ \emph {et~al.}(2010)\citenamefont {Celani},
  \citenamefont {Musacchio},\ and\ \citenamefont {Vincenzi}}]{Celani2010}\BibitemOpen
  \bibfield  {author} {\bibinfo {author} {\bibfnamefont {A.}~\bibnamefont
  {Celani}}, \bibinfo {author} {\bibfnamefont {S.}~\bibnamefont {Musacchio}},\
  and\ \bibinfo {author} {\bibfnamefont {D.}~\bibnamefont {Vincenzi}},\ }\href
  {https://doi.org/10.1103/physrevlett.104.184506} {\bibfield  {journal}
  {\bibinfo  {journal} {Physical Review Letters}\ }\textbf {\bibinfo {volume}
  {104}},\ \bibinfo {pages} {184506} (\bibinfo {year} {2010})}\BibitemShut
  {NoStop}\bibitem [{\citenamefont {Falkovich}(1994)}]{Falkovich1994}\BibitemOpen
  \bibfield  {author} {\bibinfo {author} {\bibfnamefont {G.}~\bibnamefont
  {Falkovich}},\ }\href {https://doi.org/10.1063/1.868255} {\bibfield
  {journal} {\bibinfo  {journal} {Physics of Fluids}\ }\textbf {\bibinfo
  {volume} {6}},\ \bibinfo {pages} {1411} (\bibinfo {year} {1994})}\BibitemShut
  {NoStop}\bibitem [{\citenamefont {Küchler}\ \emph {et~al.}(2019)\citenamefont
  {Küchler}, \citenamefont {Bewley},\ and\ \citenamefont
  {Bodenschatz}}]{Kuchler2019}\BibitemOpen
  \bibfield  {author} {\bibinfo {author} {\bibfnamefont {C.}~\bibnamefont
  {Küchler}}, \bibinfo {author} {\bibfnamefont {G.}~\bibnamefont {Bewley}},\
  and\ \bibinfo {author} {\bibfnamefont {E.}~\bibnamefont {Bodenschatz}},\
  }\href {https://doi.org/10.1007/s10955-019-02251-1} {\bibfield  {journal}
  {\bibinfo  {journal} {Journal of Statistical Physics}\ }\textbf {\bibinfo
  {volume} {175}},\ \bibinfo {pages} {617} (\bibinfo {year}
  {2019})}\BibitemShut {NoStop}\bibitem [{\citenamefont {Lohse}\ and\ \citenamefont
  {Müller-Groeling}(1995)}]{Lohse1995}\BibitemOpen
  \bibfield  {author} {\bibinfo {author} {\bibfnamefont {D.}~\bibnamefont
  {Lohse}}\ and\ \bibinfo {author} {\bibfnamefont {A.}~\bibnamefont
  {Müller-Groeling}},\ }\href {https://doi.org/10.1103/physrevlett.74.1747}
  {\bibfield  {journal} {\bibinfo  {journal} {Physical Review Letters}\
  }\textbf {\bibinfo {volume} {74}},\ \bibinfo {pages} {1747} (\bibinfo {year}
  {1995})}\BibitemShut {NoStop}\bibitem [{\citenamefont {Donzis}\ and\ \citenamefont
  {Sreenivasan}(2010)}]{Donzis2010}\BibitemOpen
  \bibfield  {author} {\bibinfo {author} {\bibfnamefont {D.~A.}\ \bibnamefont
  {Donzis}}\ and\ \bibinfo {author} {\bibfnamefont {K.~R.}\ \bibnamefont
  {Sreenivasan}},\ }\href {https://doi.org/10.1017/s0022112010001400}
  {\bibfield  {journal} {\bibinfo  {journal} {Journal of Fluid Mechanics}\
  }\textbf {\bibinfo {volume} {657}},\ \bibinfo {pages} {171} (\bibinfo {year}
  {2010})}\BibitemShut {NoStop}\bibitem [{\citenamefont {Verma}\ and\ \citenamefont
  {Donzis}(2007)}]{Verma2007}\BibitemOpen
  \bibfield  {author} {\bibinfo {author} {\bibfnamefont {M.~K.}\ \bibnamefont
  {Verma}}\ and\ \bibinfo {author} {\bibfnamefont {D.}~\bibnamefont {Donzis}},\
  }\href {https://doi.org/10.1088/1751-8113/40/16/010} {\bibfield  {journal}
  {\bibinfo  {journal} {Journal of Physics A: Mathematical and Theoretical}\
  }\textbf {\bibinfo {volume} {40}},\ \bibinfo {pages} {4401} (\bibinfo {year}
  {2007})}\BibitemShut {NoStop}\bibitem [{\citenamefont {Sreenivasan}\ and\ \citenamefont
  {Antonia}(1997)}]{Sreenivasan1997}\BibitemOpen
  \bibfield  {author} {\bibinfo {author} {\bibfnamefont {K.~R.}\ \bibnamefont
  {Sreenivasan}}\ and\ \bibinfo {author} {\bibfnamefont {R.~A.}\ \bibnamefont
  {Antonia}},\ }\href {https://doi.org/10.1146/annurev.fluid.29.1.435}
  {\bibfield  {journal} {\bibinfo  {journal} {Annual Review of Fluid
  Mechanics}\ }\textbf {\bibinfo {volume} {29}},\ \bibinfo {pages} {435}
  (\bibinfo {year} {1997})}\BibitemShut {NoStop}\bibitem [{\citenamefont {Leoni}\ \emph {et~al.}(2020)\citenamefont {Leoni},
  \citenamefont {Alexakis}, \citenamefont {Biferale},\ and\ \citenamefont
  {Buzzicotti}}]{ClarkDiLeoni2020}\BibitemOpen
  \bibfield  {author} {\bibinfo {author} {\bibfnamefont {P.~C.~D.}\
  \bibnamefont {Leoni}}, \bibinfo {author} {\bibfnamefont {A.}~\bibnamefont
  {Alexakis}}, \bibinfo {author} {\bibfnamefont {L.}~\bibnamefont {Biferale}},\
  and\ \bibinfo {author} {\bibfnamefont {M.}~\bibnamefont {Buzzicotti}},\
  }\href {https://doi.org/10.1103/physrevfluids.5.104603} {\bibfield  {journal}
  {\bibinfo  {journal} {Physical Review Fluids}\ }\textbf {\bibinfo {volume}
  {5}},\ \bibinfo {pages} {104603} (\bibinfo {year} {2020})}\BibitemShut
  {NoStop}\bibitem [{\citenamefont {Dunkel}\ \emph {et~al.}(2013)\citenamefont {Dunkel},
  \citenamefont {Heidenreich}, \citenamefont {Drescher}, \citenamefont
  {Wensink}, \citenamefont {Bär},\ and\ \citenamefont
  {Goldstein}}]{Dunkel2013}\BibitemOpen
  \bibfield  {author} {\bibinfo {author} {\bibfnamefont {J.}~\bibnamefont
  {Dunkel}}, \bibinfo {author} {\bibfnamefont {S.}~\bibnamefont {Heidenreich}},
  \bibinfo {author} {\bibfnamefont {K.}~\bibnamefont {Drescher}}, \bibinfo
  {author} {\bibfnamefont {H.~H.}\ \bibnamefont {Wensink}}, \bibinfo {author}
  {\bibfnamefont {M.}~\bibnamefont {Bär}},\ and\ \bibinfo {author}
  {\bibfnamefont {R.~E.}\ \bibnamefont {Goldstein}},\ }\href
  {https://doi.org/10.1103/physrevlett.110.228102} {\bibfield  {journal}
  {\bibinfo  {journal} {Physical Review Letters}\ }\textbf {\bibinfo {volume}
  {110}},\ \bibinfo {pages} {228102} (\bibinfo {year} {2013})}\BibitemShut
  {NoStop}\bibitem [{\citenamefont {Wensink}\ \emph {et~al.}(2012)\citenamefont
  {Wensink}, \citenamefont {Dunkel}, \citenamefont {Heidenreich}, \citenamefont
  {Drescher}, \citenamefont {Goldstein}, \citenamefont {Löwen},\ and\
  \citenamefont {Yeomans}}]{Wensink2012}\BibitemOpen
  \bibfield  {author} {\bibinfo {author} {\bibfnamefont {H.~H.}\ \bibnamefont
  {Wensink}}, \bibinfo {author} {\bibfnamefont {J.}~\bibnamefont {Dunkel}},
  \bibinfo {author} {\bibfnamefont {S.}~\bibnamefont {Heidenreich}}, \bibinfo
  {author} {\bibfnamefont {K.}~\bibnamefont {Drescher}}, \bibinfo {author}
  {\bibfnamefont {R.~E.}\ \bibnamefont {Goldstein}}, \bibinfo {author}
  {\bibfnamefont {H.}~\bibnamefont {Löwen}},\ and\ \bibinfo {author}
  {\bibfnamefont {J.~M.}\ \bibnamefont {Yeomans}},\ }\href
  {https://doi.org/10.1073/pnas.1202032109} {\bibfield  {journal} {\bibinfo
  {journal} {Proceedings of the National Academy of Sciences}\ }\textbf
  {\bibinfo {volume} {109}},\ \bibinfo {pages} {14308} (\bibinfo {year}
  {2012})}\BibitemShut {NoStop}\bibitem [{\citenamefont {Alert}\ \emph {et~al.}(2022)\citenamefont {Alert},
  \citenamefont {Casademunt},\ and\ \citenamefont {Joanny}}]{Alert2022}\BibitemOpen
  \bibfield  {author} {\bibinfo {author} {\bibfnamefont {R.}~\bibnamefont
  {Alert}}, \bibinfo {author} {\bibfnamefont {J.}~\bibnamefont {Casademunt}},\
  and\ \bibinfo {author} {\bibfnamefont {J.-F.}\ \bibnamefont {Joanny}},\
  }\href {https://doi.org/10.1146/annurev-conmatphys-082321-035957} {\bibfield
  {journal} {\bibinfo  {journal} {Annual Review of Condensed Matter Physics}\
  }\textbf {\bibinfo {volume} {13}},\ \bibinfo {pages} {143} (\bibinfo {year}
  {2022})}\BibitemShut {NoStop}\bibitem [{\citenamefont {Mart{\'{\i}}nez-Prat}\ \emph
  {et~al.}(2019)\citenamefont {Mart{\'{\i}}nez-Prat}, \citenamefont
  {Ign{\'{e}}s-Mullol}, \citenamefont {Casademunt},\ and\ \citenamefont
  {Sagu{\'{e}}s}}]{MartinezPrat2019}\BibitemOpen
  \bibfield  {author} {\bibinfo {author} {\bibfnamefont {B.}~\bibnamefont
  {Mart{\'{\i}}nez-Prat}}, \bibinfo {author} {\bibfnamefont {J.}~\bibnamefont
  {Ign{\'{e}}s-Mullol}}, \bibinfo {author} {\bibfnamefont {J.}~\bibnamefont
  {Casademunt}},\ and\ \bibinfo {author} {\bibfnamefont {F.}~\bibnamefont
  {Sagu{\'{e}}s}},\ }\href {https://doi.org/10.1038/s41567-018-0411-6}
  {\bibfield  {journal} {\bibinfo  {journal} {Nature Physics}\ }\textbf
  {\bibinfo {volume} {15}},\ \bibinfo {pages} {362} (\bibinfo {year}
  {2019})}\BibitemShut {NoStop}\bibitem [{\citenamefont {Alert}\ \emph {et~al.}(2020)\citenamefont {Alert},
  \citenamefont {Joanny},\ and\ \citenamefont {Casademunt}}]{Alert2020}\BibitemOpen
  \bibfield  {author} {\bibinfo {author} {\bibfnamefont {R.}~\bibnamefont
  {Alert}}, \bibinfo {author} {\bibfnamefont {J.-F.}\ \bibnamefont {Joanny}},\
  and\ \bibinfo {author} {\bibfnamefont {J.}~\bibnamefont {Casademunt}},\
  }\href {https://doi.org/10.1038/s41567-020-0854-4} {\bibfield  {journal}
  {\bibinfo  {journal} {Nature Physics}\ }\textbf {\bibinfo {volume} {16}},\
  \bibinfo {pages} {682–688} (\bibinfo {year} {2020})}\BibitemShut {NoStop}\bibitem [{\citenamefont {Carenza}\ \emph {et~al.}(2020)\citenamefont
  {Carenza}, \citenamefont {Biferale},\ and\ \citenamefont
  {Gonnella}}]{Carenza2020}\BibitemOpen
  \bibfield  {author} {\bibinfo {author} {\bibfnamefont {L.~N.}\ \bibnamefont
  {Carenza}}, \bibinfo {author} {\bibfnamefont {L.}~\bibnamefont {Biferale}},\
  and\ \bibinfo {author} {\bibfnamefont {G.}~\bibnamefont {Gonnella}},\ }\href
  {https://doi.org/10.1103/physrevfluids.5.011302} {\bibfield  {journal}
  {\bibinfo  {journal} {Physical Review Fluids}\ }\textbf {\bibinfo {volume}
  {5}},\ \bibinfo {pages} {011302} (\bibinfo {year} {2020})}\BibitemShut
  {NoStop}\bibitem [{\citenamefont {Słomka}\ and\ \citenamefont
  {Dunkel}(2015)}]{Slomka2015}\BibitemOpen
  \bibfield  {author} {\bibinfo {author} {\bibfnamefont {J.}~\bibnamefont
  {Słomka}}\ and\ \bibinfo {author} {\bibfnamefont {J.}~\bibnamefont
  {Dunkel}},\ }\href {https://doi.org/10.1140/epjst/e2015-02463-2} {\bibfield
  {journal} {\bibinfo  {journal} {The European Physical Journal Special
  Topics}\ }\textbf {\bibinfo {volume} {224}},\ \bibinfo {pages} {1349–1358}
  (\bibinfo {year} {2015})}\BibitemShut {NoStop}\bibitem [{\citenamefont {Rorai}\ \emph {et~al.}(2022)\citenamefont {Rorai},
  \citenamefont {Toschi},\ and\ \citenamefont {Pagonabarraga}}]{Rorai2022}\BibitemOpen
  \bibfield  {author} {\bibinfo {author} {\bibfnamefont {C.}~\bibnamefont
  {Rorai}}, \bibinfo {author} {\bibfnamefont {F.}~\bibnamefont {Toschi}},\ and\
  \bibinfo {author} {\bibfnamefont {I.}~\bibnamefont {Pagonabarraga}},\ }\href
  {https://doi.org/10.1103/physrevlett.129.218001} {\bibfield  {journal}
  {\bibinfo  {journal} {Physical Review Letters}\ }\textbf {\bibinfo {volume}
  {129}},\ \bibinfo {pages} {218001} (\bibinfo {year} {2022})}\BibitemShut
  {NoStop}\bibitem [{\citenamefont {Bratanov}\ \emph {et~al.}(2015)\citenamefont
  {Bratanov}, \citenamefont {Jenko},\ and\ \citenamefont
  {Frey}}]{Bratanov2015}\BibitemOpen
  \bibfield  {author} {\bibinfo {author} {\bibfnamefont {V.}~\bibnamefont
  {Bratanov}}, \bibinfo {author} {\bibfnamefont {F.}~\bibnamefont {Jenko}},\
  and\ \bibinfo {author} {\bibfnamefont {E.}~\bibnamefont {Frey}},\ }\href
  {https://doi.org/10.1073/pnas.1509304112} {\bibfield  {journal} {\bibinfo
  {journal} {Proceedings of the National Academy of Sciences}\ }\textbf
  {\bibinfo {volume} {112}},\ \bibinfo {pages} {15048–15053} (\bibinfo {year}
  {2015})}\BibitemShut {NoStop}\bibitem [{\citenamefont {Mukherjee}\ \emph {et~al.}(2023)\citenamefont
  {Mukherjee}, \citenamefont {Singh}, \citenamefont {James},\ and\
  \citenamefont {Ray}}]{Mukherjee2023}\BibitemOpen
  \bibfield  {author} {\bibinfo {author} {\bibfnamefont {S.}~\bibnamefont
  {Mukherjee}}, \bibinfo {author} {\bibfnamefont {R.~K.}\ \bibnamefont
  {Singh}}, \bibinfo {author} {\bibfnamefont {M.}~\bibnamefont {James}},\ and\
  \bibinfo {author} {\bibfnamefont {S.~S.}\ \bibnamefont {Ray}},\ }\href
  {https://doi.org/10.1038/s41567-023-01990-z} {\bibfield  {journal} {\bibinfo
  {journal} {Nature Physics}\ }\textbf {\bibinfo {volume} {19}},\ \bibinfo
  {pages} {891–897} (\bibinfo {year} {2023})}\BibitemShut {NoStop}\bibitem [{\citenamefont {Linkmann}\ \emph {et~al.}(2019)\citenamefont
  {Linkmann}, \citenamefont {Boffetta}, \citenamefont {Marchetti},\ and\
  \citenamefont {Eckhardt}}]{Linkmann2019}\BibitemOpen
  \bibfield  {author} {\bibinfo {author} {\bibfnamefont {M.}~\bibnamefont
  {Linkmann}}, \bibinfo {author} {\bibfnamefont {G.}~\bibnamefont {Boffetta}},
  \bibinfo {author} {\bibfnamefont {M.~C.}\ \bibnamefont {Marchetti}},\ and\
  \bibinfo {author} {\bibfnamefont {B.}~\bibnamefont {Eckhardt}},\ }\href
  {https://doi.org/10.1103/physrevlett.122.214503} {\bibfield  {journal}
  {\bibinfo  {journal} {Physical Review Letters}\ }\textbf {\bibinfo {volume}
  {122}},\ \bibinfo {pages} {214503} (\bibinfo {year} {2019})}\BibitemShut
  {NoStop}\bibitem [{\citenamefont {Kiran}\ \emph {et~al.}(2023)\citenamefont {Kiran},
  \citenamefont {Gupta}, \citenamefont {Verma},\ and\ \citenamefont
  {Pandit}}]{Kiran2023}\BibitemOpen
  \bibfield  {author} {\bibinfo {author} {\bibfnamefont {K.~V.}\ \bibnamefont
  {Kiran}}, \bibinfo {author} {\bibfnamefont {A.}~\bibnamefont {Gupta}},
  \bibinfo {author} {\bibfnamefont {A.~K.}\ \bibnamefont {Verma}},\ and\
  \bibinfo {author} {\bibfnamefont {R.}~\bibnamefont {Pandit}},\ }\href
  {https://doi.org/10.1103/physrevfluids.8.023102} {\bibfield  {journal}
  {\bibinfo  {journal} {Physical Review Fluids}\ }\textbf {\bibinfo {volume}
  {8}},\ \bibinfo {pages} {023102} (\bibinfo {year} {2023})}\BibitemShut
  {NoStop}\bibitem [{\citenamefont {Marston}\ and\ \citenamefont
  {Tobias}(2023)}]{Marston2023}\BibitemOpen
  \bibfield  {author} {\bibinfo {author} {\bibfnamefont {J.}~\bibnamefont
  {Marston}}\ and\ \bibinfo {author} {\bibfnamefont {S.}~\bibnamefont
  {Tobias}},\ }\href {https://doi.org/10.1146/annurev-fluid-120720-031006}
  {\bibfield  {journal} {\bibinfo  {journal} {Annual Review of Fluid
  Mechanics}\ }\textbf {\bibinfo {volume} {55}},\ \bibinfo {pages} {351–375}
  (\bibinfo {year} {2023})}\BibitemShut {NoStop}\bibitem [{\citenamefont {Parker}\ and\ \citenamefont
  {Krommes}(2014)}]{Parker2014}\BibitemOpen
  \bibfield  {author} {\bibinfo {author} {\bibfnamefont {J.~B.}\ \bibnamefont
  {Parker}}\ and\ \bibinfo {author} {\bibfnamefont {J.~A.}\ \bibnamefont
  {Krommes}},\ }\href {https://doi.org/10.1088/1367-2630/16/3/035006}
  {\bibfield  {journal} {\bibinfo  {journal} {New Journal of Physics}\ }\textbf
  {\bibinfo {volume} {16}},\ \bibinfo {pages} {035006} (\bibinfo {year}
  {2014})}\BibitemShut {NoStop}\bibitem [{\citenamefont {Constantinou}\ and\ \citenamefont
  {Parker}(2018)}]{Constantinou2018}\BibitemOpen
  \bibfield  {author} {\bibinfo {author} {\bibfnamefont {N.~C.}\ \bibnamefont
  {Constantinou}}\ and\ \bibinfo {author} {\bibfnamefont {J.~B.}\ \bibnamefont
  {Parker}},\ }\href {https://doi.org/10.3847/1538-4357/aace53} {\bibfield
  {journal} {\bibinfo  {journal} {The Astrophysical Journal}\ }\textbf
  {\bibinfo {volume} {863}},\ \bibinfo {pages} {46} (\bibinfo {year}
  {2018})}\BibitemShut {NoStop}\bibitem [{\citenamefont {G\"urcan}\ and\ \citenamefont
  {Diamond}(2015)}]{Gurcan2015}\BibitemOpen
  \bibfield  {author} {\bibinfo {author} {\bibfnamefont {O.~D.}\ \bibnamefont
  {G\"urcan}}\ and\ \bibinfo {author} {\bibfnamefont {P.~H.}\ \bibnamefont
  {Diamond}},\ }\href {https://doi.org/10.1088/1751-8113/48/29/293001}
  {\bibfield  {journal} {\bibinfo  {journal} {Journal of Physics A:
  Mathematical and Theoretical}\ }\textbf {\bibinfo {volume} {48}},\ \bibinfo
  {pages} {293001} (\bibinfo {year} {2015})}\BibitemShut {NoStop}\bibitem [{\citenamefont {Parker}\ and\ \citenamefont
  {Krommes}(2013)}]{Parker2013}\BibitemOpen
  \bibfield  {author} {\bibinfo {author} {\bibfnamefont {J.~B.}\ \bibnamefont
  {Parker}}\ and\ \bibinfo {author} {\bibfnamefont {J.~A.}\ \bibnamefont
  {Krommes}},\ }\bibfield  {journal} {\bibinfo  {journal} {Physics of Plasmas}\
  }\textbf {\bibinfo {volume} {20}},\ \href {https://doi.org/10.1063/1.4828717}
  {10.1063/1.4828717} (\bibinfo {year} {2013})\BibitemShut {NoStop}\bibitem [{\citenamefont {Constantinou}\ \emph {et~al.}(2014)\citenamefont
  {Constantinou}, \citenamefont {Farrell},\ and\ \citenamefont
  {Ioannou}}]{Constantinou2014}\BibitemOpen
  \bibfield  {author} {\bibinfo {author} {\bibfnamefont {N.~C.}\ \bibnamefont
  {Constantinou}}, \bibinfo {author} {\bibfnamefont {B.~F.}\ \bibnamefont
  {Farrell}},\ and\ \bibinfo {author} {\bibfnamefont {P.~J.}\ \bibnamefont
  {Ioannou}},\ }\href {https://doi.org/10.1175/jas-d-13-076.1} {\bibfield
  {journal} {\bibinfo  {journal} {Journal of the Atmospheric Sciences}\
  }\textbf {\bibinfo {volume} {71}},\ \bibinfo {pages} {1818–1842} (\bibinfo
  {year} {2014})}\BibitemShut {NoStop}\bibitem [{\citenamefont {Tuckerman}\ \emph {et~al.}(2020)\citenamefont
  {Tuckerman}, \citenamefont {Chantry},\ and\ \citenamefont
  {Barkley}}]{Tuckerman2020}\BibitemOpen
  \bibfield  {author} {\bibinfo {author} {\bibfnamefont {L.~S.}\ \bibnamefont
  {Tuckerman}}, \bibinfo {author} {\bibfnamefont {M.}~\bibnamefont {Chantry}},\
  and\ \bibinfo {author} {\bibfnamefont {D.}~\bibnamefont {Barkley}},\ }\href
  {https://doi.org/10.1146/annurev-fluid-010719-060221} {\bibfield  {journal}
  {\bibinfo  {journal} {Annual Review of Fluid Mechanics}\ }\textbf {\bibinfo
  {volume} {52}},\ \bibinfo {pages} {343} (\bibinfo {year} {2020})}\BibitemShut
  {NoStop}\bibitem [{\citenamefont {Prigent}\ \emph {et~al.}(2002)\citenamefont
  {Prigent}, \citenamefont {Gr{\'{e}}goire}, \citenamefont {Chat{\'{e}}},
  \citenamefont {Dauchot},\ and\ \citenamefont {van Saarloos}}]{Prigent2002}\BibitemOpen
  \bibfield  {author} {\bibinfo {author} {\bibfnamefont {A.}~\bibnamefont
  {Prigent}}, \bibinfo {author} {\bibfnamefont {G.}~\bibnamefont
  {Gr{\'{e}}goire}}, \bibinfo {author} {\bibfnamefont {H.}~\bibnamefont
  {Chat{\'{e}}}}, \bibinfo {author} {\bibfnamefont {O.}~\bibnamefont
  {Dauchot}},\ and\ \bibinfo {author} {\bibfnamefont {W.}~\bibnamefont {van
  Saarloos}},\ }\href {https://doi.org/10.1103/physrevlett.89.014501}
  {\bibfield  {journal} {\bibinfo  {journal} {Physical Review Letters}\
  }\textbf {\bibinfo {volume} {89}},\ \bibinfo {pages} {014501} (\bibinfo
  {year} {2002})}\BibitemShut {NoStop}\bibitem [{\citenamefont {Duguet}\ \emph {et~al.}(2010)\citenamefont {Duguet},
  \citenamefont {Schlatter},\ and\ \citenamefont {Henningson}}]{Duguet2010}\BibitemOpen
  \bibfield  {author} {\bibinfo {author} {\bibfnamefont {Y.}~\bibnamefont
  {Duguet}}, \bibinfo {author} {\bibfnamefont {P.}~\bibnamefont {Schlatter}},\
  and\ \bibinfo {author} {\bibfnamefont {D.~S.}\ \bibnamefont {Henningson}},\
  }\href {https://doi.org/10.1017/s0022112010000297} {\bibfield  {journal}
  {\bibinfo  {journal} {Journal of Fluid Mechanics}\ }\textbf {\bibinfo
  {volume} {650}},\ \bibinfo {pages} {119} (\bibinfo {year}
  {2010})}\BibitemShut {NoStop}\bibitem [{\citenamefont {Kashyap}\ \emph {et~al.}(2022)\citenamefont
  {Kashyap}, \citenamefont {Duguet},\ and\ \citenamefont
  {Dauchot}}]{Kashyap2022}\BibitemOpen
  \bibfield  {author} {\bibinfo {author} {\bibfnamefont {P.~V.}\ \bibnamefont
  {Kashyap}}, \bibinfo {author} {\bibfnamefont {Y.}~\bibnamefont {Duguet}},\
  and\ \bibinfo {author} {\bibfnamefont {O.}~\bibnamefont {Dauchot}},\ }\href
  {https://doi.org/10.1103/physrevlett.129.244501} {\bibfield  {journal}
  {\bibinfo  {journal} {Physical Review Letters}\ }\textbf {\bibinfo {volume}
  {129}},\ \bibinfo {pages} {244501} (\bibinfo {year} {2022})}\BibitemShut
  {NoStop}\bibitem [{\citenamefont {Vallis}\ and\ \citenamefont
  {Maltrud}(1993)}]{Vallis1993}\BibitemOpen
  \bibfield  {author} {\bibinfo {author} {\bibfnamefont {G.~K.}\ \bibnamefont
  {Vallis}}\ and\ \bibinfo {author} {\bibfnamefont {M.~E.}\ \bibnamefont
  {Maltrud}},\ }\href
  {https://doi.org/10.1175/1520-0485(1993)023<1346:GOMFAJ>2.0.CO;2} {\bibfield
  {journal} {\bibinfo  {journal} {Journal of Physical Oceanography}\ }\textbf
  {\bibinfo {volume} {23}},\ \bibinfo {pages} {1346–} (\bibinfo {year}
  {1993})}\BibitemShut {NoStop}\bibitem [{\citenamefont {Galtier}(2003)}]{Galtier2003}\BibitemOpen
  \bibfield  {author} {\bibinfo {author} {\bibfnamefont {S.}~\bibnamefont
  {Galtier}},\ }\href {https://doi.org/10.1103/physreve.68.015301} {\bibfield
  {journal} {\bibinfo  {journal} {Physical Review E}\ }\textbf {\bibinfo
  {volume} {68}},\ \bibinfo {pages} {015301} (\bibinfo {year}
  {2003})}\BibitemShut {NoStop}\bibitem [{\citenamefont {Galperin}\ \emph {et~al.}(2008)\citenamefont
  {Galperin}, \citenamefont {Sukoriansky},\ and\ \citenamefont
  {Dikovskaya}}]{Galperin2008}\BibitemOpen
  \bibfield  {author} {\bibinfo {author} {\bibfnamefont {B.}~\bibnamefont
  {Galperin}}, \bibinfo {author} {\bibfnamefont {S.}~\bibnamefont
  {Sukoriansky}},\ and\ \bibinfo {author} {\bibfnamefont {N.}~\bibnamefont
  {Dikovskaya}},\ }\href {https://doi.org/10.1088/0031-8949/2008/t132/014034}
  {\bibfield  {journal} {\bibinfo  {journal} {Physica Scripta}\ }\textbf
  {\bibinfo {volume} {T132}},\ \bibinfo {pages} {014034} (\bibinfo {year}
  {2008})}\BibitemShut {NoStop}\bibitem [{\citenamefont {Meyrand}\ \emph {et~al.}(2016)\citenamefont
  {Meyrand}, \citenamefont {Galtier},\ and\ \citenamefont
  {Kiyani}}]{Meyrand2016}\BibitemOpen
  \bibfield  {author} {\bibinfo {author} {\bibfnamefont {R.}~\bibnamefont
  {Meyrand}}, \bibinfo {author} {\bibfnamefont {S.}~\bibnamefont {Galtier}},\
  and\ \bibinfo {author} {\bibfnamefont {K.~H.}\ \bibnamefont {Kiyani}},\
  }\href {https://doi.org/10.1103/physrevlett.116.105002} {\bibfield  {journal}
  {\bibinfo  {journal} {Physical Review Letters}\ }\textbf {\bibinfo {volume}
  {116}},\ \bibinfo {pages} {105002} (\bibinfo {year} {2016})}\BibitemShut
  {NoStop}\bibitem [{\citenamefont {McCourt}(1990)}]{Mccourt1990}\BibitemOpen
  \bibfield  {author} {\bibinfo {author} {\bibfnamefont {F.}~\bibnamefont
  {McCourt}},\ }\href@noop {} {\emph {\bibinfo {title} {Nonequilibrium
  phenomena in polyatomic gases}}}\ (\bibinfo  {publisher} {Clarendon Press
  Oxford University Press},\ \bibinfo {address} {Oxford New York},\ \bibinfo
  {year} {1990})\BibitemShut {NoStop}\bibitem [{\citenamefont {Avron}\ \emph {et~al.}(1995)\citenamefont {Avron},
  \citenamefont {Seiler},\ and\ \citenamefont {Zograf}}]{Avron1995}\BibitemOpen
  \bibfield  {author} {\bibinfo {author} {\bibfnamefont {J.~E.}\ \bibnamefont
  {Avron}}, \bibinfo {author} {\bibfnamefont {R.}~\bibnamefont {Seiler}},\ and\
  \bibinfo {author} {\bibfnamefont {P.~G.}\ \bibnamefont {Zograf}},\ }\href
  {https://doi.org/10.1103/physrevlett.75.697} {\bibfield  {journal} {\bibinfo
  {journal} {Physical Review Letters}\ }\textbf {\bibinfo {volume} {75}},\
  \bibinfo {pages} {697–700} (\bibinfo {year} {1995})}\BibitemShut {NoStop}\bibitem [{\citenamefont {Ganeshan}\ and\ \citenamefont
  {Abanov}(2017)}]{Ganeshan2017}\BibitemOpen
  \bibfield  {author} {\bibinfo {author} {\bibfnamefont {S.}~\bibnamefont
  {Ganeshan}}\ and\ \bibinfo {author} {\bibfnamefont {A.~G.}\ \bibnamefont
  {Abanov}},\ }\href {https://doi.org/10.1103/physrevfluids.2.094101}
  {\bibfield  {journal} {\bibinfo  {journal} {Physical Review Fluids}\ }\textbf
  {\bibinfo {volume} {2}},\ \bibinfo {pages} {094101} (\bibinfo {year}
  {2017})}\BibitemShut {NoStop}\bibitem [{\citenamefont {Nakagawa}(1956)}]{Nakagawa1956}\BibitemOpen
  \bibfield  {author} {\bibinfo {author} {\bibfnamefont {Y.}~\bibnamefont
  {Nakagawa}},\ }\href {https://doi.org/10.4294/jpe1952.4.105} {\bibfield
  {journal} {\bibinfo  {journal} {Journal of Physics of the Earth}\ }\textbf
  {\bibinfo {volume} {4}},\ \bibinfo {pages} {105–111} (\bibinfo {year}
  {1956})}\BibitemShut {NoStop}\bibitem [{\citenamefont {Chapman}\ \emph {et~al.}(1990)\citenamefont
  {Chapman}, \citenamefont {Cowling}, \citenamefont {Burnett},\ and\
  \citenamefont {Cercignani}}]{ChapmanCowling}\BibitemOpen
  \bibfield  {author} {\bibinfo {author} {\bibfnamefont {S.}~\bibnamefont
  {Chapman}}, \bibinfo {author} {\bibfnamefont {T.}~\bibnamefont {Cowling}},
  \bibinfo {author} {\bibfnamefont {D.}~\bibnamefont {Burnett}},\ and\ \bibinfo
  {author} {\bibfnamefont {C.}~\bibnamefont {Cercignani}},\ }\href@noop {}
  {\emph {\bibinfo {title} {The Mathematical Theory of Non-uniform Gases: An
  Account of the Kinetic Theory of Viscosity, Thermal Conduction and Diffusion
  in Gases}}},\ Cambridge Mathematical Library\ (\bibinfo  {publisher}
  {Cambridge University Press},\ \bibinfo {year} {1990})\BibitemShut {NoStop}\bibitem [{\citenamefont {Lingam}\ \emph {et~al.}(2020)\citenamefont {Lingam},
  \citenamefont {Morrison},\ and\ \citenamefont {Wurm}}]{Lingam2020}\BibitemOpen
  \bibfield  {author} {\bibinfo {author} {\bibfnamefont {M.}~\bibnamefont
  {Lingam}}, \bibinfo {author} {\bibfnamefont {P.~J.}\ \bibnamefont
  {Morrison}},\ and\ \bibinfo {author} {\bibfnamefont {A.}~\bibnamefont
  {Wurm}},\ }\href {https://doi.org/10.1017/s0022377820001038} {\bibfield
  {journal} {\bibinfo  {journal} {Journal of Plasma Physics}\ }\textbf
  {\bibinfo {volume} {86}},\ \bibinfo {pages} {835860501} (\bibinfo {year}
  {2020})}\BibitemShut {NoStop}\bibitem [{\citenamefont {Hoyos}\ and\ \citenamefont {Son}(2012)}]{Hoyos2012}\BibitemOpen
  \bibfield  {author} {\bibinfo {author} {\bibfnamefont {C.}~\bibnamefont
  {Hoyos}}\ and\ \bibinfo {author} {\bibfnamefont {D.~T.}\ \bibnamefont
  {Son}},\ }\href {https://doi.org/10.1103/physrevlett.108.066805} {\bibfield
  {journal} {\bibinfo  {journal} {Physical Review Letters}\ }\textbf {\bibinfo
  {volume} {108}},\ \bibinfo {pages} {066805} (\bibinfo {year}
  {2012})}\BibitemShut {NoStop}\bibitem [{\citenamefont {Read}(2009)}]{Read2009}\BibitemOpen
  \bibfield  {author} {\bibinfo {author} {\bibfnamefont {N.}~\bibnamefont
  {Read}},\ }\href {https://doi.org/10.1103/physrevb.79.045308} {\bibfield
  {journal} {\bibinfo  {journal} {Physical Review B}\ }\textbf {\bibinfo
  {volume} {79}},\ \bibinfo {pages} {045308} (\bibinfo {year}
  {2009})}\BibitemShut {NoStop}\bibitem [{\citenamefont {Vollhardt}\ and\ \citenamefont
  {Wolfle}(2013)}]{Vollhardt2013}\BibitemOpen
  \bibfield  {author} {\bibinfo {author} {\bibfnamefont {D.}~\bibnamefont
  {Vollhardt}}\ and\ \bibinfo {author} {\bibfnamefont {P.}~\bibnamefont
  {Wolfle}},\ }\href@noop {} {\emph {\bibinfo {title} {The Superfluid Phases of
  Helium 3}}}\ (\bibinfo  {publisher} {Dover Publications},\ \bibinfo {year}
  {2013})\BibitemShut {NoStop}\bibitem [{\citenamefont {Wiegmann}\ and\ \citenamefont
  {Abanov}(2014)}]{Wiegmann2014}\BibitemOpen
  \bibfield  {author} {\bibinfo {author} {\bibfnamefont {P.}~\bibnamefont
  {Wiegmann}}\ and\ \bibinfo {author} {\bibfnamefont {A.~G.}\ \bibnamefont
  {Abanov}},\ }\href {https://doi.org/10.1103/PhysRevLett.113.034501}
  {\bibfield  {journal} {\bibinfo  {journal} {Physical Review Letters}\
  }\textbf {\bibinfo {volume} {113}},\ \bibinfo {pages} {034501} (\bibinfo
  {year} {2014})}\BibitemShut {NoStop}\bibitem [{\citenamefont {Zhao}\ \emph {et~al.}(2022)\citenamefont {Zhao},
  \citenamefont {Yang}, \citenamefont {Komura},\ and\ \citenamefont
  {Seto}}]{Zhao2022}\BibitemOpen
  \bibfield  {author} {\bibinfo {author} {\bibfnamefont {Z.}~\bibnamefont
  {Zhao}}, \bibinfo {author} {\bibfnamefont {M.}~\bibnamefont {Yang}}, \bibinfo
  {author} {\bibfnamefont {S.}~\bibnamefont {Komura}},\ and\ \bibinfo {author}
  {\bibfnamefont {R.}~\bibnamefont {Seto}},\ }\bibfield  {journal} {\bibinfo
  {journal} {Frontiers in Physics}\ }\textbf {\bibinfo {volume} {10}},\ \href
  {https://doi.org/10.3389/fphy.2022.951465} {10.3389/fphy.2022.951465}
  (\bibinfo {year} {2022})\BibitemShut {NoStop}\bibitem [{\citenamefont {Banerjee}\ \emph {et~al.}(2017)\citenamefont
  {Banerjee}, \citenamefont {Souslov}, \citenamefont {Abanov},\ and\
  \citenamefont {Vitelli}}]{Banerjee2017}\BibitemOpen
  \bibfield  {author} {\bibinfo {author} {\bibfnamefont {D.}~\bibnamefont
  {Banerjee}}, \bibinfo {author} {\bibfnamefont {A.}~\bibnamefont {Souslov}},
  \bibinfo {author} {\bibfnamefont {A.~G.}\ \bibnamefont {Abanov}},\ and\
  \bibinfo {author} {\bibfnamefont {V.}~\bibnamefont {Vitelli}},\ }\href
  {https://doi.org/10.1038/s41467-017-01378-7} {\bibfield  {journal} {\bibinfo
  {journal} {Nature Communications}\ }\textbf {\bibinfo {volume} {8}},\
  \bibinfo {pages} {1573} (\bibinfo {year} {2017})}\BibitemShut {NoStop}\bibitem [{\citenamefont {Han}\ \emph {et~al.}(2021)\citenamefont {Han},
  \citenamefont {Fruchart}, \citenamefont {Scheibner}, \citenamefont
  {Vaikuntanathan}, \citenamefont {de~Pablo},\ and\ \citenamefont
  {Vitelli}}]{Han2021}\BibitemOpen
  \bibfield  {author} {\bibinfo {author} {\bibfnamefont {M.}~\bibnamefont
  {Han}}, \bibinfo {author} {\bibfnamefont {M.}~\bibnamefont {Fruchart}},
  \bibinfo {author} {\bibfnamefont {C.}~\bibnamefont {Scheibner}}, \bibinfo
  {author} {\bibfnamefont {S.}~\bibnamefont {Vaikuntanathan}}, \bibinfo
  {author} {\bibfnamefont {J.~J.}\ \bibnamefont {de~Pablo}},\ and\ \bibinfo
  {author} {\bibfnamefont {V.}~\bibnamefont {Vitelli}},\ }\href
  {https://doi.org/10.1038/s41567-021-01360-7} {\bibfield  {journal} {\bibinfo
  {journal} {Nature Physics}\ }\textbf {\bibinfo {volume} {17}},\ \bibinfo
  {pages} {1260–1269} (\bibinfo {year} {2021})}\BibitemShut {NoStop}\bibitem [{\citenamefont {Markovich}\ and\ \citenamefont
  {Lubensky}(2021)}]{Markovich2021}\BibitemOpen
  \bibfield  {author} {\bibinfo {author} {\bibfnamefont {T.}~\bibnamefont
  {Markovich}}\ and\ \bibinfo {author} {\bibfnamefont {T.~C.}\ \bibnamefont
  {Lubensky}},\ }\href {https://doi.org/10.1103/physrevlett.127.048001}
  {\bibfield  {journal} {\bibinfo  {journal} {Physical Review Letters}\
  }\textbf {\bibinfo {volume} {127}},\ \bibinfo {pages} {048001} (\bibinfo
  {year} {2021})}\BibitemShut {NoStop}\bibitem [{\citenamefont {Fruchart}\ \emph {et~al.}(2022)\citenamefont
  {Fruchart}, \citenamefont {Han}, \citenamefont {Scheibner},\ and\
  \citenamefont {Vitelli}}]{Fruchart2022}\BibitemOpen
  \bibfield  {author} {\bibinfo {author} {\bibfnamefont {M.}~\bibnamefont
  {Fruchart}}, \bibinfo {author} {\bibfnamefont {M.}~\bibnamefont {Han}},
  \bibinfo {author} {\bibfnamefont {C.}~\bibnamefont {Scheibner}},\ and\
  \bibinfo {author} {\bibfnamefont {V.}~\bibnamefont {Vitelli}},\ }\href@noop
  {} {\bibinfo {title} {The odd ideal gas: Hall viscosity and thermal
  conductivity from non-hermitian kinetic theory}} (\bibinfo {year} {2022}),\
  \Eprint {https://arxiv.org/abs/2202.02037} {arXiv:2202.02037} \BibitemShut
  {NoStop}\bibitem [{\citenamefont {Tsai}\ \emph {et~al.}(2005)\citenamefont {Tsai},
  \citenamefont {Ye}, \citenamefont {Rodriguez}, \citenamefont {Gollub},\ and\
  \citenamefont {Lubensky}}]{Tsai2005}\BibitemOpen
  \bibfield  {author} {\bibinfo {author} {\bibfnamefont {J.-C.}\ \bibnamefont
  {Tsai}}, \bibinfo {author} {\bibfnamefont {F.}~\bibnamefont {Ye}}, \bibinfo
  {author} {\bibfnamefont {J.}~\bibnamefont {Rodriguez}}, \bibinfo {author}
  {\bibfnamefont {J.~P.}\ \bibnamefont {Gollub}},\ and\ \bibinfo {author}
  {\bibfnamefont {T.~C.}\ \bibnamefont {Lubensky}},\ }\href
  {https://doi.org/10.1103/physrevlett.94.214301} {\bibfield  {journal}
  {\bibinfo  {journal} {Physical Review Letters}\ }\textbf {\bibinfo {volume}
  {94}},\ \bibinfo {pages} {214301} (\bibinfo {year} {2005})}\BibitemShut
  {NoStop}\bibitem [{\citenamefont {Grzybowski}\ \emph {et~al.}(2000)\citenamefont
  {Grzybowski}, \citenamefont {Stone},\ and\ \citenamefont
  {Whitesides}}]{Grzybowski2000}\BibitemOpen
  \bibfield  {author} {\bibinfo {author} {\bibfnamefont {B.~A.}\ \bibnamefont
  {Grzybowski}}, \bibinfo {author} {\bibfnamefont {H.~A.}\ \bibnamefont
  {Stone}},\ and\ \bibinfo {author} {\bibfnamefont {G.~M.}\ \bibnamefont
  {Whitesides}},\ }\href {https://doi.org/10.1038/35016528} {\bibfield
  {journal} {\bibinfo  {journal} {Nature}\ }\textbf {\bibinfo {volume} {405}},\
  \bibinfo {pages} {1033–1036} (\bibinfo {year} {2000})}\BibitemShut
  {NoStop}\bibitem [{\citenamefont {Yan}\ \emph {et~al.}(2015)\citenamefont {Yan},
  \citenamefont {Bae},\ and\ \citenamefont {Granick}}]{Yan2015}\BibitemOpen
  \bibfield  {author} {\bibinfo {author} {\bibfnamefont {J.}~\bibnamefont
  {Yan}}, \bibinfo {author} {\bibfnamefont {S.~C.}\ \bibnamefont {Bae}},\ and\
  \bibinfo {author} {\bibfnamefont {S.}~\bibnamefont {Granick}},\ }\href
  {https://doi.org/10.1039/C4SM01962H} {\bibfield  {journal} {\bibinfo
  {journal} {Soft Matter}\ }\textbf {\bibinfo {volume} {11}},\ \bibinfo {pages}
  {147} (\bibinfo {year} {2015})}\BibitemShut {NoStop}\bibitem [{\citenamefont {Bililign}\ \emph {et~al.}(2021)\citenamefont
  {Bililign}, \citenamefont {Balboa~Usabiaga}, \citenamefont {Ganan},
  \citenamefont {Poncet}, \citenamefont {Soni}, \citenamefont {Magkiriadou},
  \citenamefont {Shelley}, \citenamefont {Bartolo},\ and\ \citenamefont
  {Irvine}}]{Bililign2021}\BibitemOpen
  \bibfield  {author} {\bibinfo {author} {\bibfnamefont {E.~S.}\ \bibnamefont
  {Bililign}}, \bibinfo {author} {\bibfnamefont {F.}~\bibnamefont
  {Balboa~Usabiaga}}, \bibinfo {author} {\bibfnamefont {Y.~A.}\ \bibnamefont
  {Ganan}}, \bibinfo {author} {\bibfnamefont {A.}~\bibnamefont {Poncet}},
  \bibinfo {author} {\bibfnamefont {V.}~\bibnamefont {Soni}}, \bibinfo {author}
  {\bibfnamefont {S.}~\bibnamefont {Magkiriadou}}, \bibinfo {author}
  {\bibfnamefont {M.~J.}\ \bibnamefont {Shelley}}, \bibinfo {author}
  {\bibfnamefont {D.}~\bibnamefont {Bartolo}},\ and\ \bibinfo {author}
  {\bibfnamefont {W.~T.~M.}\ \bibnamefont {Irvine}},\ }\href
  {https://doi.org/10.1038/s41567-021-01429-3} {\bibfield  {journal} {\bibinfo
  {journal} {Nature Physics}\ ,\ \bibinfo {pages} {212–218}} (\bibinfo {year}
  {2021})}\BibitemShut {NoStop}\bibitem [{\citenamefont {Tan}\ \emph {et~al.}(2022)\citenamefont {Tan},
  \citenamefont {Mietke}, \citenamefont {Li}, \citenamefont {Chen},
  \citenamefont {Higinbotham}, \citenamefont {Foster}, \citenamefont {Gokhale},
  \citenamefont {Dunkel},\ and\ \citenamefont {Fakhri}}]{Tan2021}\BibitemOpen
  \bibfield  {author} {\bibinfo {author} {\bibfnamefont {T.~H.}\ \bibnamefont
  {Tan}}, \bibinfo {author} {\bibfnamefont {A.}~\bibnamefont {Mietke}},
  \bibinfo {author} {\bibfnamefont {J.}~\bibnamefont {Li}}, \bibinfo {author}
  {\bibfnamefont {Y.}~\bibnamefont {Chen}}, \bibinfo {author} {\bibfnamefont
  {H.}~\bibnamefont {Higinbotham}}, \bibinfo {author} {\bibfnamefont {P.~J.}\
  \bibnamefont {Foster}}, \bibinfo {author} {\bibfnamefont {S.}~\bibnamefont
  {Gokhale}}, \bibinfo {author} {\bibfnamefont {J.}~\bibnamefont {Dunkel}},\
  and\ \bibinfo {author} {\bibfnamefont {N.}~\bibnamefont {Fakhri}},\ }\href
  {https://doi.org/10.1038/s41586-022-04889-6} {\bibfield  {journal} {\bibinfo
  {journal} {Nature}\ }\textbf {\bibinfo {volume} {607}},\ \bibinfo {pages}
  {287–293} (\bibinfo {year} {2022})}\BibitemShut {NoStop}\bibitem [{\citenamefont {Petroff}\ \emph {et~al.}(2015)\citenamefont
  {Petroff}, \citenamefont {Wu},\ and\ \citenamefont
  {Libchaber}}]{Petroff2015}\BibitemOpen
  \bibfield  {author} {\bibinfo {author} {\bibfnamefont {A.~P.}\ \bibnamefont
  {Petroff}}, \bibinfo {author} {\bibfnamefont {X.-L.}\ \bibnamefont {Wu}},\
  and\ \bibinfo {author} {\bibfnamefont {A.}~\bibnamefont {Libchaber}},\ }\href
  {https://doi.org/10.1103/physrevlett.114.158102} {\bibfield  {journal}
  {\bibinfo  {journal} {Physical Review Letters}\ }\textbf {\bibinfo {volume}
  {114}},\ \bibinfo {pages} {158102} (\bibinfo {year} {2015})}\BibitemShut
  {NoStop}\bibitem [{\citenamefont {Ivlev}\ \emph {et~al.}(2012)\citenamefont {Ivlev},
  \citenamefont {Löwen}, \citenamefont {Morfill},\ and\ \citenamefont
  {Royall}}]{Ivlev2012}\BibitemOpen
  \bibfield  {author} {\bibinfo {author} {\bibfnamefont {A.}~\bibnamefont
  {Ivlev}}, \bibinfo {author} {\bibfnamefont {H.}~\bibnamefont {Löwen}},
  \bibinfo {author} {\bibfnamefont {G.}~\bibnamefont {Morfill}},\ and\ \bibinfo
  {author} {\bibfnamefont {C.~P.}\ \bibnamefont {Royall}},\ }\href@noop {}
  {\emph {\bibinfo {title} {Complex Plasmas And Colloidal Dispersions}}}\
  (\bibinfo  {publisher} {World Scientific Publishing Company},\ \bibinfo
  {year} {2012})\BibitemShut {NoStop}\bibitem [{\citenamefont {Ivlev}\ \emph {et~al.}(2015)\citenamefont {Ivlev},
  \citenamefont {Bartnick}, \citenamefont {Heinen}, \citenamefont {Du},
  \citenamefont {Nosenko},\ and\ \citenamefont {Löwen}}]{Ivlev2015}\BibitemOpen
  \bibfield  {author} {\bibinfo {author} {\bibfnamefont {A.}~\bibnamefont
  {Ivlev}}, \bibinfo {author} {\bibfnamefont {J.}~\bibnamefont {Bartnick}},
  \bibinfo {author} {\bibfnamefont {M.}~\bibnamefont {Heinen}}, \bibinfo
  {author} {\bibfnamefont {C.-R.}\ \bibnamefont {Du}}, \bibinfo {author}
  {\bibfnamefont {V.}~\bibnamefont {Nosenko}},\ and\ \bibinfo {author}
  {\bibfnamefont {H.}~\bibnamefont {Löwen}},\ }\href
  {https://doi.org/10.1103/physrevx.5.011035} {\bibfield  {journal} {\bibinfo
  {journal} {Physical Review X}\ }\textbf {\bibinfo {volume} {5}},\ \bibinfo
  {pages} {011035} (\bibinfo {year} {2015})}\BibitemShut {NoStop}\bibitem [{\citenamefont {Denk}\ \emph {et~al.}(2016)\citenamefont {Denk},
  \citenamefont {Huber}, \citenamefont {Reithmann},\ and\ \citenamefont
  {Frey}}]{Denk2016}\BibitemOpen
  \bibfield  {author} {\bibinfo {author} {\bibfnamefont {J.}~\bibnamefont
  {Denk}}, \bibinfo {author} {\bibfnamefont {L.}~\bibnamefont {Huber}},
  \bibinfo {author} {\bibfnamefont {E.}~\bibnamefont {Reithmann}},\ and\
  \bibinfo {author} {\bibfnamefont {E.}~\bibnamefont {Frey}},\ }\href
  {https://doi.org/10.1103/physrevlett.116.178301} {\bibfield  {journal}
  {\bibinfo  {journal} {Physical Review Letters}\ }\textbf {\bibinfo {volume}
  {116}},\ \bibinfo {pages} {178301} (\bibinfo {year} {2016})}\BibitemShut
  {NoStop}\bibitem [{\citenamefont {Liebchen}\ and\ \citenamefont
  {Levis}(2017)}]{Liebchen2017}\BibitemOpen
  \bibfield  {author} {\bibinfo {author} {\bibfnamefont {B.}~\bibnamefont
  {Liebchen}}\ and\ \bibinfo {author} {\bibfnamefont {D.}~\bibnamefont
  {Levis}},\ }\href {https://doi.org/10.1103/physrevlett.119.058002} {\bibfield
   {journal} {\bibinfo  {journal} {Physical Review Letters}\ }\textbf {\bibinfo
  {volume} {119}},\ \bibinfo {pages} {058002} (\bibinfo {year}
  {2017})}\BibitemShut {NoStop}\bibitem [{\citenamefont {Connaughton}\ \emph {et~al.}(2015)\citenamefont
  {Connaughton}, \citenamefont {Nazarenko},\ and\ \citenamefont
  {Quinn}}]{Connaughton2015}\BibitemOpen
  \bibfield  {author} {\bibinfo {author} {\bibfnamefont {C.}~\bibnamefont
  {Connaughton}}, \bibinfo {author} {\bibfnamefont {S.}~\bibnamefont
  {Nazarenko}},\ and\ \bibinfo {author} {\bibfnamefont {B.}~\bibnamefont
  {Quinn}},\ }\href {https://doi.org/10.1016/j.physrep.2015.10.009} {\bibfield
  {journal} {\bibinfo  {journal} {Physics Reports}\ }\textbf {\bibinfo {volume}
  {604}},\ \bibinfo {pages} {1–71} (\bibinfo {year} {2015})}\BibitemShut
  {NoStop}\bibitem [{\citenamefont {Boffetta}\ \emph {et~al.}(2002)\citenamefont
  {Boffetta}, \citenamefont {Lillo},\ and\ \citenamefont
  {Musacchio}}]{Boffetta2002}\BibitemOpen
  \bibfield  {author} {\bibinfo {author} {\bibfnamefont {G.}~\bibnamefont
  {Boffetta}}, \bibinfo {author} {\bibfnamefont {F.~D.}\ \bibnamefont
  {Lillo}},\ and\ \bibinfo {author} {\bibfnamefont {S.}~\bibnamefont
  {Musacchio}},\ }\href {https://doi.org/10.1209/epl/i2002-00180-y} {\bibfield
  {journal} {\bibinfo  {journal} {Europhysics Letters (EPL)}\ }\textbf
  {\bibinfo {volume} {59}},\ \bibinfo {pages} {687–693} (\bibinfo {year}
  {2002})}\BibitemShut {NoStop}\bibitem [{\citenamefont {Tassi}\ \emph {et~al.}(2009)\citenamefont {Tassi},
  \citenamefont {Chandre},\ and\ \citenamefont {Morrison}}]{Tassi2009}\BibitemOpen
  \bibfield  {author} {\bibinfo {author} {\bibfnamefont {E.}~\bibnamefont
  {Tassi}}, \bibinfo {author} {\bibfnamefont {C.}~\bibnamefont {Chandre}},\
  and\ \bibinfo {author} {\bibfnamefont {P.~J.}\ \bibnamefont {Morrison}},\
  }\bibfield  {journal} {\bibinfo  {journal} {Physics of Plasmas}\ }\textbf
  {\bibinfo {volume} {16}},\ \href {https://doi.org/10.1063/1.3194275}
  {10.1063/1.3194275} (\bibinfo {year} {2009})\BibitemShut {NoStop}\bibitem [{\citenamefont {Hasegawa}\ and\ \citenamefont
  {Mima}(1977)}]{Hasegawa1977}\BibitemOpen
  \bibfield  {author} {\bibinfo {author} {\bibfnamefont {A.}~\bibnamefont
  {Hasegawa}}\ and\ \bibinfo {author} {\bibfnamefont {K.}~\bibnamefont
  {Mima}},\ }\href {https://doi.org/10.1103/physrevlett.39.205} {\bibfield
  {journal} {\bibinfo  {journal} {Physical Review Letters}\ }\textbf {\bibinfo
  {volume} {39}},\ \bibinfo {pages} {205–208} (\bibinfo {year}
  {1977})}\BibitemShut {NoStop}\bibitem [{\citenamefont {Charney}(1971)}]{Charney1971}\BibitemOpen
  \bibfield  {author} {\bibinfo {author} {\bibfnamefont {J.~G.}\ \bibnamefont
  {Charney}},\ }\href
  {https://doi.org/10.1175/1520-0469(1971)028<1087:gt>2.0.co;2} {\bibfield
  {journal} {\bibinfo  {journal} {Journal of the Atmospheric Sciences}\
  }\textbf {\bibinfo {volume} {28}},\ \bibinfo {pages} {1087–1095} (\bibinfo
  {year} {1971})}\BibitemShut {NoStop}\bibitem [{\citenamefont {Horton}(1999)}]{Horton1999}\BibitemOpen
  \bibfield  {author} {\bibinfo {author} {\bibfnamefont {W.}~\bibnamefont
  {Horton}},\ }\href {https://doi.org/10.1103/revmodphys.71.735} {\bibfield
  {journal} {\bibinfo  {journal} {Reviews of Modern Physics}\ }\textbf
  {\bibinfo {volume} {71}},\ \bibinfo {pages} {735–778} (\bibinfo {year}
  {1999})}\BibitemShut {NoStop}\bibitem [{\citenamefont {Pedlosky}(1979)}]{Pedlosky1979}\BibitemOpen
  \bibfield  {author} {\bibinfo {author} {\bibfnamefont {J.}~\bibnamefont
  {Pedlosky}},\ }\href {https://doi.org/10.1007/978-1-4684-0071-7} {\emph
  {\bibinfo {title} {Geophysical Fluid Dynamics}}}\ (\bibinfo  {publisher}
  {Springer US},\ \bibinfo {year} {1979})\BibitemShut {NoStop}\bibitem [{\citenamefont {Rhines}(1975)}]{Rhines1975}\BibitemOpen
  \bibfield  {author} {\bibinfo {author} {\bibfnamefont {P.~B.}\ \bibnamefont
  {Rhines}},\ }\href {https://doi.org/10.1017/s0022112075001504} {\bibfield
  {journal} {\bibinfo  {journal} {Journal of Fluid Mechanics}\ }\textbf
  {\bibinfo {volume} {69}},\ \bibinfo {pages} {417} (\bibinfo {year}
  {1975})}\BibitemShut {NoStop}\bibitem [{\citenamefont {Burns}\ \emph {et~al.}(2020)\citenamefont {Burns},
  \citenamefont {Vasil}, \citenamefont {Oishi}, \citenamefont {Lecoanet},\ and\
  \citenamefont {Brown}}]{Burns2020}\BibitemOpen
  \bibfield  {author} {\bibinfo {author} {\bibfnamefont {K.~J.}\ \bibnamefont
  {Burns}}, \bibinfo {author} {\bibfnamefont {G.~M.}\ \bibnamefont {Vasil}},
  \bibinfo {author} {\bibfnamefont {J.~S.}\ \bibnamefont {Oishi}}, \bibinfo
  {author} {\bibfnamefont {D.}~\bibnamefont {Lecoanet}},\ and\ \bibinfo
  {author} {\bibfnamefont {B.~P.}\ \bibnamefont {Brown}},\ }\href
  {https://doi.org/10.1103/physrevresearch.2.023068} {\bibfield  {journal}
  {\bibinfo  {journal} {Physical Review Research}\ }\textbf {\bibinfo {volume}
  {2}},\ \bibinfo {pages} {023068} (\bibinfo {year} {2020})}\BibitemShut
  {NoStop}\bibitem [{\citenamefont {Balk}(1991)}]{Balk1991b}\BibitemOpen
  \bibfield  {author} {\bibinfo {author} {\bibfnamefont {A.}~\bibnamefont
  {Balk}},\ }\href {https://doi.org/10.1016/0375-9601(91)90501-x} {\bibfield
  {journal} {\bibinfo  {journal} {Physics Letters A}\ }\textbf {\bibinfo
  {volume} {155}},\ \bibinfo {pages} {20–24} (\bibinfo {year}
  {1991})}\BibitemShut {NoStop}\bibitem [{\citenamefont {Balk}\ \emph {et~al.}(1991)\citenamefont {Balk},
  \citenamefont {Nazarenko},\ and\ \citenamefont {Zakharov}}]{Balk1991a}\BibitemOpen
  \bibfield  {author} {\bibinfo {author} {\bibfnamefont {A.}~\bibnamefont
  {Balk}}, \bibinfo {author} {\bibfnamefont {S.}~\bibnamefont {Nazarenko}},\
  and\ \bibinfo {author} {\bibfnamefont {V.}~\bibnamefont {Zakharov}},\ }\href
  {https://doi.org/10.1016/0375-9601(91)90105-h} {\bibfield  {journal}
  {\bibinfo  {journal} {Physics Letters A}\ }\textbf {\bibinfo {volume}
  {152}},\ \bibinfo {pages} {276–280} (\bibinfo {year} {1991})}\BibitemShut
  {NoStop}\bibitem [{\citenamefont {Nazarenko}\ and\ \citenamefont
  {Quinn}(2009)}]{Nazarenko2009}\BibitemOpen
  \bibfield  {author} {\bibinfo {author} {\bibfnamefont {S.}~\bibnamefont
  {Nazarenko}}\ and\ \bibinfo {author} {\bibfnamefont {B.}~\bibnamefont
  {Quinn}},\ }\href {https://doi.org/10.1103/physrevlett.103.118501} {\bibfield
   {journal} {\bibinfo  {journal} {Physical Review Letters}\ }\textbf {\bibinfo
  {volume} {103}},\ \bibinfo {pages} {118501} (\bibinfo {year}
  {2009})}\BibitemShut {NoStop}\bibitem [{\citenamefont {Sahoo}\ \emph {et~al.}(2017)\citenamefont {Sahoo},
  \citenamefont {Alexakis},\ and\ \citenamefont {Biferale}}]{Sahoo2017}\BibitemOpen
  \bibfield  {author} {\bibinfo {author} {\bibfnamefont {G.}~\bibnamefont
  {Sahoo}}, \bibinfo {author} {\bibfnamefont {A.}~\bibnamefont {Alexakis}},\
  and\ \bibinfo {author} {\bibfnamefont {L.}~\bibnamefont {Biferale}},\ }\href
  {https://doi.org/10.1103/physrevlett.118.164501} {\bibfield  {journal}
  {\bibinfo  {journal} {Physical Review Letters}\ }\textbf {\bibinfo {volume}
  {118}},\ \bibinfo {pages} {164501} (\bibinfo {year} {2017})}\BibitemShut
  {NoStop}\bibitem [{\citenamefont {Ascher}\ \emph {et~al.}(1997)\citenamefont {Ascher},
  \citenamefont {Ruuth},\ and\ \citenamefont {Spiteri}}]{Ascher1997}\BibitemOpen
  \bibfield  {author} {\bibinfo {author} {\bibfnamefont {U.~M.}\ \bibnamefont
  {Ascher}}, \bibinfo {author} {\bibfnamefont {S.~J.}\ \bibnamefont {Ruuth}},\
  and\ \bibinfo {author} {\bibfnamefont {R.~J.}\ \bibnamefont {Spiteri}},\
  }\href {https://doi.org/10.1016/s0168-9274(97)00056-1} {\bibfield  {journal}
  {\bibinfo  {journal} {Applied Numerical Mathematics}\ }\textbf {\bibinfo
  {volume} {25}},\ \bibinfo {pages} {151–167} (\bibinfo {year}
  {1997})}\BibitemShut {NoStop}\bibitem [{\citenamefont {Smoluchowski}(1916)}]{Smoluchowski1916}\BibitemOpen
  \bibfield  {author} {\bibinfo {author} {\bibfnamefont {M.}~\bibnamefont
  {Smoluchowski}},\ }\href {http://eudml.org/doc/215805} {\bibfield  {journal}
  {\bibinfo  {journal} {Zeitschrift fur Physik}\ }\textbf {\bibinfo {volume}
  {17}},\ \bibinfo {pages} {557} (\bibinfo {year} {1916})}\BibitemShut
  {NoStop}\bibitem [{\citenamefont {Kolmogorov}(1941)}]{Kolmogorov1941}\BibitemOpen
  \bibfield  {author} {\bibinfo {author} {\bibfnamefont {A.~N.}\ \bibnamefont
  {Kolmogorov}},\ }in\ \href@noop {} {\emph {\bibinfo {booktitle} {Dokl. Akad.
  Nauk SSSR}}},\ Vol.~\bibinfo {volume} {31}\ (\bibinfo {year} {1941})\
  p.~\bibinfo {pages} {99}\BibitemShut {NoStop}\bibitem [{\citenamefont {Gorokhovski}\ and\ \citenamefont
  {Herrmann}(2008)}]{Gorokhovski2008}\BibitemOpen
  \bibfield  {author} {\bibinfo {author} {\bibfnamefont {M.}~\bibnamefont
  {Gorokhovski}}\ and\ \bibinfo {author} {\bibfnamefont {M.}~\bibnamefont
  {Herrmann}},\ }\href {https://doi.org/10.1146/annurev.fluid.40.111406.102200}
  {\bibfield  {journal} {\bibinfo  {journal} {Annual Review of Fluid
  Mechanics}\ }\textbf {\bibinfo {volume} {40}},\ \bibinfo {pages} {343–366}
  (\bibinfo {year} {2008})}\BibitemShut {NoStop}\bibitem [{\citenamefont {Brilliantov}\ \emph {et~al.}(2015)\citenamefont
  {Brilliantov}, \citenamefont {Krapivsky}, \citenamefont {Bodrova},
  \citenamefont {Spahn}, \citenamefont {Hayakawa}, \citenamefont {Stadnichuk},\
  and\ \citenamefont {Schmidt}}]{Brilliantov2015}\BibitemOpen
  \bibfield  {author} {\bibinfo {author} {\bibfnamefont {N.}~\bibnamefont
  {Brilliantov}}, \bibinfo {author} {\bibfnamefont {P.~L.}\ \bibnamefont
  {Krapivsky}}, \bibinfo {author} {\bibfnamefont {A.}~\bibnamefont {Bodrova}},
  \bibinfo {author} {\bibfnamefont {F.}~\bibnamefont {Spahn}}, \bibinfo
  {author} {\bibfnamefont {H.}~\bibnamefont {Hayakawa}}, \bibinfo {author}
  {\bibfnamefont {V.}~\bibnamefont {Stadnichuk}},\ and\ \bibinfo {author}
  {\bibfnamefont {J.}~\bibnamefont {Schmidt}},\ }\href
  {https://doi.org/10.1073/pnas.1503957112} {\bibfield  {journal} {\bibinfo
  {journal} {Proceedings of the National Academy of Sciences}\ }\textbf
  {\bibinfo {volume} {112}},\ \bibinfo {pages} {9536–9541} (\bibinfo {year}
  {2015})}\BibitemShut {NoStop}\bibitem [{\citenamefont {Cheng}\ and\ \citenamefont
  {Redner}(1990)}]{Cheng1990}\BibitemOpen
  \bibfield  {author} {\bibinfo {author} {\bibfnamefont {Z.}~\bibnamefont
  {Cheng}}\ and\ \bibinfo {author} {\bibfnamefont {S.}~\bibnamefont {Redner}},\
  }\href {https://doi.org/10.1088/0305-4470/23/7/028} {\bibfield  {journal}
  {\bibinfo  {journal} {Journal of Physics A: Mathematical and General}\
  }\textbf {\bibinfo {volume} {23}},\ \bibinfo {pages} {1233–1258} (\bibinfo
  {year} {1990})}\BibitemShut {NoStop}\bibitem [{\citenamefont {Brilliantov}\ \emph {et~al.}(2021)\citenamefont
  {Brilliantov}, \citenamefont {Otieno},\ and\ \citenamefont
  {Krapivsky}}]{Brilliantov2021}\BibitemOpen
  \bibfield  {author} {\bibinfo {author} {\bibfnamefont {N.~V.}\ \bibnamefont
  {Brilliantov}}, \bibinfo {author} {\bibfnamefont {W.}~\bibnamefont
  {Otieno}},\ and\ \bibinfo {author} {\bibfnamefont {P.~L.}\ \bibnamefont
  {Krapivsky}},\ }\href {https://doi.org/10.1103/physrevlett.127.250602}
  {\bibfield  {journal} {\bibinfo  {journal} {Physical Review Letters}\
  }\textbf {\bibinfo {volume} {127}},\ \bibinfo {pages} {250602} (\bibinfo
  {year} {2021})}\BibitemShut {NoStop}\bibitem [{\citenamefont {Leyvraz}(2003)}]{Leyvraz2003}\BibitemOpen
  \bibfield  {author} {\bibinfo {author} {\bibfnamefont {F.}~\bibnamefont
  {Leyvraz}},\ }\href {https://doi.org/10.1016/s0370-1573(03)00241-2}
  {\bibfield  {journal} {\bibinfo  {journal} {Physics Reports}\ }\textbf
  {\bibinfo {volume} {383}},\ \bibinfo {pages} {95–212} (\bibinfo {year}
  {2003})}\BibitemShut {NoStop}\bibitem [{\citenamefont {Wattis}(2006)}]{Wattis2006}\BibitemOpen
  \bibfield  {author} {\bibinfo {author} {\bibfnamefont {J.~A.}\ \bibnamefont
  {Wattis}},\ }\href {https://doi.org/10.1016/j.physd.2006.07.024} {\bibfield
  {journal} {\bibinfo  {journal} {Physica D: Nonlinear Phenomena}\ }\textbf
  {\bibinfo {volume} {222}},\ \bibinfo {pages} {1–20} (\bibinfo {year}
  {2006})}\BibitemShut {NoStop}\bibitem [{\citenamefont {Ramkrishna}\ and\ \citenamefont
  {Singh}(2014)}]{Ramkrishna2014}\BibitemOpen
  \bibfield  {author} {\bibinfo {author} {\bibfnamefont {D.}~\bibnamefont
  {Ramkrishna}}\ and\ \bibinfo {author} {\bibfnamefont {M.~R.}\ \bibnamefont
  {Singh}},\ }\href {https://doi.org/10.1146/annurev-chembioeng-060713-040241}
  {\bibfield  {journal} {\bibinfo  {journal} {Annual Review of Chemical and
  Biomolecular Engineering}\ }\textbf {\bibinfo {volume} {5}},\ \bibinfo
  {pages} {123–146} (\bibinfo {year} {2014})}\BibitemShut {NoStop}\bibitem [{\citenamefont {Connaughton}\ \emph {et~al.}(2004)\citenamefont
  {Connaughton}, \citenamefont {Rajesh},\ and\ \citenamefont
  {Zaboronski}}]{Connaughton2004}\BibitemOpen
  \bibfield  {author} {\bibinfo {author} {\bibfnamefont {C.}~\bibnamefont
  {Connaughton}}, \bibinfo {author} {\bibfnamefont {R.}~\bibnamefont
  {Rajesh}},\ and\ \bibinfo {author} {\bibfnamefont {O.}~\bibnamefont
  {Zaboronski}},\ }\href {https://doi.org/10.1103/physreve.69.061114}
  {\bibfield  {journal} {\bibinfo  {journal} {Physical Review E}\ }\textbf
  {\bibinfo {volume} {69}},\ \bibinfo {pages} {061114} (\bibinfo {year}
  {2004})}\BibitemShut {NoStop}\bibitem [{\citenamefont {Connaughton}\ \emph {et~al.}(2006)\citenamefont
  {Connaughton}, \citenamefont {Rajesh},\ and\ \citenamefont
  {Zaboronski}}]{Connaughton2006}\BibitemOpen
  \bibfield  {author} {\bibinfo {author} {\bibfnamefont {C.}~\bibnamefont
  {Connaughton}}, \bibinfo {author} {\bibfnamefont {R.}~\bibnamefont
  {Rajesh}},\ and\ \bibinfo {author} {\bibfnamefont {O.}~\bibnamefont
  {Zaboronski}},\ }\href {https://doi.org/10.1016/j.physd.2006.08.005}
  {\bibfield  {journal} {\bibinfo  {journal} {Physica D: Nonlinear Phenomena}\
  }\textbf {\bibinfo {volume} {222}},\ \bibinfo {pages} {97–115} (\bibinfo
  {year} {2006})}\BibitemShut {NoStop}\bibitem [{\citenamefont {Connaughton}\ \emph {et~al.}(2018)\citenamefont
  {Connaughton}, \citenamefont {Dutta}, \citenamefont {Rajesh}, \citenamefont
  {Siddharth},\ and\ \citenamefont {Zaboronski}}]{Connaughton2018}\BibitemOpen
  \bibfield  {author} {\bibinfo {author} {\bibfnamefont {C.}~\bibnamefont
  {Connaughton}}, \bibinfo {author} {\bibfnamefont {A.}~\bibnamefont {Dutta}},
  \bibinfo {author} {\bibfnamefont {R.}~\bibnamefont {Rajesh}}, \bibinfo
  {author} {\bibfnamefont {N.}~\bibnamefont {Siddharth}},\ and\ \bibinfo
  {author} {\bibfnamefont {O.}~\bibnamefont {Zaboronski}},\ }\href
  {https://doi.org/10.1103/physreve.97.022137} {\bibfield  {journal} {\bibinfo
  {journal} {Physical Review E}\ }\textbf {\bibinfo {volume} {97}},\ \bibinfo
  {pages} {022137} (\bibinfo {year} {2018})}\BibitemShut {NoStop}\bibitem [{\citenamefont {Srivastava}(1971)}]{Srivastava1971}\BibitemOpen
  \bibfield  {author} {\bibinfo {author} {\bibfnamefont {R.~C.}\ \bibnamefont
  {Srivastava}},\ }\href
  {https://doi.org/10.1175/1520-0469(1971)028<0410:sdorgb>2.0.co;2} {\bibfield
  {journal} {\bibinfo  {journal} {Journal of the Atmospheric Sciences}\
  }\textbf {\bibinfo {volume} {28}},\ \bibinfo {pages} {410–415} (\bibinfo
  {year} {1971})}\BibitemShut {NoStop}\bibitem [{\citenamefont {Testik}\ and\ \citenamefont
  {Gebremichael}(2013)}]{Testik2013}\BibitemOpen
  \bibfield  {author} {\bibinfo {author} {\bibfnamefont {F.}~\bibnamefont
  {Testik}}\ and\ \bibinfo {author} {\bibfnamefont {M.~E.}\ \bibnamefont
  {Gebremichael}},\ }\href {https://doi.org/10.1029/gm191} {\emph {\bibinfo
  {title} {Rainfall: State of the Science}}},\ Geophysical Monograph Series\
  (\bibinfo  {publisher} {Wiley},\ \bibinfo {year} {2013})\BibitemShut
  {NoStop}\bibitem [{\citenamefont {Pumir}\ and\ \citenamefont
  {Wilkinson}(2016)}]{Pumir2016}\BibitemOpen
  \bibfield  {author} {\bibinfo {author} {\bibfnamefont {A.}~\bibnamefont
  {Pumir}}\ and\ \bibinfo {author} {\bibfnamefont {M.}~\bibnamefont
  {Wilkinson}},\ }\href
  {https://doi.org/10.1146/annurev-conmatphys-031115-011538} {\bibfield
  {journal} {\bibinfo  {journal} {Annual Review of Condensed Matter Physics}\
  }\textbf {\bibinfo {volume} {7}},\ \bibinfo {pages} {141–170} (\bibinfo
  {year} {2016})}\BibitemShut {NoStop}\bibitem [{\citenamefont {Babler}\ \emph {et~al.}(2012)\citenamefont {Babler},
  \citenamefont {Biferale},\ and\ \citenamefont {Lanotte}}]{Babler2012}\BibitemOpen
  \bibfield  {author} {\bibinfo {author} {\bibfnamefont {M.~U.}\ \bibnamefont
  {Babler}}, \bibinfo {author} {\bibfnamefont {L.}~\bibnamefont {Biferale}},\
  and\ \bibinfo {author} {\bibfnamefont {A.~S.}\ \bibnamefont {Lanotte}},\
  }\href {https://doi.org/10.1103/physreve.85.025301} {\bibfield  {journal}
  {\bibinfo  {journal} {Physical Review E}\ }\textbf {\bibinfo {volume} {85}},\
  \bibinfo {pages} {025301} (\bibinfo {year} {2012})}\BibitemShut {NoStop}\bibitem [{\citenamefont {Grabowski}\ and\ \citenamefont
  {Wang}(2013)}]{Grabowski2013}\BibitemOpen
  \bibfield  {author} {\bibinfo {author} {\bibfnamefont {W.~W.}\ \bibnamefont
  {Grabowski}}\ and\ \bibinfo {author} {\bibfnamefont {L.-P.}\ \bibnamefont
  {Wang}},\ }\href {https://doi.org/10.1146/annurev-fluid-011212-140750}
  {\bibfield  {journal} {\bibinfo  {journal} {Annual Review of Fluid
  Mechanics}\ }\textbf {\bibinfo {volume} {45}},\ \bibinfo {pages} {293–324}
  (\bibinfo {year} {2013})}\BibitemShut {NoStop}\bibitem [{\citenamefont {Villermaux}(2007)}]{Villermaux2007}\BibitemOpen
  \bibfield  {author} {\bibinfo {author} {\bibfnamefont {E.}~\bibnamefont
  {Villermaux}},\ }\href
  {https://doi.org/10.1146/annurev.fluid.39.050905.110214} {\bibfield
  {journal} {\bibinfo  {journal} {Annual Review of Fluid Mechanics}\ }\textbf
  {\bibinfo {volume} {39}},\ \bibinfo {pages} {419–446} (\bibinfo {year}
  {2007})}\BibitemShut {NoStop}\bibitem [{\citenamefont {Falkovich}\ \emph {et~al.}(2002)\citenamefont
  {Falkovich}, \citenamefont {Fouxon},\ and\ \citenamefont
  {Stepanov}}]{Falkovich2002}\BibitemOpen
  \bibfield  {author} {\bibinfo {author} {\bibfnamefont {G.}~\bibnamefont
  {Falkovich}}, \bibinfo {author} {\bibfnamefont {A.}~\bibnamefont {Fouxon}},\
  and\ \bibinfo {author} {\bibfnamefont {M.~G.}\ \bibnamefont {Stepanov}},\
  }\href {https://doi.org/10.1038/nature00983} {\bibfield  {journal} {\bibinfo
  {journal} {Nature}\ }\textbf {\bibinfo {volume} {419}},\ \bibinfo {pages}
  {151–154} (\bibinfo {year} {2002})}\BibitemShut {NoStop}\bibitem [{\citenamefont {Bec}\ \emph {et~al.}(2010)\citenamefont {Bec},
  \citenamefont {Biferale}, \citenamefont {Cencini}, \citenamefont {Lanotte},\
  and\ \citenamefont {Toschi}}]{Bec2010}\BibitemOpen
  \bibfield  {author} {\bibinfo {author} {\bibfnamefont {J.}~\bibnamefont
  {Bec}}, \bibinfo {author} {\bibfnamefont {L.}~\bibnamefont {Biferale}},
  \bibinfo {author} {\bibfnamefont {M.}~\bibnamefont {Cencini}}, \bibinfo
  {author} {\bibfnamefont {A.~S.}\ \bibnamefont {Lanotte}},\ and\ \bibinfo
  {author} {\bibfnamefont {F.}~\bibnamefont {Toschi}},\ }\href
  {https://doi.org/10.1017/s0022112010000029} {\bibfield  {journal} {\bibinfo
  {journal} {Journal of Fluid Mechanics}\ }\textbf {\bibinfo {volume} {646}},\
  \bibinfo {pages} {527–536} (\bibinfo {year} {2010})}\BibitemShut {NoStop}\bibitem [{\citenamefont {Rackauckas}\ and\ \citenamefont
  {Nie}(2017)}]{Rackauckas2017}\BibitemOpen
  \bibfield  {author} {\bibinfo {author} {\bibfnamefont {C.}~\bibnamefont
  {Rackauckas}}\ and\ \bibinfo {author} {\bibfnamefont {Q.}~\bibnamefont
  {Nie}},\ }\href@noop {} {\bibfield  {journal} {\bibinfo  {journal} {Journal
  of Open Research Software}\ }\textbf {\bibinfo {volume} {5}} (\bibinfo {year}
  {2017})}\BibitemShut {NoStop}\bibitem [{\citenamefont {Ball}\ \emph {et~al.}(2012)\citenamefont {Ball},
  \citenamefont {Connaughton}, \citenamefont {Jones}, \citenamefont {Rajesh},\
  and\ \citenamefont {Zaboronski}}]{Ball2012}\BibitemOpen
  \bibfield  {author} {\bibinfo {author} {\bibfnamefont {R.~C.}\ \bibnamefont
  {Ball}}, \bibinfo {author} {\bibfnamefont {C.}~\bibnamefont {Connaughton}},
  \bibinfo {author} {\bibfnamefont {P.~P.}\ \bibnamefont {Jones}}, \bibinfo
  {author} {\bibfnamefont {R.}~\bibnamefont {Rajesh}},\ and\ \bibinfo {author}
  {\bibfnamefont {O.}~\bibnamefont {Zaboronski}},\ }\href
  {https://doi.org/10.1103/physrevlett.109.168304} {\bibfield  {journal}
  {\bibinfo  {journal} {Physical Review Letters}\ }\textbf {\bibinfo {volume}
  {109}},\ \bibinfo {pages} {168304} (\bibinfo {year} {2012})}\BibitemShut
  {NoStop}\bibitem [{\citenamefont {Matveev}\ \emph {et~al.}(2017)\citenamefont
  {Matveev}, \citenamefont {Krapivsky}, \citenamefont {Smirnov}, \citenamefont
  {Tyrtyshnikov},\ and\ \citenamefont {Brilliantov}}]{Matveev2017}\BibitemOpen
  \bibfield  {author} {\bibinfo {author} {\bibfnamefont {S.~A.}\ \bibnamefont
  {Matveev}}, \bibinfo {author} {\bibfnamefont {P.~L.}\ \bibnamefont
  {Krapivsky}}, \bibinfo {author} {\bibfnamefont {A.~P.}\ \bibnamefont
  {Smirnov}}, \bibinfo {author} {\bibfnamefont {E.~E.}\ \bibnamefont
  {Tyrtyshnikov}},\ and\ \bibinfo {author} {\bibfnamefont {N.~V.}\ \bibnamefont
  {Brilliantov}},\ }\href {https://doi.org/10.1103/physrevlett.119.260601}
  {\bibfield  {journal} {\bibinfo  {journal} {Physical Review Letters}\
  }\textbf {\bibinfo {volume} {119}},\ \bibinfo {pages} {260601} (\bibinfo
  {year} {2017})}\BibitemShut {NoStop}\bibitem [{\citenamefont {Politi}\ and\ \citenamefont
  {Misbah}(2006)}]{Politi2006}\BibitemOpen
  \bibfield  {author} {\bibinfo {author} {\bibfnamefont {P.}~\bibnamefont
  {Politi}}\ and\ \bibinfo {author} {\bibfnamefont {C.}~\bibnamefont
  {Misbah}},\ }\href {https://doi.org/10.1103/physreve.73.036133} {\bibfield
  {journal} {\bibinfo  {journal} {Physical Review E}\ }\textbf {\bibinfo
  {volume} {73}},\ \bibinfo {pages} {036133} (\bibinfo {year}
  {2006})}\BibitemShut {NoStop}\bibitem [{\citenamefont {Ginot}\ \emph {et~al.}(2018)\citenamefont {Ginot},
  \citenamefont {Theurkauff}, \citenamefont {Detcheverry}, \citenamefont
  {Ybert},\ and\ \citenamefont {Cottin-Bizonne}}]{Ginot2018}\BibitemOpen
  \bibfield  {author} {\bibinfo {author} {\bibfnamefont {F.}~\bibnamefont
  {Ginot}}, \bibinfo {author} {\bibfnamefont {I.}~\bibnamefont {Theurkauff}},
  \bibinfo {author} {\bibfnamefont {F.}~\bibnamefont {Detcheverry}}, \bibinfo
  {author} {\bibfnamefont {C.}~\bibnamefont {Ybert}},\ and\ \bibinfo {author}
  {\bibfnamefont {C.}~\bibnamefont {Cottin-Bizonne}},\ }\bibfield  {journal}
  {\bibinfo  {journal} {Nature Communications}\ }\textbf {\bibinfo {volume}
  {9}},\ \href {https://doi.org/10.1038/s41467-017-02625-7}
  {10.1038/s41467-017-02625-7} (\bibinfo {year} {2018})\BibitemShut {NoStop}\bibitem [{\citenamefont {Brauns}\ \emph {et~al.}(2021)\citenamefont {Brauns},
  \citenamefont {Weyer}, \citenamefont {Halatek}, \citenamefont {Yoon},\ and\
  \citenamefont {Frey}}]{Brauns2021}\BibitemOpen
  \bibfield  {author} {\bibinfo {author} {\bibfnamefont {F.}~\bibnamefont
  {Brauns}}, \bibinfo {author} {\bibfnamefont {H.}~\bibnamefont {Weyer}},
  \bibinfo {author} {\bibfnamefont {J.}~\bibnamefont {Halatek}}, \bibinfo
  {author} {\bibfnamefont {J.}~\bibnamefont {Yoon}},\ and\ \bibinfo {author}
  {\bibfnamefont {E.}~\bibnamefont {Frey}},\ }\href
  {https://doi.org/10.1103/physrevlett.126.104101} {\bibfield  {journal}
  {\bibinfo  {journal} {Physical Review Letters}\ }\textbf {\bibinfo {volume}
  {126}},\ \bibinfo {pages} {104101} (\bibinfo {year} {2021})}\BibitemShut
  {NoStop}\bibitem [{\citenamefont {Ferraro}\ \emph {et~al.}(2023)\citenamefont
  {Ferraro}, \citenamefont {Mangini}, \citenamefont {Zitelli},\ and\
  \citenamefont {Wabnitz}}]{Ferraro2023}\BibitemOpen
  \bibfield  {author} {\bibinfo {author} {\bibfnamefont {M.}~\bibnamefont
  {Ferraro}}, \bibinfo {author} {\bibfnamefont {F.}~\bibnamefont {Mangini}},
  \bibinfo {author} {\bibfnamefont {M.}~\bibnamefont {Zitelli}},\ and\ \bibinfo
  {author} {\bibfnamefont {S.}~\bibnamefont {Wabnitz}},\ }\href
  {https://doi.org/10.1080/23746149.2023.2228018} {\bibfield  {journal}
  {\bibinfo  {journal} {Advances in Physics: X}\ }\textbf {\bibinfo {volume}
  {8}},\ \bibinfo {pages} {2228018} (\bibinfo {year} {2023})},\ \Eprint
  {https://arxiv.org/abs/2304.12006} {2304.12006} \BibitemShut {NoStop}\bibitem [{\citenamefont {Loman}\ \emph {et~al.}(2023)\citenamefont {Loman},
  \citenamefont {Ma}, \citenamefont {Ilin}, \citenamefont {Gowda},
  \citenamefont {Korsbo}, \citenamefont {Yewale}, \citenamefont {Rackauckas},\
  and\ \citenamefont {Isaacson}}]{Loman2022}\BibitemOpen
  \bibfield  {author} {\bibinfo {author} {\bibfnamefont {T.}~\bibnamefont
  {Loman}}, \bibinfo {author} {\bibfnamefont {Y.}~\bibnamefont {Ma}}, \bibinfo
  {author} {\bibfnamefont {V.}~\bibnamefont {Ilin}}, \bibinfo {author}
  {\bibfnamefont {S.}~\bibnamefont {Gowda}}, \bibinfo {author} {\bibfnamefont
  {N.}~\bibnamefont {Korsbo}}, \bibinfo {author} {\bibfnamefont
  {N.}~\bibnamefont {Yewale}}, \bibinfo {author} {\bibfnamefont {C.~V.}\
  \bibnamefont {Rackauckas}},\ and\ \bibinfo {author} {\bibfnamefont {S.~A.}\
  \bibnamefont {Isaacson}},\ }\href
  {https://doi.org/10.1371/journal.pcbi.1011530} {\bibfield  {journal}
  {\bibinfo  {journal} {PLoS Computational Biology}\ }\textbf {\bibinfo
  {volume} {19}},\ \bibinfo {pages} {e1011530} (\bibinfo {year}
  {2023})}\BibitemShut {NoStop}\bibitem [{\citenamefont {Gaspard}(2022)}]{Gaspard2022}\BibitemOpen
  \bibfield  {author} {\bibinfo {author} {\bibfnamefont {P.}~\bibnamefont
  {Gaspard}},\ }\href@noop {} {\emph {\bibinfo {title} {The Statistical
  Mechanics of Irreversible Phenomena}}}\ (\bibinfo  {publisher} {Cambridge
  University Press},\ \bibinfo {year} {2022})\BibitemShut {NoStop}\bibitem [{\citenamefont {Kondepudi}\ and\ \citenamefont
  {Prigogine}(2014)}]{Kondepudi2014}\BibitemOpen
  \bibfield  {author} {\bibinfo {author} {\bibfnamefont {D.}~\bibnamefont
  {Kondepudi}}\ and\ \bibinfo {author} {\bibfnamefont {I.}~\bibnamefont
  {Prigogine}},\ }\href {https://doi.org/10.1002/9781118698723} {\emph
  {\bibinfo {title} {Modern Thermodynamics}}}\ (\bibinfo  {publisher} {Wiley},\
  \bibinfo {year} {2014})\BibitemShut {NoStop}\bibitem [{\citenamefont {Schnakenberg}(1976)}]{Schnakenberg1976}\BibitemOpen
  \bibfield  {author} {\bibinfo {author} {\bibfnamefont {J.}~\bibnamefont
  {Schnakenberg}},\ }\href {https://doi.org/10.1103/revmodphys.48.571}
  {\bibfield  {journal} {\bibinfo  {journal} {Reviews of Modern Physics}\
  }\textbf {\bibinfo {volume} {48}},\ \bibinfo {pages} {571–585} (\bibinfo
  {year} {1976})}\BibitemShut {NoStop}\bibitem [{\citenamefont {Rao}\ and\ \citenamefont {Esposito}(2016)}]{Rao2016}\BibitemOpen
  \bibfield  {author} {\bibinfo {author} {\bibfnamefont {R.}~\bibnamefont
  {Rao}}\ and\ \bibinfo {author} {\bibfnamefont {M.}~\bibnamefont {Esposito}},\
  }\href {https://doi.org/10.1103/physrevx.6.041064} {\bibfield  {journal}
  {\bibinfo  {journal} {Physical Review X}\ }\textbf {\bibinfo {volume} {6}},\
  \bibinfo {pages} {041064} (\bibinfo {year} {2016})}\BibitemShut {NoStop}\bibitem [{\citenamefont {Taylor}(1923)}]{Taylor1923}\BibitemOpen
  \bibfield  {author} {\bibinfo {author} {\bibfnamefont {G.~I.}\ \bibnamefont
  {Taylor}},\ }\href {https://doi.org/10.1098/rspa.1923.0103} {\bibfield
  {journal} {\bibinfo  {journal} {Proceedings of the Royal Society of London}\
  }\textbf {\bibinfo {volume} {104}},\ \bibinfo {pages} {213} (\bibinfo {year}
  {1923})}\BibitemShut {NoStop}\bibitem [{\citenamefont {Proudman}(1916)}]{Proudman1916}\BibitemOpen
  \bibfield  {author} {\bibinfo {author} {\bibfnamefont {J.}~\bibnamefont
  {Proudman}},\ }\href {https://doi.org/10.1098/rspa.1916.0026} {\bibfield
  {journal} {\bibinfo  {journal} {Proceedings of the Royal Society of London}\
  }\textbf {\bibinfo {volume} {92}},\ \bibinfo {pages} {408} (\bibinfo {year}
  {1916})}\BibitemShut {NoStop}\bibitem [{{\relax DLMF}()}]{NIST_DLMF}\BibitemOpen
  {\relax DLMF},\ \href {https://dlmf.nist.gov/} {\bibinfo {title} {{\it NIST
  Digital Library of Mathematical Functions}}},\ \bibinfo {howpublished}
  {\url{https://dlmf.nist.gov/}, Release 1.1.10 of 2023-06-15} (\bibinfo {year}
  {2023}),\ \bibinfo {note} {f.~W.~J. Olver, A.~B. {Olde Daalhuis}, D.~W.
  Lozier, B.~I. Schneider, R.~F. Boisvert, C.~W. Clark, B.~R. Miller, B.~V.
  Saunders, H.~S. Cohl, and M.~A. McClain, eds.}\BibitemShut {Stop}\end{thebibliography}
\end{document}